\documentclass[aps,pra,reprint,notitlepage,superscriptaddress,nofootinbib, 10pt, onecolumn]{revtex4-1}
\pdfoutput=1

\usepackage[utf8]{inputenc}
\usepackage{amsmath}
\usepackage{url}
\usepackage{braket}
\usepackage{natbib}
\setcitestyle{numbers,sort&compress}
\usepackage{array}   % for \newcolumntype macro
\newcolumntype{L}{>{$}l<{$}} % math-mode version of "l" column type

\usepackage{xcolor}
\usepackage{hyperref}
\usepackage{natbib}

\usepackage{qcircuit}
\usepackage{mathdots}
\usepackage{hhline}
\usepackage{siunitx}
\usepackage{graphicx}
\setkeys{Gin}{width=\linewidth,totalheight=\textheight,keepaspectratio}
\graphicspath{{figures/}}

\usepackage{amssymb, amsthm, bm}
\usepackage[ruled]{algorithm2e}

\usepackage{empheq}
\usepackage[most]{tcolorbox}
\newtcbox{\mymath}[1][]{%
    nobeforeafter, math upper, tcbox raise base,
    enhanced, colframe=blue!30!black,
    colback=white, boxrule=1pt,
    #1}
	
\usepackage{acronym}
\newacro{QSP}{quantum signal processing}
\newacro{SOCP}{second order cone program}
\newacro{QSVT}{quantum singular value transform}
\newacro{SVD}{singular value decomposition}
\newacro{QRAM}{quantum random access memory}

\newcommand{\bigo}[1]{\mathcal{O}(#1)}

\newcommand{\acos}{\cos ^{-1}}

\newcommand{\polylog}{\text{polylog}}
\newcommand{\controlslash}{*!<0em,0em>-=-<0em>{\oslash}}
\newcommand{\ctrlslash}[1]{\controlslash \qwx[#1] \qw}

\newcommand{\outputgroupv}[6]{\POS"#1,#3"."#2,#3"."#1,#3"."#2,#3"!C*+<#4>\frm{\}}, \POS"#1,#3"."#2,#3"."#1,#3"."#2,#3"*!C!<-1.7em,#5>=<0em>{#6}}
\newcommand{\shiftright}[2]{\makebox[#1][r]{\makebox[0pt][l]{#2}}}
\newcommand{\shiftleft}[2]{\makebox[0pt][r]{\makebox[#1][l]{#2}}}

\newcommand{\numq}[2]{\raisebox{#1}{/${}^{#2}$}}
\newcommand{\rb}[2]{\raisebox{#1}{#2}}
\newcommand{\rotb}[2]{\rotatebox{#1}{#2}}
\newcommand{\mltg}[2]{\multigate{#1}{#2}}
\newcommand{\LOADSS}{\text{LOAD}_\text{ss}}
\newcommand{\LOADBB}{\text{LOAD}_\text{bb}}
\newcommand{\Ryj}[1]{R_y(\theta^{({#1})})}
\newcommand{\SwapD}{$\text{Swap}^\dagger$}

\newcommand{\qvdots}{% \vdots for qcircuit
  \raisebox{0.6em}{\ensuremath{\vdots}}%
}

\DeclareMathOperator{\sgn}{sgn}

\begin{document}

\title{Quantum Resources Required to Block-Encode a Matrix of Classical Data}

\author{B.~David Clader}
\affiliation{Goldman Sachs \& Co., New York, NY, USA}
\email{dave.clader@gs.com}

\author{Alexander M.~Dalzell}
\affiliation{AWS Center for Quantum Computing, Pasadena, CA, USA}
\affiliation{California Institute of Technology, Pasadena, CA, USA}

\author{Nikitas Stamatopoulos}
\affiliation{Goldman Sachs \& Co., New York, NY, USA}

\author{Grant Salton}
\affiliation{Amazon Quantum Solutions Lab, Seattle, WA, USA}
\affiliation{AWS Center for Quantum Computing, Pasadena, CA, USA}
\affiliation{California Institute of Technology, Pasadena, CA, USA}

\author{Mario Berta}
\affiliation{AWS Center for Quantum Computing, Pasadena, CA, USA}
\affiliation{California Institute of Technology, Pasadena, CA, USA}
\affiliation{Department of Computing, Imperial College London, London, UK}

\author{William J.~Zeng}
\affiliation{Goldman Sachs \& Co., New York, NY, USA}

\begin{abstract}
We provide modular circuit-level implementations and resource estimates for several methods of block-encoding a dense $N\times N$ matrix of classical data to precision $\epsilon$; the minimal-depth method achieves a $T$-depth of $\bigo{\log (N/\epsilon)},$ while the minimal-count method achieves a $T$-count of $\bigo{N\log(1/\epsilon)}$. We examine resource tradeoffs between the different approaches, and we explore implementations of two separate models of quantum random access memory (QRAM). As part of this analysis, we provide a novel state preparation routine with $T$-depth $\bigo{\log (N/\epsilon)}$, improving on previous constructions with scaling $\bigo{\log^2 (N/\epsilon)}$. Our results go beyond simple query complexity and provide a clear picture into the resource costs when large amounts of classical data are assumed to be accessible to quantum algorithms.
\end{abstract}

\maketitle

%%%%%%%%%%%%%%%%%%%%%%%%%%%%%%%%%%%%%%%%%%%%
%                                          %
%              INTRODUCTION                %
%                                          %
%%%%%%%%%%%%%%%%%%%%%%%%%%%%%%%%%%%%%%%%%%%%

\section{Introduction}\label{sec:introduction}

\subsection{Motivation}\label{sec:introduction-motivation}

A commonly used access model for quantum algorithms requiring classical data is a so-called ``\emph{block-encoding}'' of the classical data into a unitary operator. Costs associated with these algorithms are often quoted in terms of the asymptotic scaling of the number of queries to the block-encoding oracle. However, if we are to make assessments about the potential of \ac{QRAM}-based algorithms, we must take into account the cost of each block-encoding query as well. In this work, we provide a detailed implementation and resource comparison of different methods one might use to block-encode a dense matrix $A$ representing classical data.

Block-encodings for $N \times N$ matrices\footnote{Block-encodings for non-square matrices are a straightforward extension of the square case. We show how to do this in App.~\ref{app:off-diagonal_matrices}.} are useful because they can, in principle, be implemented in exponentially small time (i.e., with quantum circuits of depth only $\polylog(N)$ \cite{chakraborty2019}), albeit still with space cost of $\bigo{N^2}$ qubits. The block-encoding framework has proven useful for quantum algorithms for a variety of applications. Together with insights from adiabatic quantum computing, block-encoding has been crucial to the discovery of quantum linear system solvers with provably optimal scaling with respect to relevant parameters \cite{gilyen2019,PhysRevLett.122.060504, costa2021optimal,an2022}. The framework has led to faster algorithms for phase estimation \cite{Rall2021fastercoherent}, quantum gradients \cite{gilyen2019optimizing}, and improved Hamiltonian simulation and regression techniques \cite{low2019hamiltonian,chakraborty2019}. Furthermore, block-encoding has been used to analyze quantum optimization algorithms \cite{Kerenidis2020lpsdp,augustino2021quantum} and related algorithms for portfolio optimization \cite{kerenidis2019quantum}. Many of these algorithms make use of the \ac{QSVT} algorithm \cite{gilyen2019,martyn2021}, which uses a technique known as \ac{QSP} \cite{low2016qsp,low2019hamiltonian} that performs polynomial transformations on the block-encoded matrix.

%%%%%%%%%%%%%%%%%%%%%%%%%%%%%%%%%%%%%%%%%%%%%%%%%%%%%%%%%%%%%%%%%%%%%%%%%%%%%%%%

\subsection{Model}\label{sec:introduction-model}

A unitary matrix $U_A$ block-encodes the matrix $A \in \mathbb{R}^{N\times N}$ when the top-left block of $U_A$ is proportional to $A$, i.e.
\begin{equation}
\label{eq:block_encoding}
U_A = \begin{pmatrix} A/\alpha & \cdot \\ \cdot & \cdot \end{pmatrix},
\end{equation}
where $\alpha \ge ||A||$ is a normalization constant. We use $||\cdot||$ to denote the operator norm. The other blocks in $U_A$ are irrelevant, but they must be encoded such that $U_A$ is unitary. Note that we focus on real matrices $A$, but the extension to complex matrices is straightforward.

There is a variety of techniques available to block-encode a matrix \cite{gilyen2019,chakraborty2019}. For applications where we wish to operate on dense, classical data, such as is often encountered in machine learning or finance applications, a particularly relevant method is the \ac{QRAM} input model and its many proposed implementations \cite{giovannetti2008qram, giovannetti2008qrampra,matteo2020,paler2020,hann2021}. More concretely, we refer to \ac{QRAM} as the quantum circuit that allows query access to classical data in superposition
\begin{equation}
\label{eq:qram_query}
\sum_j \alpha_j \ket{j}\ket{0} \overset{\text{QRAM}}{\longrightarrow} \sum_j \alpha_j \ket{j}\ket{b_j},
\end{equation}
where $j$ is the address in superposition with amplitude $\alpha_j$ and $\ket{b_j}$ is the classical data loaded into a quantum state. In this work, we discuss two different circuit-level approaches to accomplish the \ac{QRAM} operation, which are optimized for different objectives. The first is a select-swap (SS) model, which is particularly efficient in terms of $T$-gate utilization \cite{low2018trading}. The second is a bucket-brigade (BB) model \cite{giovannetti2008qram}, which has reduced susceptibility to errors when operated on potentially faulty hardware \cite{hann2021}. We also provide a variant of the select-swap \ac{QRAM} approach that performs a slightly different version of Eq.~\eqref{eq:qram_query}, where a more general single-qubit state can be loaded controlled by a flag qubit.

The block-encoding constructions presented in our work compile all circuits into Clifford gates and $T$ gates, where attempts are made to minimize either the total number of qubits, the total number of $T$ gates---henceforth called the $T$-count---or the minimum number of layers of parallel $T$ gates---henceforth called the $T$-depth. Our estimates, as overviewed in the next subsection, illustrate the trade-offs between these three metrics. We focus on $T$ gates since, in many proposals for fault-tolerant quantum computation based on quantum error-correcting codes like the surface code, Clifford gates are transversal and can be performed essentially for free (or in some cases by simply updating the Pauli frame for the decoder completely in software). Meanwhile, $T$ gates require the expensive process of magic state distillation \cite{Knill2004magic,Bravyi2005magic}. Indeed, the motivating model for our work is one in which the logical quantum circuits we describe are carried out with lattice surgery \cite{Horsman2012latticesurgery,Litinski2018latticesurgery,Litinski2019gameofsurfacecodes,chamberland2022universal} on many physical qubits encoded with the surface code. We additionally assume that logical CNOT gates can be performed between arbitrary qubits in constant time, and that the fanout-CNOT operation, i.e.~the product of many CNOTs with the same control but different targets, can be performed in constant time \cite{Litinski2018latticesurgery}. As such, our calculations can be seen as optimistic and could be underestimates if there are additional hardware-specific constraints on logical topology and connectivity. 

%%%%%%%%%%%%%%%%%%%%%%%%%%%%%%%%%%%%%%%%%%%%%%%%%%%%%%%%%%%%%%%%%%%%%%%%%%%%%%%%

\subsection{Overview of results}\label{sec:introduction-results}

The main contributions of our work are summarized as follows:

\begin{itemize}
	\item Clifford+$T$ circuits with minimal $T$-depth, or alternatively, with minimal $T$-count, to perform a block-encoding unitary $U_A$ as given in Eq.~\eqref{eq:block_encoding} for $N \times N$ matrices $A$
	\item Resource calculations for the block-encoding, including all constant factors, expressed in terms of a small number of tunable parameters
	\item Clifford+$T$ circuits for QRAM and state preparation sub-routines, along with associated resource calculations
	\item Conceptual improvements to state preparation procedure yielding quadratic speedup in $T$-depth complexity.
\end{itemize}

Our results for the circuit resource costs in terms of the number of qubits, $T$-count, and $T$-depth are shown in the top part of Tab.~\ref{tab:min_T_depth}, including constant factors. We note that the cost of a having the $T$-depth be exponentially smaller than the time needed even to write down all of the matrix entries is a large overhead of $\bigo{N^2}$ ancilla qubits and total $T$-count. It is possible to reduce the qubit and $T$-count to $\bigo{N}$ at the cost of making $T$-depth grow also to $\bigo{N}$.

In addition to parameterized resource estimates, we compute the exact costs to block-encode matrices of sizes $[16\times 16, 256\times 256,4096\times 4096]$ with precision $\epsilon = 0.01$. To estimate a matrix norm, needed for the scale factor $\alpha$, we took the entries to be uniformly distributed random numbers on the interval $[5.0, 105.0]$ as representative of the type of data we might choose to block-encode (e.g., financial market data). Example numbers are shown in the bottom part of Tab.~\ref{tab:min_T_depth}.

%%%%%%%%%%%% START TABLE %%%%%%%%%%%%%%%%%%

\begin{table}
\caption{Quantum resources required to block-encode an $N \times N$ matrix to precision $\epsilon$, where we assume that $N=2^n$. The top part of the table contains parameterized estimates, while the bottom shows realistic values obtained for chosen values of the parameters. The parameter $R_y$ is the number of $T$ gates needed to synthesize a single qubit rotation about the $Y$ axis by an arbitrary angle, and $t$ is the number of bits precision to store the classical data, where both $t$ and $R_y$ scale as $\bigo{\log(1/\epsilon)}$. In the bottom part of the table, we computed the costs to block-encode random real matrices of sizes $[16\times 16, 256\times 256,4096\times 4096]$ up to precision $\epsilon = 0.01$.}
\label{tab:min_T_depth}
\begin{center}
\renewcommand{\arraystretch}{1.4}
\begin{tabular}{l|l|l}
\textbf{Resource} & \textbf{Minimum Depth} & \textbf{Minimum Count}\\
\hhline{=|=|=}
\textbf{\# Qubits}  & $4 N^2 - 3 N + 2 n-1$ & $N(t+1) + 3 n -t + 1$\\ 
\hline                                               
\textbf{$T$-Depth}  & $10n+8R_y -4$ & $8 N + 16 n + 4 R_{y} n t - 8$ \\
\hline
\textbf{$T$-Count}  & $(4R_y + 32)N^2 - 24N -4 R_y- 32n -8  $ & $8(2t+3)N - 16 t (n+1) + 4 R_{y} n t - 24$\\
\hhline{=|=|=}
\textbf{\# Qubits}  & [\num{1e3}, \num{3e5}, \num{7e7}] & [\num{4e2}, \num{7e3}, \num{1e5}]\\
\hline                                                                                                                            
\textbf{$T$-Depth}  & [\num{5e2}, \num{7e2}, \num{8e2}] & [\num{2e4}, \num{7e4}, \num{2e5}]\\
\hline                                               
\textbf{$T$-Count}  & [\num{7e4}, \num{2e7}, \num{7e9}] & [\num{3e4}, \num{2e5}, \num{2e6}]\\
\end{tabular}
\end{center}
\end{table}

%%%%%%%%%%%% END TABLE %%%%%%%%%%%%%%%%%%

%%%%%%%%%%%%%%%%%%%%%%%%%%%%%%%%%%%%%%%%%%%%%%%%%%%%%%%%%%%%%%%%%%%%%%%%%%%%%%%%

\subsection{Block-encoding strategy}\label{sec:introduction-strategy}

In order to make the definition from Eq.~\eqref{eq:block_encoding} precise, we provide quantum circuits that perform a unitary $\tilde{U}_A$ on $n + \ell$ qubits, where $N=2^n$ for which 
\begin{equation}
\left\lVert A - \alpha (\bra{0}_\ell \otimes I )\tilde{U}_A(\ket{0}_\ell \otimes I)\right\rVert = \alpha \left\lVert (\bra{0}_\ell \otimes I )(U_A-\tilde{U}_A)(\ket{0}_\ell \otimes I)\right\rVert \leq \epsilon,
\end{equation}
where the parameter $\ell$ denotes the number of ancillas, $I$ denotes the identity on $n$ qubits, and $\lVert \cdot \rVert$ denotes the operator norm.\footnote{If $N$ is not a power of 2, we can pad the matrix with zeros such that this is the case.} Subscripts on bras and kets are included for clarity to indicate the associated number of qubits in the state. Following standard convention, we call $\tilde{U}_A$ a $(\alpha,\ell,\epsilon)$-block-encoding of $A$. The normalization factor for the block-encoding we treat is $\alpha = ||A||_F$, where $||\cdot||_F$ denotes the Frobenius norm.\footnote{One can use a different norm (the q-norm), which we discuss in App.~\ref{app:q-norm_encoding}.}

\begin{figure}[t!]
\centering
\mbox{\Qcircuit @C=0.8em @R=0.8em {
& {\raisebox{0.18cm}{/${}^{n}$}}         \qw &\multigate{2}{U_A} & \qw & \qw &   & &\qw & {\raisebox{0.18cm}{/${}^{n}$}}         \qw &\multigate{2}{U_L} &\qw &\qw &\qw &\multigate{2}{U_R^{\dagger}} &\qw &\qw &   & &\qw & {\raisebox{0.18cm}{/${}^{n}$}}         \qw &\multigate{1}{\substack{\text{SP} \\ \{\ket{\phi_k}\}}} &\qswap \qwx[2] &\qw &\qw &\qw &\multigate{1}{\substack{\text{SP}^\dagger \\ \{\ket{\psi_j}\}}} &\qw \\
& {\raisebox{0.18cm}{/${}^{\ell'}$}} \qw &\ghost{U_A}        & \qw & \qw & = & &\qw & {\raisebox{0.18cm}{/${}^{\ell'}$}} \qw &\ghost{U_L}        &\qw &\qw &\qw &\ghost{U_R^{\dagger}}        &\qw &\qw & = & &\qw & {\raisebox{0.18cm}{/${}^{\ell'}$}} \qw &\ghost{\substack{\text{SP} \\ \{\ket{\phi_k}\}}}        &\qw            &\qw &\qw &\qw &\ghost{\substack{\text{SP} \\ \{\ket{\psi_j}\}}}                  &\qw \\
& {\raisebox{0.18cm}{/${}^{n}$}}         \qw &\ghost{U_A}        & \qw & \qw &   & &\qw & {\raisebox{0.18cm}{/${}^{n}$}}         \qw &\ghost{U_L}        &\qw &\qw &\qw &\ghost{U_R^{\dagger}}        &\qw &\qw &   & &\qw & {\raisebox{0.18cm}{/${}^{n}$}}         \qw & \ctrlslash{-1}                                                         &\qswap         &\qw &\qw &\qw &\ctrlslash{-1}                                                                &\qw \\
}
}
\caption{Reduction from block-encoding to controlled-state preparation. The block-encoding unitary $U_A$ is the product of two controlled-state preparation procedures, $U_L$ and $U_R^{\dagger}$. Given a family of quantum states $\{\ket{\phi_k}\}_{k=0}^{N-1}$ the controlled-$\text{SP}$ gate denotes the unitary that performs $\ket{0}_n \mapsto \ket{\phi_k}_n$ on the first register conditioned on the final register being $\ket{k}$ and using $\ell' = \ell-n$ ancillas on the second register that begin and end in the state $\ket{0}_{\ell'}$. One can obtain the $A_{jk}$ matrix element from the unitary by evaluating the $\bra{00j}\cdot\ket{00k}$ matrix element, as in Eq.~\eqref{eq:Ajk-mat-el}. The $\oslash$ indicates that the register acts as a control and may produce non-trivial action on the target for any control setting (in contrast to $\bullet$ and $\circ$). The controlled-state-preparation gates can be implemented either with the circuit from Fig.~\ref{fig:controlled_state_prep} or the circuit from Fig.~\ref{fig:controlled_state_prep_prerotated}, discussed in Sec.~\ref{sec:state_prep}.}
\label{fig:U_A}
\end{figure}
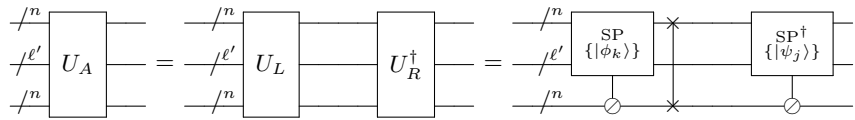

All of our constructions follow the prescription laid out in Refs.~\cite{kerenidis2016quantum,gilyen2019,chakrabarti2020quantum}, which calls for forming $U_A$ as the product of a pair of controlled-state preparation unitaries $U_L$ and $U_R$. In this prescription,
\begin{equation}
U_A = U_R^{\dagger} U_L\,,
\end{equation}
where, controlled on an $n$-qubit register in the state $\ket{j}_n$, $U_R$ prepares the $n$-qubit state $\ket{\psi_j}_n$, and $U_L$ prepares the state $\ket{\phi_j}_n$ with the assistance of $\ell'$ \ac{QRAM} ancilla qubits\footnote{The unitary $U_L$ additionally swaps the two $n$-qubit registers, cf.~Fig.~\ref{fig:U_A}.}. That is, we have
\begin{align}
U_R \ket{0}_n\ket{0}_{\ell'}\ket{j}_{n}= \ket{\psi_j}_n   \ket{0}_{\ell'} \ket{j}_n\quad\text{and}\quad U_L \ket{0}_n\ket{0}_{\ell'}\ket{k}_{n}= \ket{k}_n        \ket{0}_{\ell'} \ket{\phi_k}_n\,. 
\end{align}
We then choose
\begin{align}\label{eq:fro-states}
\ket{\psi_j}_n= \sum_{k=0}^{N-1} \frac{A_{jk}}{||A_{j,\cdot}||} \ket{k}_n\quad\text{and}\quad \ket{\phi_k}_n= \sum_{j=0}^{N-1} \frac{||A_{j,\cdot}||}{||A||_F} \ket{j}_n \quad\text{(independent of $k$),}
\end{align}
\noindent where $A_{j,\cdot}$ denotes the $j$-th row of $A$ and $\|A_{j,\cdot}\|$ the standard Euclidean norm of that vector. Since $\ket{\phi_k}_n$ is independent of $k$, $U_L$ is simply state preparation, rather than controlled-state preparation. However, in the q-norm version of the block-encoding presented in App.~\ref{app:q-norm_encoding}, $\ket{\phi_k}_n$ will depend on $k$. It is easily verified that $U_A$ is an exact block-encoding of $A$, i.e.,
\begin{equation}\label{eq:Ajk-mat-el}
(\bra{0}_\ell \otimes \bra{j}_n)U_A(\ket{0}_\ell \otimes \ket{k}_n) = A_{jk}\quad\text{with $\ell = n+\ell'$.}
\end{equation}
We refer to Fig.~\ref{fig:U_A} for an overview of this reduction from block-encoding to controlled-state preparation. To implement the controlled-state preparation unitaries $U_R$ and $U_L$, we combine a QRAM-like data-loading step with a protocol for state preparation of $n$-qubit states.

%%%%%%%%%%%% START FIGURE %%%%%%%%%%%%%%%%%%

%%%%%%%%%%%% END FIGURE %%%%%%%%%%%%%%%%%%

%%%%%%%%%%%%%%%%%%%%%%%%%%%%%%%%%%%%%%%%%%%%%%%%%%%%%%%%%%%%%%%%%%%%%%%%%%%%%%%%

\subsection{State preparation}\label{sec:introduction-preparation}

The state preparation problem has been studied in a series of prior literature \cite{grover2000synthesis,grover2002creating,Kaye2001statePrep,soklakov2006stateprep,mottonen2004transformation,plesch2011stateprep,kerenidis2016quantum,low2018trading,sanders2019blackbox,bausch2020fast,wang2021fast,araujo2021divide}. To prepare an $n$-qubit state, one basic approach is to perform a sequence of $n$ single-qubit rotations on successive qubits, where the angle of rotation is controlled on previously rotated qubits. That is, for $p=0,\ldots,n-1$, a rotation is performed on qubit $p$ by one of $2^{p}$ possible rotation angles determined by the setting of qubits $0,\ldots,p-1$. Thus, there are $\sum_{p=0}^{n-1} 2^p = N-1$ angles (where $N=2^n$ is the dimension of the Hilbert space) that might be used at some point in the procedure. In Ref.~\cite{low2018trading}, it was shown that given an $n$-qubit state $\ket{\psi}_n$, this approach can produce a state $\ket{\tilde{\psi}}_n$ for which
\begin{equation}
\|\ket{\psi}_n - \ket{\tilde{\psi}}_n\|\leq \epsilon\quad \text{in $T$-depth $\bigo{\log^2(N/\epsilon)}$ and using $\bigo{N}$ total qubits.}
\end{equation}
This scaling comes from the need to do the $n=\log(N)$ controlled-rotations in series\footnote{All logarithms in this paper are taken base 2.}, and to perform each controlled-rotation by (i) leveraging a $\bigo{n}$-depth controlled-swap network to copy in a $\bigo{\log(1/\epsilon)}$-bit description of the relevant rotation angle, and (ii) for each bit applying an $\epsilon$-accurate Clifford+$T$ gate decomposition of the corresponding controlled-rotation, costing $\bigo{\log(1/\epsilon)}$ gates per bit.\footnote{A slightly more efficient approach applies the controlled-rotation using a controlled-adder and a pre-prepared Fourier state, rather than gate syntheses for each bit separately \cite{low2018trading,Gidney2018halvingcostof}.} In Sec.~\ref{sec:preloaded}, we implement a version of this approach, which we call the ``\emph{fixed-precision}'' method.

The extension to controlled-state preparation calls for creating one of $N = 2^n$ $n$-qubit states $\{\ket{\phi_k}\}_{k=0}^{N-1}$ controlled on an $n$-qubit register. As there are now $N(N-1)$ rotation angles that might be needed, the first step is to load the correct subset (which depends on the setting of the control $k$) of $N-1$ angles into an ancilla register. This loading step can be understood as a more general form of the QRAM query from Eq.~\eqref{eq:qram_query}. In Sec.~\ref{sec:qram} we provide details of this initial loading step and illustrate two separate circuit-level approaches compatible with the fixed-precision approach: (i) the select-swap (SS) approach, which is based on Ref.~\cite{low2018trading}, and (ii) the bucket-brigade (BB) approach, which is based on Refs.~\cite{giovannetti2008qram,giovannetti2008qrampra,hann2021}. Both approaches allow the loading step to be accomplished in as little as $\bigo{\log(N)}$ $T$-depth. The select-swap approach has a better constant factor, while the bucket-brigade approach has enhanced natural noise resilience \cite{hann2021} that may lead to better overall performance when quantum error correction overhead is considered. After loading, a normal state preparation procedure is performed, followed by unloading of the angles to reset the ancillas to $\ket{0}$. As QRAM is a common subroutine in quantum algorithms, our estimates in Sec.~\ref{sec:qram} could be useful beyond the application of block-encoding. 

%%%%%%%%%%%%%%%%%%%%%%%%%%%%%%%%%%%%%%%%%%%%%%%%%%%%%%%%%%%%%%%%%%%%%%%%%%%%%%%%

\subsection{Improved state preparation}\label{sec:introduction-state}

In addition to the state preparation procedure presented above, we also present a distinct approach to state preparation that we call the ``\emph{pre-rotated}'' approach. Pre-rotated state preparation may be of independent interest, as it quadratically improves the $T$-depth for the task of state preparation and controlled-state preparation from $\bigo{\log^2(N/\epsilon)}$ to $\bigo{\log(N/\epsilon)}$. The pre-rotated approach achieves logarithmic $T$-depth by (a) pre-applying all possible single-qubit rotations onto $N-1$ ancilla qubits in parallel using $\bigo{\log(1/\epsilon)}$ depth, and (b) enacting each of the $\log(N)$ controlled-rotations in series using a constant-depth controlled-swap network that injects the appropriate ancilla into the state register. Our controlled-swap networks require only constant depth per controlled-rotation because we wait to uncompute garbage until after all $\log(N)$ controlled-rotations have been completed. Note that the fixed-precision method can implement step (b), but not step (a). The pre-rotated method utilizes both steps and also requires a flag mechanism for uncomputing the ancillas that were not injected. These results of this approach are summarized in Tab.~\ref{tab:block_encoding_pre}. A similar idea of pre-rotation for state preparation was also explored in Ref.~\cite{araujo2021divide}, but their construction is not garbage-free. 

%%%%%%%%%%%%%%%%%%%%%%%%%%%%%%%%%%%%%%%%%%%%%%%%%%%%%%%%%%%%%%%%%%%%%%%%%%%%%%%%

\subsection{Variations and extensions}\label{sec:introduction-extensions}

%%%%%%%%%%%% START FIGURE %%%%%%%%%%%%%%%%%%

\begin{figure}[t!]
\centering
\includegraphics[width=0.8\textwidth]{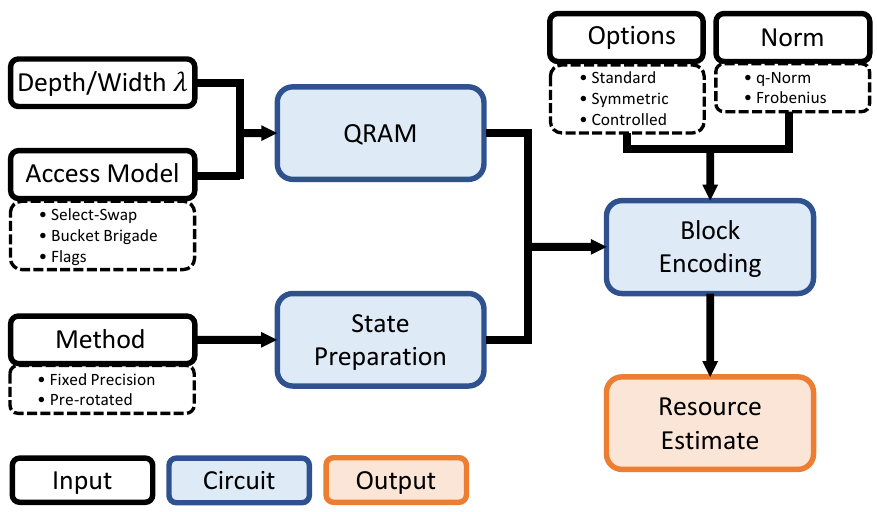}
\caption{Schematic demonstrating our resource estimation procedure for block-encoding. The user specifies a \ac{QRAM} access model, a parameter $\lambda$ that allows one to trade depth for width, the method for state preparation routine, along with the block-encoding normalization factor and whether one wants a standard block-encoding, a controlled version, and/or a symmetric encoding. Note that not all possible combinations are allowed, e.g., the state-preparation method {\it pre-rotated} is only available together with the access model {\it flags} and the choice $\lambda = n$. Details are provided in the main text.}
\label{fig:block_diagram}
\end{figure}

%%%%%%%%%%%% END FIGURE %%%%%%%%%%%%%%%%%%

We discuss further variations of block-encodings that include, in particular, trade-offs for $T$-count, and versions for improved noise resilience when using imperfect quantum hardware. The available variations are illustrated in Fig.~\ref{fig:block_diagram}.
The technical contributions of this work are as follows:
\begin{itemize}
\item We demonstrate how to implement a mechanism suggested in Refs.~\cite{low2018trading, Babbush2018unary, hann2021} that trades $T$-depth for $T$-count, parameterized by an integer $\lambda \in \{0,1,\ldots,\log(N)\}$. The $T$-count for block-encoding is minimized at $\bigo{N}$ using the select-swap QRAM with $\lambda = 0$, 
as reported in Tab.~\ref{tab:min_T_depth}. However, the corresponding $T$-depth is also $\bigo{N}$, which is exponentially worse than the minimal depth case. Our results for general $\lambda$ for the bucket-brigade and the select-swap approaches appear later in Tab.~\ref{tab:block_encoding_fixed}. We report exact expressions including constant pre-factors and sub-leading terms for all cases, which allows us to find a minimal $T$-count construction. These ideas are discussed in Secs.~\ref{sec:select_swap} and \ref{sec:bucket_brigade}.
\item We explain how our construction can be adapted to perform a controlled-block-encoding of $A$, i.e.~the operation $\ket{0}\bra{0} \otimes I + \ket{1}\bra{1} \otimes U_A$. These controlled block-encodings are important e.g.~when composing block-encodings of multiple matrices \cite{gilyen2019} and are discussed in Sec.~\ref{sec:controlled_block_encoding}.
\item We construct block-encodings of symmetrized matrices
\begin{equation}
\bar{A} = \begin{pmatrix} 0 & A \\ A^{T} & 0\end{pmatrix} \quad \text{where} \quad \bar{A} \in \mathbb{R}^{(M+N)\times (M+N)}\,,
\end{equation}
which are useful in some applications, and which can be block-encoded more efficiently than the general construction owing to their structure. Symmetrized block-encodings are discussed in App.~\ref{app:off-diagonal_matrices}.
\item We give a variation of the symmetrized encoding for which the normalization factor is the $q$-norm, which can be smaller than the Frobenius norm; the resource counts are similar. This variation is discussed in App.~\ref{app:q-norm_encoding}.
\end{itemize}

%%%%%%%%%%%%%%%%%%%%%%%%%%%%%%%%%%%%%%%%%%%%%%%%%%%%%%%%%%%%%%%%%%%%%%%%%%%%%%%%

\subsection{Outlook}

While implementation of QRAM-based algorithms with a large amount of classical data on actual hardware is quite a ways off, we hope our detailed circuit layouts and resource counts challenge the hardware community to think through these requirements and investigate whether or not specialized hardware can meet them with reduced resources. Just as modern computer random-access memory (RAM) is distinct from processing elements and permanent memory, it would be desirable for future quantum computers to have hybrid architectures in which the QRAM elements are distinct from the quantum processor and optimized for the requirements at hand. We hope that our results provide a blueprint for these requirements, as having an efficient means to load classical data into a quantum computer could open up many applications in business, finance, and beyond.

%%%%%%%%%%%%%%%%%%%%%%%%%%%%%%%%%%%%%%%%%%%%%%%%%%%%%%%%%%%%%%%%%%%%%%%%%%%%%%%%

\subsection{Outline}

The remainder of our manuscript is structured as follows. In Sec.~\ref{sec:qram}, we give circuits and resource estimates for performing the QRAM operation of Eq.~\eqref{eq:qram_query}, as well as a more general data-loading operation that we use in our minimal depth block-encoding construction. In Sec.~\ref{sec:state_prep}, we elaborate on the conceptual approach to state preparation and controlled-state preparation and give circuits and resource estimates for these tasks. We present two distinct approaches, which we call fixed-precision and pre-rotated, where the former relies on the SS-QRAM or BB-QRAM, while the latter relies on the generalized QRAM operation. In Sec.~\ref{sec:resource_estimates}, we compute and report the overall resource estimates for block-encoding. In Sec.~\ref{sec:error}, we give an error analysis for finite precision implementations. Finally, in Sec.~\ref{sec:conclusion}, we offer some concluding remarks. 
 Various technical details and variants are deferred to App.~\ref{app:alternate-encodings}--\ref{app:decompositions}.

%%%%%%%%%%%%%%%%%%%%%%%%%%%%%%%%%%%%%%%%%%%%
%                                          %
%                 QRAM                     %
%                                          %
%%%%%%%%%%%%%%%%%%%%%%%%%%%%%%%%%%%%%%%%%%%%

\section{Data Loading and QRAM Implementations}\label{sec:qram}

\subsection{Overview}

As overviewed in the previous section, block-encoding relies on controlled-state preparation, which necessitates an initial data loading step resembling the QRAM query of Eq.~\eqref{eq:qram_query}. We present multiple options for QRAM implementations as depicted in the upper-left block of the flow chart in Fig.~\ref{fig:block_diagram}. We explain each of these options and provide non-Clifford resource counts.  Since the QRAM operation is an important primitive in many quantum algorithms, this resource analysis could be of interest independent from our block-encoding estimates.

Note that in this section and throughout the paper we make extensive use of the symbol $\oslash$ when depicting $(a+b)$-qubit gates in which the $a$-qubit register acts as a control. That is, when the gate can be decomposed as
\begin{equation}
\sum_{j=0}^{2^a-1}\ket{j}\bra{j}_a \otimes U_j
\end{equation}
for some set of $b$-qubit unitaries $\{U_j\}$, we draw $\oslash$ on the $a$-qubit register in the circuit diagram. When $U_j$ is identity for all $j$ besides $j= \ket{1}^{\otimes b}$, we replace $\oslash$ with $\bullet$, and when $U_j$ is identity for all $j$ besides $j= \ket{0}^{\otimes b}$, we replace it with $\circ$. A similar usage of $\oslash$ appears in Ref.~\cite{low2018trading}.

%%%%%%%%%%%%%%%%%%%%%%%%%%%%%%%%%%%%%%%%%%%%%%%%%%%%%%%%%%%%%%%%%%%%%%%%%%%%%%%%

\subsection{Select-swap data loading}\label{sec:select_swap}

The select-swap QRAM access-model allows one to trade circuit depth for width via a tunable integer parameter $\lambda$, where $0 \leq \lambda \leq n$ \cite{low2018trading}. The model also allows one to utilize qubits prepared in arbitrary initial states (so-called ``dirty'' qubits) at the cost of a constant factor increase to the circuit depth. While dirty qubits could lead to resource savings when considering a full architectural implementation of an algorithm, we do not make explicit use of dirty qubits; we assume that the memory is initialized to the $\ket{0}$ state for all circuits in this paper, except where otherwise stated.

%%%%%%%%%%%% START FIGURE %%%%%%%%%%%%%%%%%%

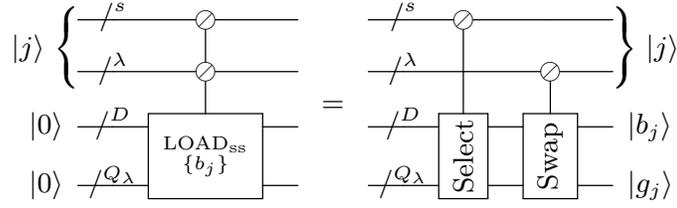
\begin{figure}[t!]
\centering
\scalebox{1.2}{
\mbox{
\Qcircuit @C=1.2em @R=1.1em {
               & \numq{0.15cm}{s} \qw             & \ctrlslash{1}                           &\qw  &          && \numq{0.15cm}{s} \qw             & \ctrlslash{2}               & \qw                  & \qw  \\
               & \numq{0.15cm}{\lambda} \qw       & \ctrlslash{1}                           &\qw  & \rb{-0.8cm}{=} && \numq{0.15cm}{\lambda} \qw       & \qw                         & \ctrlslash{1}            & \qw  \\
\lstick{\ket{0}} & \numq{0.15cm}{D} \qw             & \mltg{1}{\substack{\LOADSS \\ \{b_j\}}} &\qw  &          && \numq{0.15cm}{D} \qw             & \mltg{1}{\rotb{90}{Select}} & \mltg{1}{\rotatebox{90}{Swap}} & \rstick{\ket{b_j}} \qw\\
\lstick{\ket{0}} & \rb{0.15cm}{$/^{Q_\lambda}$} \qw & \ghost{\substack{\LOADSS \\ \{b_j\}}}   &\qw  &        && \rb{0.15cm}{$/^{Q_\lambda}$} \qw & \ghost{\rotb{90}{Select}}   & \ghost{\rotatebox{90}{Swap}}   & \rstick{\ket{g_j}} \qw\\
{\inputgroupv{1}{2}{0.8em}{1.0em}{\raisebox{-0.2cm}{$\ket{j}$}}}
{\outputgroupv{1}{2}{10}{0.8em}{1.0em}{\raisebox{-0.2cm}{$\ket{j}$}}}
}}
}
\caption{\label{fig:select-swap-high-level}Select-Swap circuit with variable depth/width, which coherently loads classical data into a $D$-qubit data register. The $n=s+\lambda$ qubit control register is divided up into $s$ qubits for the select control and $\lambda$ qubits for the swap control. The size of the ancilla register is $Q_\lambda = (2^\lambda -1)D$. Select iterates through all $2^s$ settings of the first $s$ bits and writes the corresponding $2^\lambda D$ bits of classical data to the final two registers, costing $\bigo{2^s}$ depth. Swap moves the correct $D$ bit register into the first of the $2^{\lambda}$ positions in $\bigo{\lambda}$ depth. When select is implemented with unary iteration to minimize $T$-count and $T$-depth, an additional $s-1$ ancilla qubits are needed (not shown) \cite{Babbush2018unary}. Implementation details for select and swap are given in Fig.~\ref{fig:select-swap-simple} of App.~\ref{app:select_swap}.}
\end{figure}

%%%%%%%%%%%% END FIGURE %%%%%%%%%%%%%%%%%%

The select-swap formalism---introduced in Ref.~\cite{low2018trading}---realizes the QRAM data-loading operation of Eq.~\eqref{eq:qram_query} for a $D$-bit data register, except that it leaves additional garbage states in a $(\Lambda-1)D$-qubit ancilla register, where $\Lambda = 2^\lambda$. That is, select-swap realizes the unitary $\text{LOAD}_\text{ss}$ operation defined by the equation
\begin{equation}\label{eq:qram_query_garbage}
\text{LOAD}_\text{ss}\left(\sum_{j=0}^{N-1}\alpha_j \ket{j}_\lambda\ket{0}_D\ket{0}_{(\Lambda-1)D}\right) = \sum_{j=0}^{N-1}\alpha_\lambda \ket{j}_n\ket{b_j}_D\ket{g_j}_{(\Lambda-1)D},
\end{equation}
where $\ket{b_j}_D$ is the $D$-bit classical data associated with address $j$, and $\ket{g_j}_{(\Lambda-1)D}$ is a garbage state that contains shuffled and possibly phase-flipped versions of classical data at other addresses. Note that the garbage could be uncomputed by copying the data into an ancilla register and then running $\text{LOAD}_\text{ss}$ in reverse, but this will not be necessary in our application. 

The idea behind select-swap is as follows. The $n=s+\lambda$ qubit control register is divided into $s$ qubits for the select control and $\lambda$ qubits for the swap control, as shown schematically in Fig.~\ref{fig:select-swap-high-level}. This division corresponds to a partition of the $2^n$ $D$-bit entries in the classical database into $2^s$ subsets of size $2^\lambda=\Lambda$. The first portion of $\text{LOAD}_\text{ss}$ is a select subroutine, which implements unary iteration \cite{Babbush2018unary} through the $s$ high-bits of the control register and, controlled on each setting, writes the corresponding subset of classical data into $\Lambda$ $D$-qubit registers. The depth of select is exponential in the number of control bits $s$. The next portion is the Swap subroutine, which, controlled on the $\lambda$ low bits of the control, swaps the appropriate ancilla register into the top register, leaving the other $\Lambda-1$ registers in some garbage state, as defined by the following equation. 
\begin{equation}\label{eq:SwapToTop}
\text{Swap}\left(\sum_{j=0}^{\Lambda-1}\alpha_j \ket{j}_n \ket{\xi_0}_D\ket{\xi_1}_D\ldots \ket{\xi_{\Lambda-1}}_D\right) = \sum_{j=0}^{\Lambda-1}\alpha_j \ket{j}_n \ket{\xi_j}_D\ket{h_j}_{(\Lambda-1)D}
\end{equation}
where the states $\ket{\xi_i}_D$ are arbitrary and $\ket{h_j}_{(\Lambda-1)D}$ is a garbage state. The depth of Swap is linear in the number of control bits $\lambda$. More details of these two subroutines are given in App.~\ref{app:select_swap}. 

Note that during the select subroutine, fanout CNOT gates are used to write the classical data, where the presence or absence of a target on each qubit is determined by the classical data at compile time. Fanout CNOT is an architectural primitive for surface-code based quantum computers, so we assume their cost is minimal relative to non-Clifford gates \cite{Litinski2018latticesurgery}. If single-target CNOT gates are required by the architecture instead, the fanout CNOT can be decomposed into a logarithmic depth network of normal CNOT gates \cite{low2018trading}. Moreover, the swap subroutine requires controlled-swap gates, which can be implemented fairly efficiently with a phase-incorrect parallel controlled-swap operation, shown in Fig.~\ref{fig:multi_cswap} of App.~\ref{app:decompositions} (the incorrect phases can be absorbed into the garbage states $\ket{g_j}$).

%%%%%%%%%%%%%%%%%%%%%%%%%%%%%%%%%%%%%%%%%%%%%%%%%%%%%%%%%%%%%%%%%%%%%%%%%%%%%%%%

\subsection{Bucket-brigade data loading}\label{sec:bucket_brigade}

The bucket-brigade QRAM was initially proposed in Ref.~\cite{giovannetti2008qram}. The primary motivation behind studying this approach is that it has improved noise-resilience compared to the select-swap approach \cite{giovannetti2008qram, giovannetti2008qrampra, Hong2012qram, Arunachalam2015qram, matteo2020, paler2020, hann2021}. We will not reprise this argument here; we refer the reader to the references, especially Refs.~\cite{matteo2020, hann2021}. If this noise-resilience can be realized in a physical machine, it can reduce the resources required to error correct the data-loading operation, leading to physical resource savings. Whether SS-QRAM or BB-QRAM is preferred will depend upon the underlying architecture and details on the overhead required to implement quantum error correction. 

%%%%%%%%%%%% START FIGURE %%%%%%%%%%%%%%%%%%

\begin{figure}[ht!]
\centering
\scalebox{1.2}{
\mbox{
\Qcircuit @C=0.6em @R=1.1em {
        & \qw &\numq{0.15cm}{s} \qw                                     &\ctrlslash{1}                           &\qw && &&& {\rb{0.15cm}{$/^s$}} \qw                                & \qw      & \ctrlslash{3} & \qw             & \qw            & \qw                     & \qw             & \qw             & \ctrlslash{3} & \qw &                       && \ctrlslash{3} & \qw                & \qw             & \qw                        & \qw             & 
\qw             & \ctrlslash{3} & \qw      &\qw &                       \\
        & \qw &\numq{0.15cm}{\lambda} \qw                               &\ctrlslash{1}                           &\qw && &&& {\rb{0.15cm}{$/^\lambda$}} \qw                          & \qw      & \qw           & \qswap          & \qw            & \qw                     & \qw             & \qswap          & \qw           & \qw &                       && \qw           & \qswap             & \qw             & \qw                        & \qw             &
\qswap          & \qw           & \qw      &\qw &                       \\
\lstick{\ket{0}}& \qw & {\rb{0.15cm}{$/^D$}} \qw                                &\mltg{4}{\substack{\LOADBB \\ \{b_j\}}} &\qw && &&& {\rb{0.15cm}{$/^D$}} \qw                                & \gate{H} & \qw           & \qw             & \qswap                & \qw                     & \qswap          & \qw             & \qw           & \qw & \rb{0.75cm}{$\cdots$} && \qw           & \qw                & \qswap          & \qw                        & \qswap          & \qw             & \qw           & \gate{H} &\qw & \rstick{\ket{b_j}} \qw\\
\lstick{\ket{0}}& \qw &\qw                                                      &\ghost{\substack{\LOADBB \\ \{b_j\}}}   &\qw &&=&&& \qw                                           & \qw      & \targ         & \ctrl{-2}       & \ctrl{-1}                & \qw                     & \ctrl{-1}       & \ctrl{-2}       & \targ         & \qw &               && \targ         & \ctrl{-2}          & \ctrl{-1}       & \qw                        & \ctrl{-1}       & \ctrl{-2}       & \targ         & \qw      &\qw & \rstick{\ket{0}} \qw  \\
\lstick{\ket{0}}& \qw &{\rb{0.15cm}{$/^D$}} \qw                                 &\ghost{\substack{\LOADBB \\ \{b_j\}}}   &\qw && &&& {\rb{0.15cm}{$/^D$}} \qw                                & \qw      & \qw           & \qw             & \qswap \qwx[-1]            & \mltg{2}{\text{BB}_{0}} & \qswap \qwx[-1] & \qw             & \qw           & \qw &                     && \qw           & \qw                & \qswap \qwx[-1] &\mltg{2}{\text{BB}_{2^s-1}} & \qswap \qwx[-1] & \qw             & \qw           & \qw      &\qw & \rstick{\ket{0}} \qw  \\
\lstick{\ket{0}}& \qw &  {\rb{0.15cm}{$/^\lambda$}} \qw                         &\ghost{\substack{\LOADBB \\ \{b_j\}}}   &\qw && &&& {\rb{0.15cm}{$/^\lambda$}} \qw                          & \qw      & \qw           & \qswap \qwx[-2] & \qw                & \ghost{\text{BB}_{0}}   & \qw             & \qswap \qwx[-2] & \qw           & \qw & \rb{0.75cm}{$\cdots$} && \qw           & \qswap \qwx[-2]    & \qw             & \ghost{\text{BB}_{2^s-1}}  & \qw             & \qswap \qwx[-2] & \qw           & \qw      &\qw & \rstick{\ket{0}} \qw  \\
\lstick{\ket{0}}& \qw & {\rb{0.15cm}{\shiftleft{0.4cm}{$/^{q_\lambda}$}}} \qw   &\ghost{\substack{\LOADBB \\ \{b_j\}}}   &\qw && &&& {\rb{0.15cm}{\shiftright{-0.3cm}{$/^{q_\lambda}$}}} \qw & \qw      & \qw           & \qw             & \qw                & \ghost{\text{BB}_{0}}   & \qw             & \qw             & \qw           & \qw &                   && \qw           & \qw                & \qw             & \ghost{\text{BB}_{2^s-1}}  & \qw             & \qw             & \qw           & \qw      &\qw & \rstick{\ket{0}} \qw  \\
{\inputgroupv{1}{2}{0.8em}{1.0em}{\raisebox{-0.2cm}{$\ket{j}$}}}
{\outputgroupv{1}{2}{30}{0.8em}{1.0em}{\raisebox{-0.2cm}{$\ket{j}$}}}
%{\inputgroupv{3}{4}{0.8em}{1.0em}{\raisebox{-7.3cm}{$\Lambda$}}}
}}}
\caption{\label{fig:bb-high-level}Bucket brigade circuit with variable circuit depth/width, which coherently loads classical data into a $D$-qubit data register. The circuit iterates through all $2^s$ settings of the first $s$ bits of the address register, with the first and last iteration shown. In each iteration, a multiply controlled Toffoli gate (where some bits are controlled on $\ket{0}$ and some on $\ket{1}$, denoted generically with $\oslash$) computes whether the $s$ bits agree with the address setting $\ket{j}$. If so, the remaining $\lambda = n-s$ address qubits and the $D$-qubit bus (third register) in the state $\ket{+}^{\otimes D}$ are swapped into ancilla registers, where the bucket-brigade circuit (denoted by $\text{BB}_i$, see Fig.~\ref{fig:bucket-brigade} of App.~\ref{app:bucket_brigade_qram} and Ref.~\cite{hann2021}) is performed to load the state $H^{\otimes D}\ket{b_j}_D$ into the bus. No garbage is produced, since during all other iterations the ancillas are in $\ket{0}$, which is a $+1$ eigenvector of the bucket brigade circuit. When the sequence of multiply controlled Toffoli gates is implemented with unary iteration to minimize $T$-count and $T$-depth, an additional $s-1$ ancilla qubits are needed (not shown) \cite{Babbush2018unary}, which, combined with the $q_\lambda = (2^\lambda-1)(2D+1)-1$ ancillas needed for the bucket-brigade, gives a total of $Q_\lambda = (2D+1)2^\lambda+n-D-2$ ancillas for the $\text{LOAD}_\text{bb}$ operation. 
}
\end{figure}
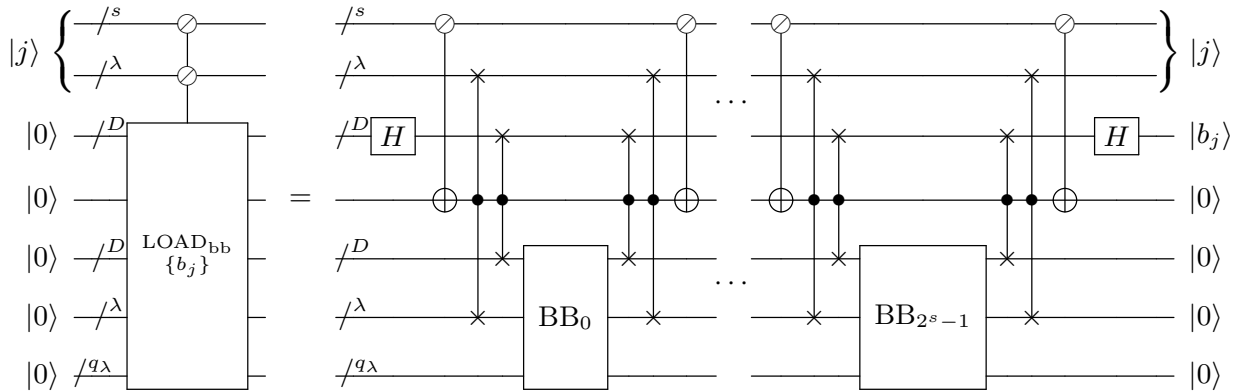

%%%%%%%%%%%% END FIGURE %%%%%%%%%%%%%%%%%%

Unlike the SS-QRAM, the BB-QRAM does not leave garbage and the unitary $\LOADBB$ can be defined by the equation
\begin{equation}\label{eq:qram_query_bb}
\text{LOAD}_\text{bb}\left(\sum_{j=0}^{N-1}\alpha_j \ket{j}_n\ket{0}_D\ket{0}_{Q_\lambda}\right) = \sum_{j=0}^{N-1}\alpha_j \ket{j}_n\ket{b_j}_D \ket{0}_{Q_\lambda},
\end{equation}
where the last register contains $Q_{\lambda}$ ancilla states that begin and end in $\ket{0}$. Similar to the swap portion of the SS-QRAM circuit (see Fig.~\ref{fig:select-swap-simple} of App.~\ref{app:select_swap}), the BB-QRAM circuit is composed primarily of controlled swap gates, with an arrangement that acts to minimize entanglement and enhance noise resilience. Additionally, as in the SS-QRAM case, we can trade circuit depth for width using a parameter $\lambda$ \cite{hann2021} as shown in Fig.~\ref{fig:bb-high-level}, where the implementation of the gates labeled $\text{BB}_i$ is described in more detail in App.~\ref{app:bucket_brigade_qram}. However, reducing circuit width does not reduce the overall $T$-count as it does with the SS implementation, which is a drawback of this scheme. Note that only the portion of the circuit labeled $\text{BB}_i$ will retain the noise resilience, and thus the noise resilience is lost when we choose $\lambda = 0$.

%%%%%%%%%%%%%%%%%%%%%%%%%%%%%%%%%%%%%%%%%%%%%%%%%%%%%%%%%%%%%%%%%%%%%%%%%%%%%%%%

\subsection{QRAM operation with flags}\label{sec:QRAM_flags}

We introduce a third and final data-loading operation, which generalizes the previous approaches by allowing data of the form $\cos(\theta/2)\ket{0} + \sin(\theta/2)\ket{1}$ to be loaded for any $\theta$, rather than just classical bits $\ket{0}$ or $\ket{1}$. Additionally, this load operation has a flag qubit and only acts non-trivially if the flag is set to 1, which is necessary to implement the minimal depth ``pre-rotated'' version of state preparation in Sec.~\ref{sec:pre-rotated}.

Suppose we have a set of $N = 2^n$ angles $\{\theta^{(j)}\}_{j=0}^{N-1}$. Let $R_y(\theta)$ denote the unitary rotation about the $Y$-axis by angle $\theta$, that is
\begin{equation}\label{eq:Ry_def}
R_y(\theta) \ket{0} = \cos(\theta/2)\ket{0} + \sin(\theta/2)\ket{1}, \qquad R_y(\theta) \ket{1} = -\sin(\theta/2)\ket{0} + \cos(\theta/2)\ket{1}\\
\end{equation} 
The data-loading with flags operation, which we call LOADF, acts on four registers: an $n$-qubit ``address'' register, a single-qubit ``flag'' register, an $N$-qubit ``angle'' register, and an $N$-qubit ancilla register. The unitary LOADF operation is defined by the equation
\begin{align}\label{eq:LOAD}
\text{LOADF}&\left(\sum_{j=0}^{N-1} \sum_{f=0}^1 \alpha_{jf} \ket{j}_n \ket{f}_1 \ket{0}_N\ket{0}_{N}\right) \\
={}&\;\;\;
\sum_{j=0}^{N-1} \sum_{f=0}^1 \alpha_{jf} \ket{j}_n \ket{f}_1 \left[ R_y(f\theta^{(j)})\ket{0}_1 \right]\ket{0}_{N-1}\ket{0}_{N}
\end{align}
When the first address register is in the state $\ket{j}$, the unitary $R_y(\theta^{(j)})$ is applied to the first position of the angle register if and only if the flag bit $f$ is set to $1$.

%%%%%%%%%%%% START FIGURE %%%%%%%%%%%%%%%%%%

\begin{figure}[t!]
\centering
\scalebox{1.2}{
\mbox{
\Qcircuit @C=1.2em @R=1.1em {
\lstick{\text{address  } \ket{j}}     & \numq{0.15cm}{n} \qw   &\ctrlslash{1}                                         &\qw&              &&\numq{0.15cm}{n} \qw  & \qw    &\ctrlslash{4}                        & \qw             
&\ctrlslash{4}                        &\ctrlslash{2}                        &\rstick{\ket{j}}   \qw                   \\
\lstick{\text{flag  } \ket{f}}        & \qw                    &\ctrl{1}                                              &\qw&              &&\qw                   & \qw    & \qw                                 & \ctrl{1}       
& \qw                                 & \qw                                 &\rstick{\ket{f}} \qw                     \\
                                      & \qw                    &\mltg{3}{\substack{\text{LOADF} \\ \{\theta^{(j)}\}}} &\qw&              &&\qw                   & \qw    & \qw                                 & \mltg{1}{V}      
& \qw                                 &\mltg{1}{\rotatebox{90}{Swap}}       &\rstick{R_y(f\theta^{(j)})\ket{0}}\qw    \\
                                      & \numq{0.15cm}{N-1} \qw &\ghost{\substack{\text{LOADF} \\ \{\theta^{(j)}\}}}   &\qw&\rb{0cm}{=}   &&\numq{0.15cm}{N-1} \qw& \qw    & \qw                                 & \ghost{V}     
& \qw                                 & \ghost{\rotatebox{90}{Swap}}        &\rstick{\ket{0}}\qw  \\
                                      & \qw                    &\ghost{\substack{\text{LOADF} \\ \{\theta^{(j)}\}}}   &\qw&              &&\qw                   &\gate{X}&\mltg{1}{\rotatebox{90}{\SwapD}}     & \ctrlslash{-1} 
&\mltg{1}{\rotatebox{90}{Swap}}       & \gate{X}                                 & \qw                   \\
                                      & \numq{0.15cm}{N-1} \qw &\ghost{\substack{\text{LOADF} \\ \{\theta^{(j)}\}}}   &\qw&              &&\numq{0.15cm}{N-1} \qw& \qw    & \ghost{\rotatebox{90}{\SwapD}}      & \ctrlslash{-1} 
& \ghost{\rotatebox{90}{Swap}}        & \qw                                 &  \qw                   \\
{\inputgroupv{3}{4}{0.8em}{1.0em}{\rb{-0.2cm}{\shiftleft{1cm}{angle $\ket{0}$}}}}
{\inputgroupv{5}{6}{0.8em}{1.0em}{\rb{-0.2cm}{\shiftleft{1.4cm}{ancilla $\ket{0}$}}}}
{\outputgroupv{5}{6}{13}{0.8em}{1.0em}{\rb{-0.4cm}{$\ket{0}$}}}
}}
}
\caption{\label{fig:LOADF}Circuit that implements LOADF, which loads a single-qubit state of the form $\cos(\theta/2)\ket{0} + \sin(\theta/2) \ket{1}$, where the angle $\theta$ depends on a control register. The controlled-Swap gates on registers of size $N$ have the action of swapping the register in position $j$ to position $1$; the adjoint of swap moves position 1 to position $j$. The gate $V$ applies the rotation $R_y(\theta^{(k)})$ to the $k$-th angle register controlled on the $k$-th ancilla and flag qubits set to $\ket{1}$, for every $k$. The controlled-Swap gates have depth $\bigo{n}$, and the $V$ gate has depth $\bigo{\log(1/\delta)}$ if we wish to perform each $R_y(\theta^{(k)})$ unitary up to precision $\delta$. A complete example of LOADF for $n=2$ is shown in Fig.~\ref{fig:load_min_depth}}
\end{figure}
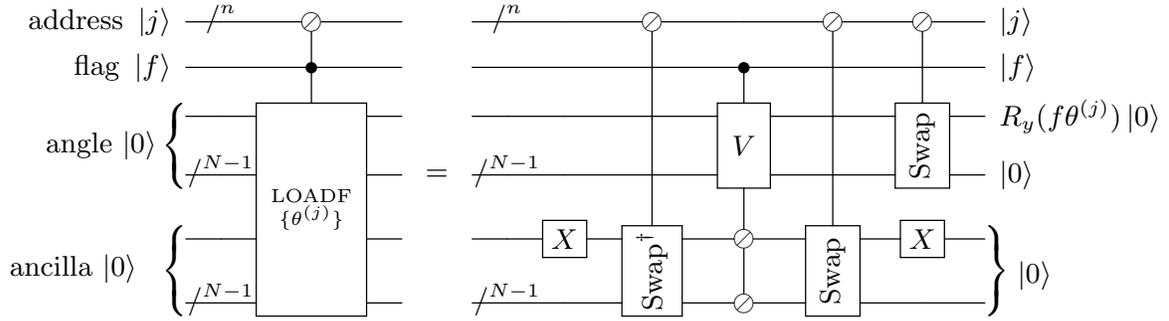

%%%%%%%%%%%% END FIGURE %%%%%%%%%%%%%%%%%%

Our implementation of LOADF involves applying doubly-controlled-$R_y$ gates, which can be done using the decomposition in Fig.~\ref{fig:controlled-ry} of App.~\ref{app:decompositions} and adding an extra control. Indeed, we will need to apply a gate we call $V$, which enacts a unitary specified by the equation
\begin{equation}
V\bigg(\ket{f}\ket{0}_N\ket{x_0}_1\ket{x_1}_1\ldots \ket{x_{N-1}}_1\bigg) = \ket{f}\left[\bigotimes_{k=0}^{N-1}R_y(fx_k\theta^{(k)})\ket{0}_1\right]\ket{x_0}_1\ket{x_1}_1\ldots \ket{x_{N-1}}_1\ket{0}_N\,.
\end{equation}
In other words, for each $k = 0, \ldots , N-1$, $V$ performs a $R_y(\theta^{(k)})$ rotation  on the $k$-th qubit of the angle register if the flag bit is $1$ and the $k$-th qubit of the ancilla register is $1$. This can be implemented in $\bigo{1}$ total $T$-depth using the doubly-controlled-rotation decomposition of Fig.~\ref{fig:controlled-ry} with the parallel-Toffoli construction implicit from Fig.~\ref{fig:cswap2}. 

%%%%%%%%%%%% START FIGURE %%%%%%%%%%%%%%%%%%

\begin{figure}[ht!]
\centering
\scalebox{1.2}{
\mbox{
\Qcircuit @C=0.6em @R=1.0em {
\lstick{\ket{j_0}} & \qw    & \qw      & \qw      & \ctrl{10} & \ctrl{8} & \qw            & \qw            & \qw            & \qw            & \ctrl{8} & \ctrl{10}& \qw       & \qw      & \ctrl{4} & \ctrl{6} & \qw      & \rstick{\ket{j_0}} \qw                   \\
\lstick{\ket{j_1}} & \qw    & \qw      & \ctrl{8} & \qw       & \qw      & \qw            & \qw            & \qw            & \qw            & \qw      & \qw      & \ctrl{8}  & \qw      & \qw      & \qw      & \ctrl{4} & \rstick{\ket{j_1}} \qw                   \\
\lstick{\ket{f}}   & \qw    & \qw      & \qw      & \qw       & \qw      & \ctrl{1}       & \ctrl{2}       & \ctrl{3}       & \ctrl{4}       & \qw      & \qw      & \qw       & \qw      & \qw      & \qw      & \qw      & \rstick{\ket{f}}   \qw                   \\
\lstick{\ket{0}}   & \qw    & \qw      & \qw      & \qw       & \qw      & \gate{\Ryj{0}} & \qw            & \qw            & \qw            & \qw      & \qw      & \qw       & \qw      & \qswap   & \qw      & \qswap   & \rstick{R_y(f\theta^{(j)})\ket{0}} \qw                   \\ 
\lstick{\ket{0}}   & \qw    & \qw      & \qw      & \qw       & \qw      & \qw            & \gate{\Ryj{1}} & \qw            & \qw            & \qw      & \qw      & \qw       & \qw      & \qswap   & \qw      & \qw      & \rstick{\ket{0}}\qw     \\ 
\lstick{\ket{0}}   & \qw    & \qw      & \qw      & \qw       & \qw      & \qw            & \qw            & \gate{\Ryj{2}} & \qw            & \qw      & \qw      & \qw       & \qw      & \qw      & \qswap   & \qswap   & \rstick{\ket{0}}\qw     \\ 
\lstick{\ket{0}}   & \qw    & \qw      & \qw      & \qw       & \qw      & \qw            & \qw            & \qw            & \gate{\Ryj{3}} & \qw      & \qw      & \qw       & \qw      & \qw      & \qswap   & \qw      & \rstick{\ket{0}}\qw     \\ 
\lstick{\ket{0}}   & \qw    & \gate{X} & \qswap   & \qw       & \qswap   & \ctrl{-4}      & \qw            & \qw            & \qw            & \qswap   & \qw      & \qswap    & \gate{X} & \qw      & \qw      & \qw      & \rstick{\ket{0}}\qw     \\ 
\lstick{\ket{0}}   & \qw    & \qw      & \qw      & \qw       & \qswap   & \qw            & \ctrl{-4}      & \qw            & \qw            & \qswap   & \qw      & \qw       & \qw      & \qw      & \qw      & \qw      & \rstick{\ket{0}}\qw     \\ 
\lstick{\ket{0}}   & \qw    & \qw      & \qswap   & \qswap    & \qw      & \qw            & \qw            & \ctrl{-4}      & \qw            & \qw      & \qswap   & \qswap    & \qw      & \qw      & \qw      & \qw      & \rstick{\ket{0}}\qw     \\ 
\lstick{\ket{0}}   & \qw    & \qw      & \qw      & \qswap    & \qw      & \qw            & \qw            & \qw            & \ctrl{-4}      & \qw      & \qswap   & \qw       & \qw      & \qw      & \qw      & \qw      & \rstick{\ket{0}}\qw     \\ 
}}
}
\caption{\label{fig:load_min_depth}Example circuit for LOADF operation on $n=2$. Each $R_y(\theta^{(k)})$ gate can be performed in $T$-depth $\bigo{1}$ plus the depth required to decompose the single-qubit rotation into Clifford + $T$, scaling as $\bigo{\log{1/\delta}}$ if the target precision is $\delta$.}
\end{figure}
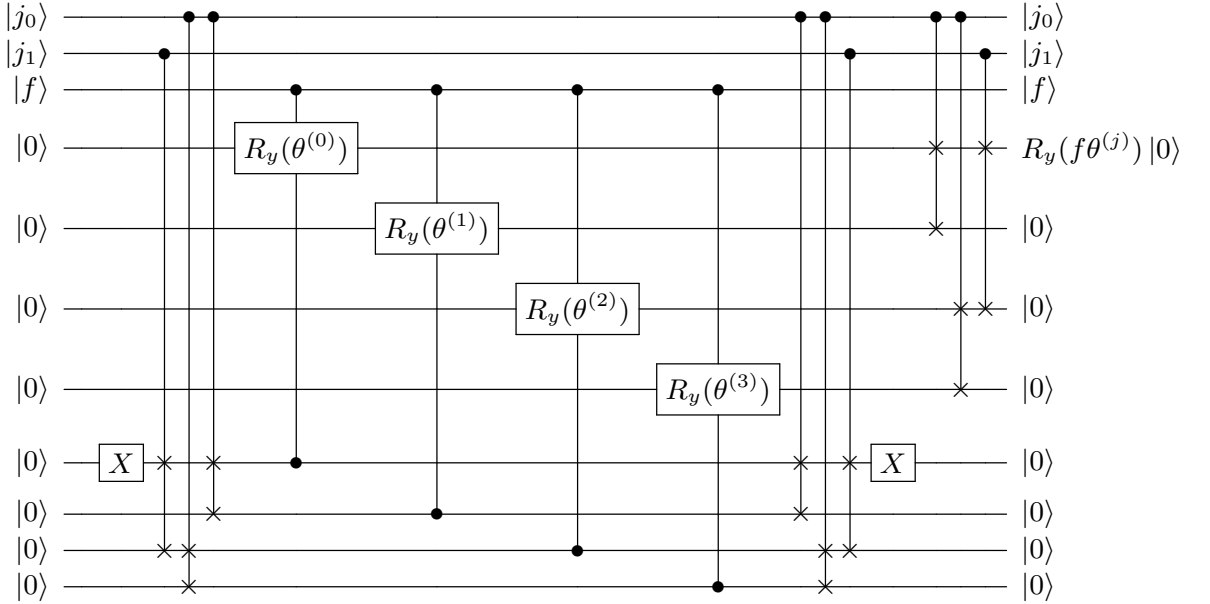

%%%%%%%%%%%% END FIGURE %%%%%%%%%%%%%%%%%%

Since LOADF only applies $R_y(\theta^{(j)})$ and no other rotations, we must first prepare the state $\ket{x_0}_1\ldots \ket{x_{N-1}}_1$ for which $x_j = 1$ and $x_k = 0$ for all $k \neq j$, so that application of $V$ applies the correct rotation. To prepare this state, we flip the first of the $N$ ancilla bits to $\ket{1}$, and then run the inverse of the controlled-swap network from the $\LOADSS$ operation in Fig.~\ref{fig:select-swap-high-level}, which moves the $\ket{1}$ state into the $j$-th ancilla. Application of $V$ computes $R_y(f\theta^{(j)})\ket{0}$ into the $j$-th position of the angle register, while leaving all other angle registers in $\ket{0}$. Next, the controlled-swap network from $\LOADSS$ is applied to move the $j$-th angle register to the first position and the $j$-th ancilla register back to the first position, so that an $X$ gate on the first ancilla leaves the entire ancilla register back in $\ket{0}$. This implementation realizes the LOADF operation and is depicted in Fig.~\ref{fig:LOADF}. A complete example for $n=2$ is given in Fig.~\ref{fig:load_min_depth}.

The LOADF operation loads a single angle conditioned on a single flag. In general, we can load $D$ states $R_y(\theta^{(j)}_r)\ket{0}_1$ for $r=1,\ldots, D$ by copying the flag, angle, and ancilla registers $D$ times and running $D$ copies of LOADF in parallel, where the controlled-Swap operations for all $D$ copies share a common control (the $n$-qubit address register).\footnote{For controlled-state preparation (see Fig.~\ref{fig:controlled_state_prep_prerotated}), we will need to load $N-1$ angles conditioned on $N-1$ separate flags, so we will take $D=N-1$.} The address qubits only ever act as controls for multi-qubit controlled swap gates, so increasing $D$ does not increase the $T$-depth. 

Note also that the LOADF operation requires controlled-swap gates with the target registers in states of the form $R_y(\theta)\ket{0}$, which does not admit the use of the phase-incorrect controlled-swap that can be used for $\LOADSS$ and $\LOADBB$. Therefore, we utilize the parallel implementation shown in Fig.~\ref{fig:cswap2}. This circuit shows a controlled swap between arbitrary two-qubit states with the assistance of two ancilla qubits. The Toffoli construction from \cite{jones2013} then allows one to implement the Toffoli with a $T$-depth of 1 and a $T$-count of 4, at the cost of a single extra clean ancilla qubit.

%%%%%%%%%%%%%%%%%%%%%%%%%%%%%%%%%%%%%%%%%%%%%%%%%%%%%%%%%%%%%%%%%%%%%%%%%%%%%%%%

\subsection{QRAM resource estimates}

We now present the non-Clifford resource estimates for the three data-loading operations implementations discussed above, summarized in Tab.~\ref{tab:load_resources}. For LOAD, we report the resources required for the hybrid scheme that allows one to vary the circuit depth and width via the parameter $\lambda \in \{0,1,\ldots, \log(N)\}$. For both $\LOADSS$ and $\LOADBB$, the $T$-depth is minimized at $\bigo{n}$ when $\lambda = n$. Taking $D = \bigo{1}$, the $\LOADSS$ operation achieves minimum $T$-count of $\bigo{2^{n/2}}$ when $\lambda \approx n/2$, while $\LOADBB$ requires $\Omega(2^n)$ $T$-count for all $\lambda$.   For $\lambda = 0$ one has unary iteration \cite{Babbush2018unary}, which yields the minimal-qubit-count but maximal-depth case. 

Our counts can be verified using the circuit diagrams in Figs.~\ref{fig:select-swap-high-level}, \ref{fig:bb-high-level}, and \ref{fig:LOADF}, along with the following additional observations, which reference gate decompositions given in App.~\ref{app:decompositions}. 
\begin{itemize}
\item The Select gate from Fig.~\ref{fig:select-swap-high-level} and the sequence of multiply-controlled Toffoli gates in Fig.~\ref{fig:bb-high-level} are accomplished with unary iteration which requires $4(2^s-1)$ (non-parallelizable) $T$ gates and $s-1$ additional ancilla qubits that are not depicted in those figures (see Ref.~\cite{Babbush2018unary} for construction). We assume intermediate measurements and subsequent classically controlled Clifford gates are allowable; if not, an additional $4(s-1)$ $T$ gates would be needed. 
\item The Swap gate that appears in Figs.~\ref{fig:select-swap-high-level} and \ref{fig:LOADF} is accomplished using $2^\lambda-1$ controlled-swap gates, occuring over $\lambda$ parallel layers. For $\LOADSS$, these parallel controlled-swap gates are implemented with the construction in Fig.~\ref{fig:multi_cswap}, costing total $T$-count $4(2^\lambda-1)$ and $T$-depth $4\lambda$. For LOADF (where $\lambda=n$), the goal is to give the minimal possible depth, so we choose to implement the parallel controlled-swap gates using the construction in Fig.~\ref{fig:cswap2} (costing $2^{n-1}$ ancilla qubits) where we perform each of the Toffolis using the depth-1 count-4 construction of Ref.~\cite{jones2013} (costing $2^{n-1}$ additional ancilla qubits and requiring intermediate measurements). The total $T$-depth for the Swap gate is $n$ and the total $T$-count is $4(2^n-1)$. The final two swap gates of Fig.~\ref{fig:LOADF} are performed in parallel. 
\item The costs of the $\text{BB}_j$ gates of Fig.~\ref{fig:bb-high-level} are evaluated by inspection of Fig.~\ref{fig:bucket-brigade} and quoted in the caption of that figure. 
\item The controlled-controlled-rotations appearing within the circuit for the gate $V$ of Fig.~\ref{fig:LOADF} are implemented with Fig.~\ref{fig:controlled-ry}, which introduces a set of parallel Toffoli gates and single-qubit rotations. Each of these parallel Toffoli gates is implemented the same way as parallel controlled-swap gates, reusing the same $2^n$ ancillas, contributing depth $1$ and count $4 \cdot 2^n$. Note that if it is known that the flags are set to 1, these Toffolis become CNOTs, which are Clifford gates. Each single-qubit rotation costs $T$-depth and count equal to $R_y$, where $R_y$ is the number of $T$ gates in the gate synthesis of a single qubit rotation unitary, for which the leading term is $3\log(1/\delta)$ when the target precision is $\delta$ (see Eq.~\eqref{eq:Ry_prerotated} of the error analysis in Sec.~\ref{sec:error}). 
\item For LOADF, Fig.~\ref{fig:LOADF} is for $D=1$; to generalize to larger $D$, the flag, angle, and ancilla registers are copied $D$ times, and all gates are performed in parallel, multiplying the $T$-count by a factor of $D$ and leaving the depth unchanged. 
\end{itemize}

%%%%%%%%%%%% START TABLE %%%%%%%%%%%%%%%%%%

\begin{table}[h]
\begin{center}
\caption{Resource counts for LOAD and LOADF operations, which coherently load one out of $2^n$ possible $D$-qubit data registers, controlled by an $n$-qubit address register. LOAD loads $D$ bits of classical data, and is accomplished using one of two models of QRAM implementation, select-swap (SS) and bucket-brigade (BB), each of which admits a depth/width tradeoff governed by integer parameter $0 \leq \lambda \leq n$.  LOADF performs a more general data loading operation where each of the $D$ data qubits is left in the state $\cos(\theta/2)\ket{0} + \sin(\theta/2)\ket{1}$ for some $\theta \in [0,2\pi]$. The parameter $R_y$ is the $T$-count of the single-qubit Clifford+$T$ gate sequence used to synthesize each single-qubit state, which scales as $\bigo{\log(1/\delta))}$ with the desired precision $\delta$ (see Eq.~\eqref{eq:Ry_prerotated}). We do not examine a depth/width tradeoff for LOADF as it is only used in our minimal depth construction. }
\label{tab:load_resources}
\renewcommand{\arraystretch}{1.4}
\begin{tabular}{l|l|l|l}
\textbf{Resource}   & \textbf{LOAD}${}_{\textbf{ss}} $                  & \textbf{LOAD}${}_{\textbf{bb}} $                                   & \textbf{LOADF} \\
\hhline{=|=|=|=}                                                                                                                                                                     
%\toprule[1.5pt]                                                                                                                                                                             
\textbf{\# Qubits}  & $D \cdot 2^\lambda + 2n-\lambda -1$               & $(2D+1) \cdot 2^\lambda + 2n -2$                                  & $4D\cdot 2^n  + n + D$           \\
\hline                                                                                                                                                                                                                                                               
\textbf{$T$-Depth}  & $4 \cdot 2^{n - \lambda} + 4\lambda - 4$          & $(48 \lambda - 36)\cdot 2^{n - \lambda} -4$                       & $2n + 2R_y + 2$                  \\
\hline                                                                                                                     
\textbf{$T$-Count}  & $\begin{aligned}[t] & 4D \cdot 2^\lambda + 4 \cdot 2^{n-\lambda}\\ &-4D-4 \end{aligned}$ & $\begin{aligned}[t] &16(D+1)\cdot 2^{n} \\ &-(8D+8\lambda+12) \cdot 2^{n-\lambda} -4 \end{aligned}$    & $(2R_y + 20)D \cdot 2^n -12D$    \\ 
\end{tabular}
\end{center}
\end{table}

%%%%%%%%%%%% END TABLE %%%%%%%%%%%%%%%%%%

%%%%%%%%%%%%%%%%%%%%%%%%%%%%%%%%%%%%%%%%%%%%
%                                          %
%              Q STATE PREP                %
%                                          %
%%%%%%%%%%%%%%%%%%%%%%%%%%%%%%%%%%%%%%%%%%%%

\section{Quantum State Preparation}\label{sec:state_prep}

\subsection{Overview}

In Sec.~\ref{sec:introduction-strategy}, we described how block-encoding is reduced to controlled-state preparation. Here we give explicit circuits for state preparation and controlled-state preparation. In Sec.~\ref{sec:state_prep_conceptual}, we describe the conceptual approach to state preparation coming from prior literature. In Sec.~\ref{sec:preloaded}, we give circuits for what we call the ``fixed-precision'' version of state preparation and controlled-state preparation. Then, in Sec.~\ref{sec:pre-rotated}, we give circuits for what we call the ``pre-rotated'' version of state preparation and controlled-state preparation. Both versions have the same conceptual framework, but differ on other aspects. The fixed-precision version of controlled-state preparation utilizes either the SS-QRAM or BB-QRAM circuit from Sec.~\ref{sec:qram} as a subroutine, and is compatible with the depth/width trade-off parameterized by $\lambda$. Choosing $\lambda=0$ with the SS-QRAM leads to our minimal $T$-count construction. Meanwhile, the pre-rotated version is used to achieve minimal $T$-depth. Indeed, the $T$-depth for preparing an $N$-dimensional state to error $\epsilon$ scales as $\bigo{\log(N/\epsilon)}$ for the pre-rotated construction, an asymptotic improvement over $\bigo{\log^2(N/\epsilon)}$ from Ref.~\cite{low2018trading}.

%%%%%%%%%%%%%%%%%%%%%%%%%%%%%%%%%%%%%%%%%%%%%%%%%%%%%%%%%%%%%%%%%%%%%%%%%%%%%%%%

\subsection{Rotation angles and binary tree data structure}\label{sec:state_prep_conceptual}

Prior to the specific circuit-level instantiations, we describe our general approach to state preparation, which has appeared throughout the literature \cite{Kaye2001statePrep,grover2002creating,kerenidis2016quantum,low2018trading}. The task is to construct a circuit that creates an $n$-qubit quantum state $\ket{\psi}_n = ||\vec{\beta}||^{-1}\sum_{j=0}^{N-1}\beta_j \ket{j}_n$ given a list of its coefficients $\{\beta_j\}$, where $N=2^n$ is a power of two and zero padding of the classical data can ensure this, and $||\vec{\beta}||$ denotes the Euclidean vector norm to ensure normalization. We use the notation $\vec{\beta}$ to indicate the vector of values $[\beta_0, \beta_1, \cdots, \beta_{N-1}]$ and $\vec{\beta}_u^v = [\beta_u, \beta_{u+1}, \cdots , \beta_{v-1}]$ to denote the vector of values between the high $(v)$ and low $(u)$ indices with $v > u$. 

\begin{figure}
\centering
\includegraphics[width=8.0cm]{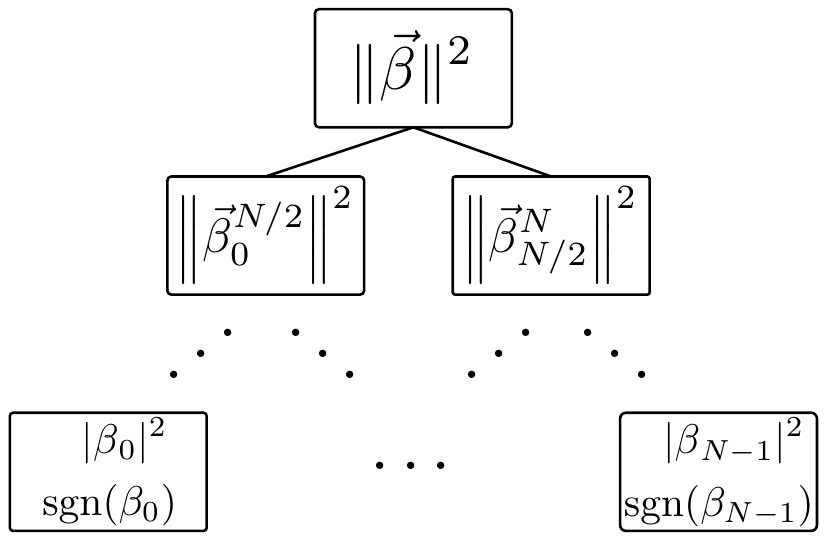}
\caption{\label{fig:generic_tree} Illustration of the binary tree data structure that is used to create the quantum state $\ket{\psi} = ||\vec{\beta}||^{-1}\sum_{j=0}^{N-1}\beta_j\ket{j}$.}
\end{figure}

We assume knowledge of the real amplitudes $\beta_j$. Given these amplitudes, we can construct a classical binary tree data structure as follows. The tree has depth $n$, and at the leaves we store the values $|\beta_j|^2$ and $\sgn(\beta_j)$ in the leaf nodes of the tree.\footnote{If we wish to allow $\beta_j$ to be complex, we can store a complex phase in place of $\sgn(\beta_j)$} Then, each internal node stores the sum of the two child nodes. This proceeds to the root of the tree, which stores the sum of the squares of all $N$ amplitudes, i.e.~$||\vec{\beta}||^2$. We show an example of such a tree in Fig. \ref{fig:generic_tree}. The binary tree data structure allows efficient updates if single amplitudes change; that is, if an amplitude is updated, only $\log(N)$ of the $2N-1$ nodes need to be recomputed. Thus, the data structure can be efficiently maintained if matrix entries arrive in an on-line fashion \cite{kerenidis2016quantum}.

%%%%%%%%%%%% START FIGURE %%%%%%%%%%%%%%%%%%

%%%%%%%%%%%% END FIGURE %%%%%%%%%%%%%%%%%%

Using the data in the binary tree, we construct a circuit that prepares the state $\ket{\psi}$. The state is initialized to the $\ket{0}_n$ state. In step 1, a $Y$-axis rotation is applied on the first of the $n$ qubits by the angle $\theta_{1} = 2 \acos ( ||\vec{\beta}_{0}^{N/2}||/||\vec{\beta}||)$, as in Eq.~\eqref{eq:Ry_def}, yielding the state
\begin{equation}\label{eq:step1_state}
\ket{\psi_1}_n = \left[ \cos(\theta_{1}/2)\ket{0} + \sin (\theta_{1}/2)\ket{1} \right]\ket{0}_{n-1}.
\end{equation}
This angle can be computed from the values stored at the root of the binary tree and its children. In step 2, a rotation is applied to the second qubit, either by the angle $\theta_{2} = 2 \acos ( ||\vec{\beta}_{0}^{N/4}||/||\vec{\beta}_{0}^{N/2}||)$ or by the angle $\theta_{3} = 2 \acos ( ||\vec{\beta}_{N/2}^{3N/4}||/||\vec{\beta}_{N/2}^{N}||)$, where the angle $\theta_{2}$ is used conditioned on the first qubit being in state $\ket{0}$ and $\theta_{3}$ is used conditioned on the first qubit being in state $\ket{1}$. Note that $\theta_2$ and $\theta_3$ can each be computed from values stored at one level-2 node and its children. This creates the state
\begin{equation}\label{eq:step2_state}
\begin{split}
\ket{\psi_2}_n & = \bigg[ \cos(\theta_{1}/2)\ket{0}\bigg(\cos(\theta_{2}/2)\ket{0} + \sin(\theta_{2}/2)\ket{1}\bigg)\bigg]\ket{0}_{n-2} \\
& + \bigg[ \sin (\theta_{1}/2)\ket{1}\bigg(\cos(\theta_{3}/2)\ket{0} + \sin(\theta_{3}/2)\ket{1}\bigg) \bigg]\ket{0}_{n-2}.
\end{split}
\end{equation}
Collecting terms, we can rewrite this state as 
\begin{equation}\label{eq:step2_state_collected}
\begin{split}
\ket{\psi_2}_n & = \big[ \alpha_{00}\ket{00} + \alpha_{01}\ket{01} + \alpha_{10}\ket{10} + \alpha_{11}\ket{11}\big]\ket{0}_{n-2},
\end{split}
\end{equation}
where
\begin{equation}\label{eq:step2_angles}
\begin{split}
\alpha_{00} &= \cos(\theta_{1}/2)\cos(\theta_{2}/2) = ||\vec{\beta}_{0}^{N/4}||/||\vec{\beta}|| \\
\alpha_{01} &= \sin(\theta_{1}/2)\sin(\theta_{2}/2) = ||\vec{\beta}_{N/4}^{N/2}||/||\vec{\beta}|| \\
\alpha_{10} &= \sin(\theta_{1}/2)\cos(\theta_{3}/2) = ||\vec{\beta}_{N/2}^{3N/4}||/||\vec{\beta}|| \\
\alpha_{01} &= \sin(\theta_{1}/2)\sin(\theta_{3}/2) = ||\vec{\beta}_{3N/4}^{N}||/||\vec{\beta}||.
\end{split}
\end{equation}
In general, for any $w  \in \{1,\ldots,n\}$ and for any $w$-bit binary string $y \in \{0,1\}^w$, we can define $\alpha_y = ||\vec{\beta}_{y2^{n-w}}^{(y+1)2^{n-w}}||/||\vec{\beta}||$, where for the purpose of multiplication on the right-hand side we interpret $y$ the integer associated with its binary string. Note that the values stored at the $2^w$ nodes of level $w$ of the binary tree are precisely $|\alpha_y|^2$ for $y \in \{0,1\}^w$.  Extrapolating from the pattern in steps 1 and 2, we assert that the state after step $w$ is given by
\begin{equation}
\ket{\psi_w}_n = \sum_{y\in \{0,1\}^w}\alpha_y \ket{y}_w \ket{0}_{n-w}\,.
\end{equation}
This formula will hold if we implement the transition from $\ket{\psi_w}_n$ to $\ket{\psi_{w+1}}_n$ as follows:
\begin{equation}\label{eq:stepw_state}
\begin{split}
\ket{\psi_w}_n = \sum_{y\in \{0,1\}^w}\alpha_y \ket{y}_w \ket{0}_{n-w} &\to \sum_{y\in \{0,1\}^w}\alpha_y \ket{y}_w\bigg(\frac{\alpha_{y0}}{\alpha_{y}}\ket{0}_1 + \frac{\alpha_{y1}}{\alpha_{y}}\ket{1}_1\bigg) \ket{0}_{n-w-1}\\
& = \sum_{y\in \{0,1\}^{w+1}}\alpha_y \ket{y}_{w+1}\ket{0}_{n-w-1} = \ket{\psi_{w+1}}\,.
\end{split}
\end{equation}
That is, we apply a rotation to qubit $w+1$ by an angle $\theta_{1y}$, where
\begin{equation}
\label{eq:state_prep_theta}
\theta_{1y} = 2\acos (\alpha_{y0}/\alpha_{y})\,,
\end{equation}
which depends on $y$.\footnote{Writing the angle as $\theta_{1y}$ ensures that when the subscript is interpreted as an integer written in binary, the angle at step 1 is $\theta_1$, the angle at step 2 is one of $\{\theta_2,\theta_3\}$, the angle at step 3 is one of $\{\theta_4,\theta_5,\theta_6,\theta_7\}$, etc.}
At the final iteration, the state becomes
\begin{equation}\label{eq:stepf_state}
\ket{\psi_n}_n = \sum_{y\in \{0,1\}^n}\alpha_y \ket{y} = \sum_{y\in \{0,1\}^n}\frac{|\beta_y|}{||\vec{\beta}||} \ket{y} \equiv \sum_{j=0}^{N-1}\frac{|\beta_j|}{||\vec{\beta}||} \ket{j},
\end{equation}
where in the final equivalence, we switched from binary notation, denoted by the variable $y$ to decimal notation denoted by the variable $j$. This is exactly the state that we set out to prepare, up to the sign of the amplitude. To set the correct sign, we apply a $(-1)$ phase to any state $\ket{j}$ for which the $j$-th leaf in the binary tree indicates a sign of $(-1)$. In practice, this is accomplished by loading the sign bit into an ancilla register and applying a Pauli-$Z$ gate, and then unloading the sign bit.\footnote{We note that complex amplitudes could also be considered; in that case, one would load the complex phase into an ancilla register and then apply a controlled rotation by that angle about the $Z$ axis.}

The binary tree data structure in Fig.~\ref{fig:generic_tree} has $2N-1$ nodes. All nodes must store the value $||\vec{\beta}_u^v||^2$ and the leaf node must also store a single bit for the sign. However, we notice that in the above construction, all that matters is the angle of rotation, and there are only $N-1$ distinct angles, which are each inferred from the values at two of the nodes. (For example, $\theta_{1} = 2 \acos ( ||\vec{\beta}_{0}^{N/2}||/||\vec{\beta}||)$.) Thus, in our resource analysis, we assume that these angles are classically pre-computed to avoid the need for any arithmetic on the quantum computer. Ultimately, the information about these angles enters the circuit through the coherent data-loading operations discussed in Sec.~\ref{sec:qram}.

For controlled-state preparation, we are tasked with preparing one of $N$ different arbitrary states, depending on the setting of a control register. To do so, we must compute the $N-1$ angles for each of these $N$ states, giving $N(N-1)$ distinct angles, which can be organized into $N$ separate binary tree data structures. 

%%%%%%%%%%%%%%%%%%%%%%%%%%%%%%%%%%%%%%%%%%%%%%%%%%%%%%%%%%%%%%%%%%%%%%%%%%%%%%%%

\subsection{Fixed-precision circuit}\label{sec:preloaded}

The high-level protocol described in the previous section calls for applying $n$ single-qubit rotations by an angle that depends on the setting of other qubits. Our ``fixed-precision''  state preparation protocol, which we describe in this section, performs each of these rotations by loading a binary representation of the correct angle up to some fixed precision (i.e.~an approximation of the angle with a pre-specified number of bits), performing a controlled $R_y$ rotation by that angle, and then uncomputing the binary description of the angle. 

The angle-loading step is essentially a QRAM query since a different angle must be loaded for each control setting as in Eq.~\eqref{eq:qram_query}. However, in this application, it is allowable to leave garbage in an ancilla register as long as the garbage is eventually uncomputed. Our implementation of the angle-loading step assumes that the initial state has a $t$-bit description of all $N-1$ angles stored in $N-1$ $t$-qubit ancilla registers. The circuit consists entirely of controlled-swap gates that shuffle these $N-1$ ancillas to move the correct angle into the first position, leaving the other $N-2$ registers in a garbage state that is entangled with the data. These circuits are very similar to the swap portion of the select-swap networks described in Sec.~\ref{sec:select_swap}.

%%%%%%%%%%%% START FIGURE %%%%%%%%%%%%%%%%%%

\begin{figure}[t!]
\centering
\scalebox{1.2}{
\begin{tabular}{l}
\Qcircuit @C=0.8em @R=0.7em {
\lstick{\ket{0}}                &  \numq{0.1cm}{n} \qw & \mltg{3}{\text{SP}} & \qw &              && \gate{R_y}    & \ctrlslash{1} & \gate{R_y}     & \ctrlslash{1} & \gate{R_y}     & \qw & 
\cdots && \ctrlslash{1} & \gate{R_y}     & \ctrlslash{1}   & \qw      & \ctrlslash{1}       & \rstick{\ket{\psi}}\qw               \\
                                &  \numq{0.1cm}{t} \qw & \ghost{\text{SP}}   & \qw &\rb{-0.8cm}{=}&& \ctrlslash{-1}& \mltg{2}{S_2} & \ctrlslash{-1} & \mltg{2}{S_3} & \ctrlslash{-1} & \qw & 
\cdots && \mltg{2}{S_n} & \ctrlslash{-1} & \mltg{2}{S_\pm} & \gate{Z} & \mltg{2}{S^\dagger} & \qw                                  \\
                                &  \numq{0.1cm}{a} \qw & \ghost{\text{SP}}   & \qw &              && \qw           & \ghost{S_2}   & \qw            & \ghost{S_3}   & \qw            & \qw & 
\cdots && \ghost{S_n}   & \qw            & \ghost{S_\pm}   & \qw      & \ghost{S^\dagger}   & \qw                                  \\
\lstick{\ket{\bar{b}}}      &  \numq{0.1cm}{N} \qw & \ghost{\text{SP}}   & \qw &              && \qw           & \ghost{S_2}   & \qw            & \ghost{S_3}   & \qw            & \qw & 
\cdots && \ghost{S_n}   & \qw            & \ghost{S_\pm}   & \qw      & \ghost{S^\dagger}   & \rstick{\ket{\bar{b}}}  \qw          \\
{\inputgroupv{2}{3}{0.8em}{1.0em}{\raisebox{-0.2cm}{$\ket{\bar{\Theta}}$}}}
{\outputgroupv{2}{3}{20}{0.8em}{1.0em}{\raisebox{-0.2cm}{$\ket{\bar{\Theta}}$}}}
} 
\end{tabular}
}
\caption{\label{fig:optimized_state_prep_circuit}Fixed-precision state preparation circuit, which approximately prepares the state $\ket{\psi}_n$ into the first $n$ qubits. The protocol requires an ancilla register initialized in a computational basis state $\ket{\bar{\Theta}}_{(N-1)t}\ket{\bar{b}}_N$ containing information about the $N-1$ rotation angles and $N$ sign bits. Note that $a=(N-2)t$ for a total of $D=(N-1)t+N$ ancillas. The circuit alternates between swapping a $t$-bit description of the next angle into the first ancilla register (denoted by $S_p$ for $p=2,\ldots,n$), and rotating a single qubit by the angle stored in that register (denoted by $R_y$ in the figure). The symbol $\oslash$ indicates that a different rotation angle is performed for each of the $2^t$ possible settings of the register. The operation $S_1$ is omitted because it is the identity, and the other $S_p$ operations can each be completed in constant $T$-depth; an example of this implementation for $n=3$ is shown in Fig.~\ref{fig:swap_circuit}. The $Z$ gate acts only on the first qubit of the $t$-qubit register, which contains the appropriate sign bit after application of $S_\pm$. }
\end{figure}
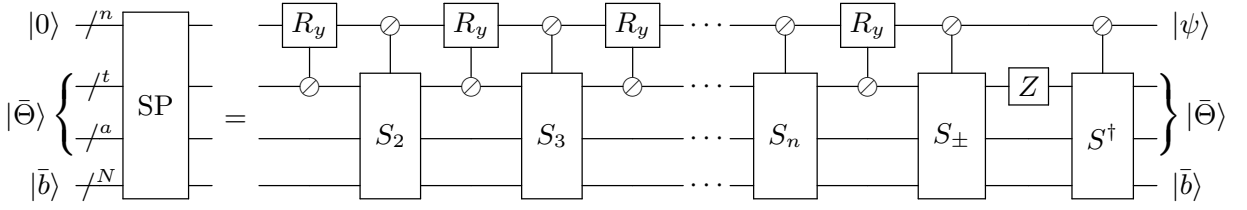

%%%%%%%%%%%% END FIGURE %%%%%%%%%%%%%%%%%%

%%%%%%%%%%%% START FIGURE %%%%%%%%%%%%%%%%%%

\begin{figure}[h!]
\centering
\scalebox{1.2}{
\mbox{
\Qcircuit @C=1.0em @R=1.4em {
\lstick{\ket{0}}                & \qw                & \gate{R_y}      & \ctrl{5} & \ctrl{7} & \ctrl{9} & \qw            & \qw             & \qw      & \qw            & \qw             & \qw       &        & \\
\lstick{\ket{0}}                & \qw                & \qw             & \qw      & \qw      & \qw      & \qw            & \gate{R_y}      & \ctrl{7} & \qw            & \qw             & \qw       &        & \\
\lstick{\ket{0}}                & \qw                & \qw             & \qw      & \qw      & \qw      & \qw            & \qw             & \qw      & \qw            & \gate{R_y}      & \qw       & \cdots & \\
\lstick{\ket{\bar{\theta}_{1}}} & \numq{0.1cm}{t}\qw & \ctrlslash{-3}  & \qw      & \qw      & \qw      & \qswap \qwx[1] & \ctrlslash{-2}  & \qw      & \qswap \qwx[3] & \ctrlslash{-1}  & \qw       &        & \\ 
\lstick{\ket{\bar{\theta}_{2}}} & \numq{0.1cm}{t}\qw & \qw             & \qswap   & \qw      & \qw      & \qswap         & \qw             & \qw      & \qw            & \qw             & \qw       &        & \\ 
\lstick{\ket{\bar{\theta}_{3}}} & \numq{0.1cm}{t}\qw & \qw             & \qswap   & \qw      & \qw      & \qw            & \qw             & \qw      & \qw            & \qw             & \qw       & \cdots & \\ 
\lstick{\ket{\bar{\theta}_{4}}} & \numq{0.1cm}{t}\qw & \qw             & \qw      & \qswap   & \qw      & \qw            & \qw             & \qswap   & \qswap         & \qw             & \qw       &        & \\ 
\lstick{\ket{\bar{\theta}_{5}}} & \numq{0.1cm}{t}\qw & \qw             & \qw      & \qswap   & \qw      & \qw            & \qw             & \qw      & \qw            & \qw             & \qw       &        & \\ 
\lstick{\ket{\bar{\theta}_{6}}} & \numq{0.1cm}{t}\qw & \qw             & \qw      & \qw      & \qswap   & \qw            & \qw             & \qswap   & \qw            & \qw             & \qw       & \cdots & \\ 
\lstick{\ket{\bar{\theta}_{7}}} & \numq{0.1cm}{t}\qw & \qw             & \qw      & \qw      & \qswap   & \qw            & \qw             & \qw      & \qw            & \qw             & \qw       &        & \\ 
{\gategroup{1}{4}{10}{7}{0.9em}{--}}
{\gategroup{2}{9}{9}{10}{0.9em}{--}}
}}
}
\caption{\label{fig:swap_circuit} Example three-qubit fixed-precision state preparation circuit. For simplicity of presentation, the sign bits and the inverse operation $S^\dagger$ are omitted. The two boxes correspond to implementations of $S_2$ and $S_3$ from Fig.~\ref{fig:optimized_state_prep_circuit}. The key point is that all controlled-swaps in the implementation of $S_p$ have a single common control, which allows them all to be implemented with $T$-depth 4, independent of $n$, as shown in Fig.~\ref{fig:multi_cswap} of App.~\ref{app:decompositions}.}
\end{figure}
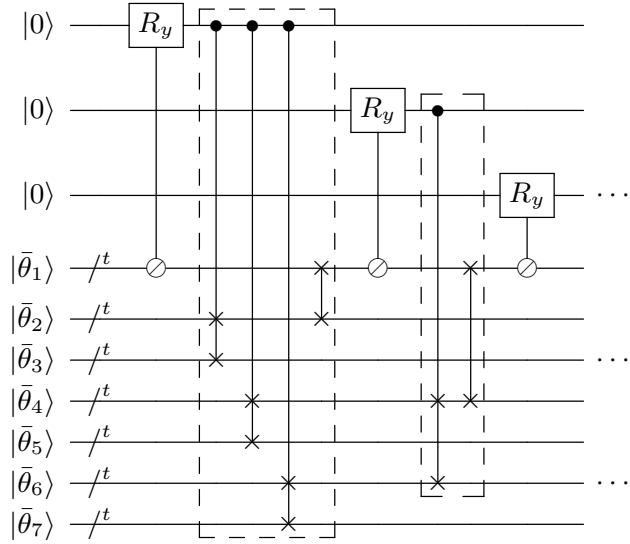

%%%%%%%%%%%% END FIGURE %%%%%%%%%%%%%%%%%%

Given the set of $N-1$ angles $\{\theta_1,\ldots,\theta_{N-1}\}$ needed to create the state $\ket{\psi}_n$, let $\ket{\bar{\theta}_j}_t$ be a computational basis state corresponding to the binary representation of the quantity $2^t\theta_j/\pi$, rounded to the nearest integer.  Moreover, let $\{s_0,\ldots,s_{N-1}\}$ denote the sign bits for each of the $N$ computational basis states. Define $\ket{\bar{\Theta}}_{(N-1)t} = \ket{\bar{\theta}_1}_t\ket{\bar{\theta}_2}_t\ket{\bar{\theta}_3}_t\ldots\ket{\bar{\theta}_{N-1}}_t$ and $\ket{\bar{s}}_N = \ket{s_0}_1\ket{s_1}_1\ldots \ket{s_{N-1}}_1$. Then, we assume that the state preparation protocol acts on an initial state of $n + D$ qubits, with $D = (N-1)t+N$, given by
\begin{equation}
\ket{0}_n\ket{\bar{\Theta}}_{(N-1)t} \ket{\bar{s}}_N\,. 
\end{equation}
This is a computational basis state and thus can be prepared in a single Clifford layer by applying $X$ gates on the appropriate qubits determined by the classical data. In the case of controlled-state preparation, a different set of $N-1$ angles and $N$ sign bits needs to be loaded depending on the setting of an $n$-qubit control register; in this case the operation $\text{LOAD}_\text{ss}$ or $\text{LOAD}_{\text{bb}}$ defined in Secs.~\ref{sec:select_swap} and \ref{sec:bucket_brigade} with $D = (N-1)t+N$ is used to load in the $D$ bits of angular and sign data. %The state preparation unitary is then defined as

Recall that the state we want to create is $\ket{\psi}_n = ||\vec{\beta}||^{-1}\sum_{j=0}^{N-1}\beta_j \ket{j}$. For each basis state $\ket{j}$ a different sequence of $n$ angles and one sign bit is actually applied. Accordingly, we define $n+1$ controlled swap networks denoted by $S_1, S_2, S_3,\ldots, S_n, S_{\pm}$ (note that $S_1$ will always be the identity operation and can be omitted from the circuit). These can each be written in the form
\begin{equation}\label{eq:S_p}
S_p = \sum_{j=0}^{N-1} \ket{j}\bra{j}_n \otimes S_p^{(j)}\,,
\end{equation}
where $S_p^{(j)}$ acts on the $D$-qubit ancilla register such that the product $S_p^{(j)}S_{p-1}^{(j)}\ldots S_1^{(j)}$ has the action of swapping the $t$-bit description of the angle associated with the $p$-th rotation for $\ket{j}$ into the first $t$-qubit ancilla register, and $S_{\pm}^{(j)}S_n^{(j)}\ldots S_1^{(j)}$ has the action of swapping the sign bit for $\ket{j}$ into the first $1$-qubit register. Importantly, it will be the case that $S_p$ is controlled only on the first $p-1$ bits of $\ket{j}$. 

To prepare $\ket{\psi}_n$, these swap networks are interleaved with controlled single-qubit rotations: for each $p=1,\ldots,n$, after the gate $S_p$, a controlled-$R_y$ rotation is implemented on qubit $p$ controlled by the $t$-bit description of the angle stored in the first register. This is accomplished with $t$ controlled-$R_y$ rotations by a fixed angle with a single bit as the control. After the gate $S_{\pm}$, a $Z$ gate is applied to the first ancilla qubit to apply the correct sign. Finally, the angles are restored to their initial positions by performing the gate $S^{\dagger} = S_1^{\dagger}S_2^{\dagger}\ldots S_p^{\dagger}S_\pm^{\dagger}$. This protocol realizes the unitary SP depicted in Fig.~\ref{fig:optimized_state_prep_circuit} and defined by the equation
\begin{equation}
\text{SP}\left(\ket{0}_n\ket{\bar{\Theta}}_{(N-1)t} \ket{\bar{s}}_N\right) = \ket{\psi}_n\ket{\bar{\Theta}_1}_{(N-1)t}\ket{\bar{s}}_N
\end{equation}

The controlled swap networks in the SS-QRAM circuit have depth $\bigo{p}$ when there are $\bigo{p}$ controls. A straightforward way to implement $S_p$ is to use these networks first to do the reverse of $S_{p-1}$ in depth $\bigo{p}$ to swap out the $(p-1)$-th angle, and second to swap in the correct angle in depth $\bigo{p}$. This would suggest the $T$-depth of performing $S_1,\ldots, S_n,S_\pm$ is $\sum_{p=0}^{n}\bigo{p} = \bigo{n^2}$.  However, we give an optimization that reduces the depth of each $S_p$ to $\bigo{1}$, and thus an overall depth of $\bigo{n}$; this is seen in the example implementations of $S_2$ and $S_3$ for $n=3$ shown in Fig.~\ref{fig:swap_circuit}. The main idea is to avoid undoing the work already accomplished by $S_{p-1}$, and note that each $S_p$ can be controlled on just one of the $n$ data qubits. 

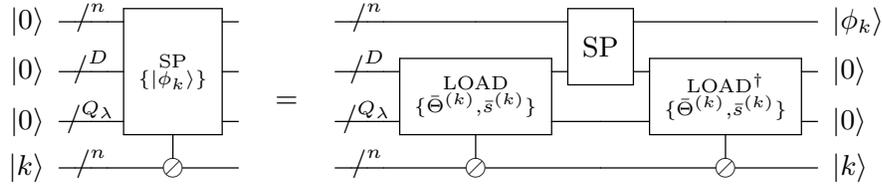
\begin{figure}[ht!]
\centering
\scalebox{1.2}{
\mbox{
\Qcircuit @C=0.55em @R=0.8em {
\lstick{\ket{0}} &\qw &\numq{0.18cm}{n}                 \qw &\qw &\mltg{2}{\substack{\text{SP} \\ \{\ket{\phi_k}\}}} & \qw & & &              & & & &\qw & {\rb{0.18cm}{/${}^{n}$}}         \qw &\qw &\qw                                         & \mltg{1}{\text{SP}} & \qw                                                                              &\rstick{\ket{\phi_k}} \qw \\
\lstick{\ket{0}} &\qw &\numq{0.18cm}{D} \qw                 &\qw &\ghost{\substack{\text{SP} \\ \{\ket{\phi_k}\}}}   & \qw & & &\rb{-0.7cm}{=}& & & &\qw & {\rb{0.18cm}{/${}^{D}$}}         \qw &\qw &\mltg{1}{\substack{\text{LOAD}\\\{\bar{\Theta}^{(k)},\bar{s}^{(k)}\}}} & \ghost{\text{SP}}   & \mltg{1}{\substack{\text{LOAD}^\dagger\\\{\bar{\Theta}^{(k)},\bar{s}^{(k)}\}}} &\rstick{\ket{0}}      \qw \\
\lstick{\ket{0}} &\qw &{\rb{0.18cm}{/${}^{Q_\lambda}$}} \qw &\qw &\ghost{\substack{\text{SP} \\ \{\ket{\phi_k}\}}}   & \qw & & &              & & & &\qw & {\rb{0.18cm}{/${}^{Q_\lambda}$}} \qw &\qw &\ghost{\substack{\text{LOAD}\\\{\bar{\Theta}^{(k)},\bar{s}^{(k)}\}}}    & \qw                 & \ghost{\substack{\text{LOAD}^\dagger\\\{\bar{\Theta}^{(k)},\bar{s}^{(k)}\}}}    &\rstick{\ket{0}}      \qw \\
\lstick{\ket{k}} &\qw &{\rb{0.18cm}{/${}^{n}$}}         \qw &\qw &\ctrlslash{-1}                                     & \qw & & &              & & & &\qw & {\rb{0.18cm}{/${}^{n}$}}         \qw &\qw &\ctrlslash{-1}                                         & \qw                 & \ctrlslash{-1}                                                                   &\rstick{\ket{k}}      \qw \\
}
}
}
\caption{\label{fig:controlled_state_prep}controlled-state preparation circuit diagram for the fixed-precision approach to state preparation.  For each of the $N$ possible settings of the bottom register, a different $n$-qubit state $\ket{\phi_k}$ is prepared in the top register. This is accomplished with three subroutines. The LOAD subroutine, which can be either $\LOADSS$ from Fig.~\ref{fig:select-swap-high-level} or $\LOADBB$ from Fig.~\ref{fig:bb-high-level}, loads in a $t$-bit description of the $N-1$ angles $\ket{\bar{\Theta}^{(k)}}_{(N-1)t}$ and the $N$ sign bits $\ket{\bar{s}^{(k)}}_N$ into the second register of size $D=(N-1)t+N$, with the assistance of $Q_\lambda$ ancillas. The SP subroutine is given in Fig.~\ref{fig:optimized_state_prep_circuit}, and prepares the state $\ket{\phi_k}_n$. Finally, the reverse of the LOAD operation is performed to reset the second register and the ancillas to $\ket{0}$. The instances of controlled-state preparation that appear in the block-encoding unitary $U_A$ of Fig.~\ref{fig:U_A} can be accomplished with this circuit. 
}
\end{figure}

The full block-encoding circuit in Fig.~\ref{fig:U_A} requires a controlled-state preparation, not state preparation. Here there are $N$ different state $\ket{\phi_k}$, each with their own angle data $\bar{\Theta}^{(k)}$ and sign data $\bar{s}^{(k)}$. To perform controlled-state preparation one simply loads this data conditioned on a control register in state $\ket{j}$ using either $\text{LOAD}_\text{ss}$ or $\text{LOAD}_\text{bb}$ defined in Sec.~\ref{sec:qram}. Then the state preparation circuit from Fig.~\ref{fig:optimized_state_prep_circuit} is performed followed by the inverse of the LOAD operation to clear the data registers. This is shown in Fig.~\ref{fig:controlled_state_prep}.

%%%%%%%%%%%% START FIGURE %%%%%%%%%%%%%%%%%% 

%%%%%%%%%%%% END FIGURE %%%%%%%%%%%%%%%%%%

%%%%%%%%%%%%%%%%%%%%%%%%%%%%%%%%%%%%%%%%%%%%%%%%%%%%%%%%%%%%%%%%%%%%%%%%%%%%%%%%

\subsection{Pre-rotated circuit}\label{sec:pre-rotated}

%%%%%%%%%%%% START FIGURE %%%%%%%%%%%%%%%%%%

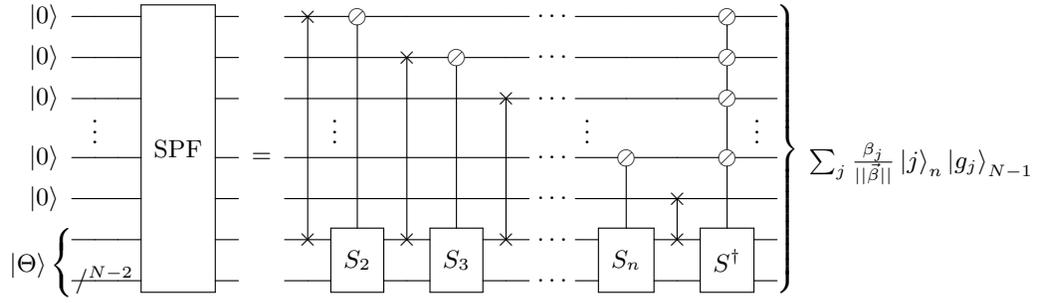
\begin{figure}[ht!]
\centering
\scalebox{1.05}{
\begin{tabular}{l}
\Qcircuit @C=0.9em @R=0.7em {
\lstick{\ket{0}} & \qw                                              & \qw & \mltg{7}{\text{SPF}} & \qw &               && \qswap \qwx[6] & \ctrlslash{6} & \qw            & \qw           & \qw            & \qw & 
\cdots && \qw           & \qw            & \ctrlslash{1}       & \qw \\
\lstick{\ket{0}} & \qw                                              & \qw & \ghost{\text{SPF}}   & \qw &               && \qw            & \qw           & \qswap \qwx[5] & \ctrlslash{5} & \qw            & \qw & 
\cdots && \qw           & \qw            & \ctrlslash{1}       & \qw \\
\lstick{\ket{0}} & \qw                                              & \qw & \ghost{\text{SPF}}   & \qw &               && \qw            & \qw           & \qw            & \qw           & \qswap \qwx[4] & \qw & 
\cdots && \qw           & \qw            & \ctrlslash{2}       & \qw \\
&\rb{0.1cm}{$\vdots$}&& &&&&\rb{0.1cm}{\shiftright{0.6cm}{$\vdots$}} &&&&&&\rb{0.1cm}{\shiftright{0.8cm}{$\vdots$}} &&&&&\rb{0.1cm}{\shiftleft{0.3cm}{$\vdots$}} \\
\lstick{\ket{0}} & \qw                                              & \qw & \ghost{\text{SPF}}   & \qw & \rb{0cm}{=}   && \qw            & \qw           & \qw            & \qw           & \qw            & \qw & 
\cdots && \ctrlslash{2} & \qw            & \ctrlslash{2}       & \qw \\
\lstick{\ket{0}} & \qw                                              & \qw & \ghost{\text{SPF}}   & \qw &               && \qw            & \qw           & \qw            & \qw           & \qw            & \qw & 
\cdots && \qw           & \qswap \qwx[1] & \qw                 & \qw \\
                 & \qw                                              & \qw & \ghost{\text{SPF}}   & \qw &               && \qswap         & \mltg{1}{S_2} & \qswap         & \mltg{1}{S_3} & \qswap         & \qw & 
\cdots && \mltg{1}{S_n} & \qswap         & \mltg{1}{S^\dagger} & \qw \\
                 & {\rb{0.1cm}{\shiftright{-0.5cm}{$/^{N-2}$}}} \qw & \qw & \ghost{\text{SPF}}   & \qw &               && \qw            & \ghost{S_2}   & \qw            & \ghost{S_3}   & \qw            & \qw & 
\cdots && \ghost{S_n}   & \qw            & \ghost{S^\dagger}   & \qw \\
{\inputgroupv{7}{8}{0.8em}{1.0em}{\raisebox{-0.2cm}{$\ket{\Theta}$}}}
{\outputgroupv{1}{8}{19}{0.8em}{1.0em}{\rb{-3.4cm}{\shiftright{-0.3cm}{$\sum_j \frac{\beta_j}{||\vec{\beta}||} \ket{j}_n\ket{g_j}_{N-1}$}}}}
} 
\end{tabular}
}
\caption{\label{fig:optimized_state_prep_circuit_prerotated}Pre-rotated state preparation circuit, which approximately prepares the state $\ket{\psi}_n$ into the first $n$ qubits with garbage. That is, if $\ket{\psi}_n = \sum_j \frac{\beta_j}{||\vec{\beta}||} \ket{j}_n$, the state that is prepared is $\sum_j \frac{\beta_j}{||\vec{\beta}||} \ket{j}_n\ket{g_j}_{N-1}$. Each $S_p$ gate can be performed in $\bigo{1}$ $T$-depth, for a total $T$-depth of $\bigo{\log(N)}$. The entanglement between the $n$-qubit data register and $(N-1)$-qubit garbage register can be uncomputed using a flag mechanism as described in the main text with additional $\bigo{\log(N/\epsilon)}$ cost, where $\epsilon$ is the error on the state prepared. }
\end{figure}

%%%%%%%%%%%% END FIGURE %%%%%%%%%%%%%%%%%%

In this subsection, we present an alternative approach to state preparation which achieves smaller $T$-depth than the fixed-precision version, both practically and asymptotically. The asymptotic $T$-depth scaling is $\bigo{\log(N/\epsilon)}$, with $\epsilon$ the error on the state prepared.  The pre-rotated version takes the idea of pre-computing the angles and signs for state preparation one step further, and encodes them as the amplitude of a single qubit. For a given quantum state $\ket{\psi}_n$, we assume that we have classically computed the $N-1$ angles $\{\theta_r\}_{r=1}^{N-1}$ and $N$ sign bits $\{s_j\}_{j=0}^{N-1}$ stored in the binary tree data structure as discussed previously. Let
\begin{align}
\ket{\theta_r}_1 &= \begin{cases}
\hspace{46pt}\cos(\theta_r/2) \ket{0}_1 + \hspace{53pt}\sin(\theta_r/2) \ket{1}_1                            & \text{if } 1\leq r < N/2 \\
(-1)^{s_{2r-N}}\cos(\theta_r/2) \ket{0}_1+(-1)^{s_{2r-N+1}}\sin(\theta_r/2) \ket{1}_1  & \text{if } N/2 \leq r < N
\end{cases}
\end{align}
and for each $r$ let $V_r$ be a Clifford+$T$ gate decomposition of an $R_y$ rotation by some angle that prepares $\ket{0}\mapsto\ket{\theta_r}$ up to error $\delta$.\footnote{Any $\ket{\theta_r}$ can be written as $R_y(\theta)\ket{0}$ for some $\theta$. Let $H_r$ be a Clifford+$T$ decomposition of $R_y(\theta/2)$ up to error $\delta/2$, so that $V_r = H_r^*H_r$ approximates $R_y(\theta)$ up to error $\delta$, where $H_r^*$ denotes the complex conjugate of $H_r$. Decomposing in this way allows us to give a $\delta$-approximate Clifford+$T$ decomposition for the controlled-$R_y(\theta)$ gate by using $H_r$ and $H_r^\dagger$ in place of $R_y(\theta/2)$ and $R_y(-\theta/2)$ in Fig.~\ref{fig:controlled-ry} of App.~\ref{app:decompositions}.} Note that if we have classically pre-computed the angle $\theta_r$, we can also classically compute a gate sequence $V_r$ that gives a $\delta$ approximation for $R_y(\theta_r)$ in classical time $\polylog(1/\delta)$ \cite{ross2016optimal}. We assume that the input state is the product-state on $n +N-1$ qubits given by
\begin{equation}\label{eq:post_load}
\ket{0}_n\ket{\Theta}_{N-1} = \ket{0}_n\left(\bigotimes_{r=1}^{N-1} \ket{\theta_{r}}_1\right)
\end{equation}
which also acts as the definition of $\ket{\Theta}_{N-1}$. This product state can be prepared by applying each of the $V_r$ to $N-1$ ancillas initially in $\ket{0}$ in parallel. In the case of controlled-state preparation, later, we will use the LOADF operation to prepare a different initial state for each setting of a control register.   

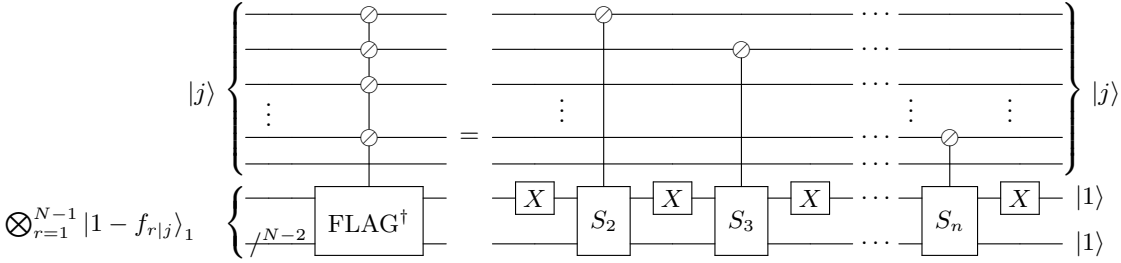
\begin{figure}[ht!]
\centering
\scalebox{1.05}{
\begin{tabular}{l}
\Qcircuit @C=0.9em @R=0.7em {
& \qw                                              & \qw & \ctrlslash{1}                 & \qw &               && \qw      & \ctrlslash{6} & \qw      & \qw           & \qw      & \qw & 
\cdots && \qw           & \qw      & \qw \\
& \qw                                              & \qw & \ctrlslash{1}                 & \qw &               && \qw      & \qw           & \qw      & \ctrlslash{5} & \qw      & \qw & 
\cdots && \qw           & \qw      & \qw \\
& \qw                                              & \qw & \ctrlslash{2}                 & \qw &               && \qw      & \qw           & \qw      & \qw           & \qw      & \qw & 
\cdots && \qw           & \qw      & \qw \\
&\rb{0.1cm}{$\vdots$}&& &&&&\rb{0.2cm}{\shiftright{0.6cm}{$\vdots$}} &&&&&&\rb{0.2cm}{\shiftright{0.8cm}{$\vdots$}} &&&&&\rb{0.2cm}{\shiftleft{1.0cm}{$\vdots$}} \\
& \qw                                              & \qw & \ctrlslash{2}                 & \qw & \rb{0cm}{=}   && \qw      & \qw           & \qw      & \qw           & \qw      & \qw & 
\cdots && \ctrlslash{2} & \qw      & \qw \\
& \qw                                              & \qw & \qw                           & \qw &               && \qw      & \qw           & \qw      & \qw           & \qw      & \qw & 
\cdots && \qw           & \qw      & \qw \\
& \qw                                              & \qw & \mltg{1}{\text{FLAG}^\dagger} & \qw &               && \gate{X} & \mltg{1}{S_2} & \gate{X} & \mltg{1}{S_3} & \gate{X} & \qw & 
\cdots && \mltg{1}{S_n} & \gate{X} & \rstick{\ket{1} }\qw \\
& {\rb{0.1cm}{\shiftright{-0.5cm}{$/^{N-2}$}}} \qw & \qw & \ghost{\text{FLAG}^\dagger}   & \qw &               && \qw      & \ghost{S_2}   & \qw      & \ghost{S_3}   & \qw      & \qw & 
\cdots && \ghost{S_n}   & \qw      & \rstick{\ket{1}} \qw \\
{\inputgroupv{1}{6}{0.8em}{1.0em}{\raisebox{-1.6cm}{$\ket{j}$}}}
{\outputgroupv{1}{6}{18}{0.8em}{1.0em}{\raisebox{-1.6cm}{$\ket{j}$}}}
{\inputgroupv{7}{8}{0.8em}{1.0em}{\raisebox{0cm}{\shiftleft{2.5cm}{$\bigotimes_{r=1}^{N-1} \ket{1-f_{r|j}}_1$}}}}
} 
\end{tabular}
}
\caption{\label{fig:FLAG} Circuit for the adjoint of the FLAG gate used to disentangle the garbage and data registers in the pre-rotated state-preparation protocol. We present $\text{FLAG}^{\dagger}$ rather than $\text{FLAG}$ to draw attention to the similarity between FLAG and the first half of SPF from Fig.~\ref{fig:controlled_state_prep_prerotated}. Controlled on the data qubits in state $\ket{j}$, FLAG switches $n$ of the $N-1$ flag qubits from $\ket{1} \mapsto \ket{0}$ at locations corresponding to the positions of the $n$ angles that are used in the synthesis of amplitude $\ket{j}$.}
\end{figure}

Given the state in Eq.~\eqref{eq:post_load} as input, we perform a nearly identical state preparation circuit to that shown in the previous subsection. The only differences are that the application of the sign bit with $S_\pm$ and a $Z$ gate can be omitted, as the sign bit is built into the state $\ket{\theta_r}$, and that the $nt$ non-Clifford controlled-$R_y$ rotations are replaced by $n$ single-qubit swap gates (which are Clifford gates). That is, rather than use the angle qubit as a control for a rotation, we inject the angle into the data with a simple swap gate, as shown in Fig.~\ref{fig:optimized_state_prep_circuit_prerotated}. For any computational basis state $\ket{j}$, $n$ of the $N-1$ states $\ket{\theta_r}$ are injected into the data. Once the state preparation procedure is complete, we reverse the swap circuit to return all the angles $\ket{\theta_r}_1$ back to their initial positions with the exception that the $n$ angles that were injected are replaced by $\ket{0}_1$. Let $f_{r|j} = 1$ if angle $\theta_r$ is an ancestor of the $j$-th leaf in the binary tree, i.e.~if $\ket{\theta_r}$ was injected into the state in the computational path associated with $\ket{j}$; otherwise $f_{r|j}=0$. Then we can state the action of our protocol, denoted by the unitary SPF and depicted in Fig.~\ref{fig:optimized_state_prep_circuit_prerotated}, by the following equation
\begin{equation}\label{eq:post_injection}
\text{SPF}\left(\ket{0}_n\ket{\Theta}_{N-1}\right) = \sum_{j} \frac{\beta_j}{||\vec{\beta}||} \ket{j}_n \ket{g_j}_{N-1}
\end{equation}
where
\begin{equation}
\ket{g_j}_{N-1} = \bigotimes_{r=1}^{N-1} \ket{(1-f_{r|j})\theta_r}_1\,.
\end{equation}
The state given by Eq.~\eqref{eq:post_injection} has the correct coefficients for each basis state $\ket{j}_n$, but has $\ket{0}_1$ states in place of $\ket{\theta_r}_1$ for the $n$ angles that were swapped into one of the first $n$ registers. In other words, there is garbage leftover that is entangled with the data. In some applications, this garbage might be allowable. However, one can also uncompute the garbage and disentangle the two parts of the state, which we do by computing the ``flag'' bits $1-f_{r|j}$ into ancilla registers, using them as a control to apply a controlled-$V_r$ operation controlled on that bit, and then uncomputing the flag bits. The $N-1$ flag bits can be computed into $N-1$ ancillas using the unitary FLAG, depicted in Fig.~\ref{fig:FLAG} and defined by the equation 
\begin{equation}
\text{FLAG}\left(\sum_{j=0}^{N-1} \alpha_j \ket{j}_n \ket{1}_{N-1}\right) = \sum_{j=0}^{N-1} \alpha_j \ket{j}_n \bigotimes_{r=1}^{N-1}\ket{1-f_{r|j}}_1
\end{equation}
The circuit for FLAG is very similar to the SPF circuit, and, in fact, FLAG can be run in parallel with the $S^\dagger$ portion of the SPF gate.

\begin{figure}[ht!]
\centering
\scalebox{1.2}{
\mbox{
\Qcircuit @C=0.55em @R=0.8em {
\lstick{\ket{0}} &\qw &\rb{0.18cm}{/${}^{n}$}   \qw &\qw &\qw                                         &\mltg{1}{\text{SPF}} & \ctrlslash{2}      & \qw            & \ctrlslash{2}              & \qw      & \rstick{\ket{\psi}} \qw \\
\lstick{\ket{0}} &\qw &\rb{0.18cm}{/${}^{N-1}$} \qw &\qw &\gate{\bigotimes_{r=1}^{N-1} R_y(\theta_r)} &\ghost{\text{SPF}}   & \qw                & \gate{\bigotimes_{r=1}^{N-1} R_y(\theta_r)}       & \qw                        & \qw      & \rstick{\ket{0}}    \qw \\
\lstick{\ket{0}} &\qw &\rb{0.18cm}{/${}^{N-1}$} \qw &\qw &\gate{X}                                     &\qw                  & \gate{\text{FLAG}} & \ctrlslash{-1} & \gate{\text{FLAG}^\dagger} & \gate{X} & \rstick{\ket{0}}    \qw \\
}
}
}
\caption{\label{fig:state_prep_prerotated} Circuit for garbage-free state preparation with pre-rotated method. An arbitrary $n$-qubit state $\ket{\psi}$ is prepared with the assistance of $2(N-1)$ ancilla qubits. The circuit for the SPF gate is given in Fig.~\ref{fig:optimized_state_prep_circuit_prerotated} and the FLAG gate is given in Fig.~\ref{fig:FLAG}. The $X$ gate denotes a Pauli-$X$ on all $N-1$ qubits. The controlled-$R_y$ gate denotes $N-1$ completely parallel controlled $R_y(\theta_r)$ gates, with each flag qubit (third register) acting as a control for a rotation on a different angle qubit (second register). To prepare $\ket{\psi}$ to precision $\epsilon$, the single-qubit rotations must be synthesized with a gate sequence of $T$-depth $\bigo{\log(1/\epsilon)}$, and meanwhile the SPF and FLAG gates incur $\bigo{\log(N)}$ $T$-depth for a total $T$-depth of $\bigo{\log(N/\epsilon)}$. 
}
\end{figure}
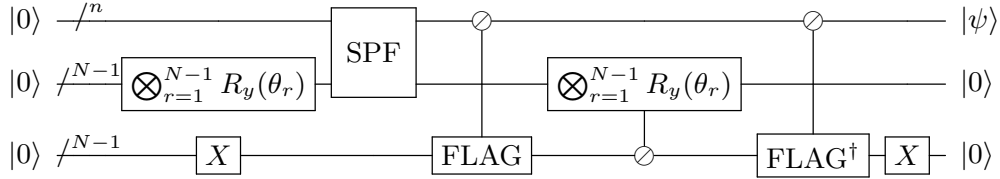

\noindent The full pre-rotated garbage-free state-preparation protocol, including uncomputation of garbage with flags, is depicted in Fig.~\ref{fig:state_prep_prerotated}.

\begin{figure}[ht!]
\centering
\scalebox{1.2}{
\mbox{
\Qcircuit @C=0.55em @R=0.8em {
\lstick{\ket{0}} &\qw &\numq{0.18cm}{n}           \qw &\qw &\mltg{3}{\substack{\text{SP} \\ \{\ket{\phi_k}\}}} & \qw & & &              & & & & \qw &\qw      &\qw                                                        &
 \mltg{1}{\text{SPF}} & \ctrlslash{3}      & \qw                                                               & \ctrlslash{3}              & \qw      & \rstick{\ket{\phi_k}} \qw \\
\lstick{\ket{0}} &\qw &\rb{0.18cm}{/${}^{N-1}$}   \qw &\qw &\ghost{\substack{\text{SP} \\ \{\ket{\phi_k}\}}}   & \qw & & &              & & & & \qw &\qw      &\mltg{1}{\substack{\text{LOADF}\\\\\{\Theta^{(k)}\}}} &
 \ghost{\text{SPF}}   & \qw                & \mltg{1}{\substack{\text{LOADF}^\dagger\\\\\{\Theta^{(k)}\}}} & \qw                        & \qw      & \rstick{\ket{0}}      \qw \\
\lstick{\ket{0}} &\qw &{\rb{0.24cm}{/${}^{Q_F}$}} \qw &\qw &\ghost{\substack{\text{SP} \\ \{\ket{\phi_k}\}}}   & \qw & & &\rb{0cm}{=}   & & & & \qw &\qw      &\ghost{\substack{\text{LOADF}\\\\\{\Theta^{(k)}\}}}   &
 \qw                  & \qw                & \ghost{\substack{\text{LOADF}^\dagger\\\\\{\Theta^{(k)}\}}}   & \qw                        & \qw      & \rstick{\ket{0}}      \qw \\
\lstick{\ket{0}} &\qw &\rb{0.18cm}{/${}^{N-1}$}   \qw &\qw &\ghost{\substack{\text{SP} \\ \{\ket{\phi_k}\}}}   & \qw & & &              & & & & \qw &\gate{X} &\ctrlslash{-1}                                            &
 \qw                  & \gate{\text{FLAG}} & \ctrlslash{-1}                                                    & \gate{\text{FLAG}^\dagger} & \gate{X} & \rstick{\ket{0}}      \qw \\
\lstick{\ket{k}} &\qw &{\rb{0.18cm}{/${}^{n}$}}   \qw &\qw &\ctrlslash{-1}                                     & \qw & & &              & & & & \qw &\qw      &\ctrlslash{-1}                                            &
 \qw                  & \qw                & \ctrlslash{-1}                                                    & \qw                        & \qw      & \rstick{\ket{k}}      \qw \\
}
}
}
\caption{\label{fig:controlled_state_prep_prerotated}controlled-state preparation circuit diagram for the pre-rotated approach to state preparation.  For each of the $N$ possible settings of the bottom register, a different $n$-qubit state $\ket{\phi_k}$ is prepared in the top register, which is accomplished with three subroutines. First, an $X^{\otimes(N-1)}$ gate sets the $N-1$ flag qubits (fourth register) to $1$, and $N-1$ parallel copies of LOADF (Fig.~\ref{fig:LOADF}) load the state $\ket{\Theta^{(k)}}$ into the second register, conditioned on the last register being $\ket{k}$ (note that since all the flags are set to 1, the doubly-controlled rotations that appear in the circuit for LOADF can be replaced with singly-controlled rotations in this instance). Each of these copies uses the same $n$-qubit control, but its own ancilla space of $Q_F$ qubits, so the total ancilla count is $Q_F = (N-1)(2N-1)$. Second, the state is computed into the first register using the SPF operation (Fig.~\ref{fig:optimized_state_prep_circuit_prerotated}), which leaves garbage in the second register. Third, the garbage is uncomputed by setting the flags for the angles that were injected into the circuit to 0 using the FLAG gate (Fig.~\ref{fig:FLAG}), running LOADF in reverse, and then returning the rest of the flags to $\ket{0}$ with $\text{FLAG}^{\dagger}$. The depth of LOADF is $\bigo{\log(N/\epsilon)}$ and the depth of SPF and FLAG is $\bigo{\log(N)}$ for a total depth of $\bigo{\log(N/\epsilon)}$, where $\epsilon$ is the error on the state $\ket{\phi_k}$.  The instances of controlled-state preparation that appear in the block-encoding unitary $U_A$ of Fig.~\ref{fig:U_A} can be accomplished with this circuit. 
}
\end{figure}
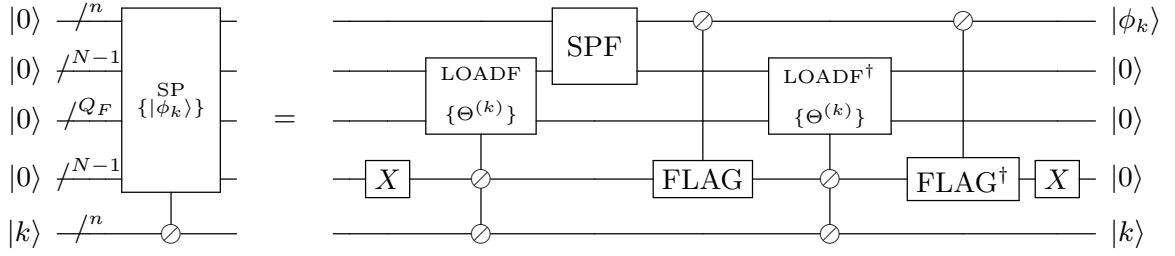

To perform controlled-state preparation, one must first load one of $N$ initial states $\ket{\Theta^{(k)}}_{N-1}$ into an ancilla register, depending on the setting of an $n$-qubit control register in the state $\ket{k}$. This is accomplished by the LOADF operation depicted in Fig.~\ref{fig:LOADF}. In particular, we perform LOADF with $D=N-1$, which is equivalent to $N-1$ copies of LOADF to load each of the states $\{\ket{\theta^{(k)}_r}\}_{r=1}^{N-1}$. Each of these LOADF operations has its own flag qubit, which we initialize to $\ket{1}_1$, but all share the same $n$-qubit control, and use their own $Q_F$-qubit ancilla space. Once $\ket{\Theta^{(k)}}_{N-1}$ is loaded, application of SPF yields the correct state with garbage. To disentangle the garbage and reset all ancillas to $\ket{0}$ we follow a three step process: first, we apply FLAG, which flips flag bits from 1 to 0 in positions that were injected into the data; second, we run the reverse of LOADF, which sends $\ket{\theta_r}_1\mapsto\ket{0}_1$ in all the positions where the control bit is 1, which is precisely the positions where it is not already $\ket{0}_1$; third, we apply FLAG again, followed by an $X$ gate to return all flag bits to $\ket{0}_1$. This process is depicted in Fig.~\ref{fig:controlled_state_prep_prerotated}.

%%%%%%%%%%%% START FIGURE %%%%%%%%%%%%%%%%%%

%%%%%%%%%%%% END FIGURE %%%%%%%%%%%%%%%%%%

%%%%%%%%%%%%%%%%%%%%%%%%%%%%%%%%%%%%%%%%%%%%%%%%%%%%%%%%%%%%%%%%%%%%%%%%%%%%%%%%

\subsection{State preparation resource estimates}\label{sec:state-prep-resource}

Here, we summarize the non-Clifford resources required for state preparation. As with the \ac{QRAM} estimates, we utilize the phase-incorrect controlled-swap gates for the fixed-precision version and the phase-correct version for the pre-rotated case. The phase-correct version requires additional ancilla qubits not shown in these circuits. Details of the controlled-swap circuits for both cases are summarized in App.~\ref{app:decompositions}. The state preparation resource counts are provided in Tab.~\ref{tab:state-prep-compare}. For controlled-state preparation, one simply prepends the state preparation routine with a LOAD operation followed by a $\text{LOAD}^\dagger$ operation (LOADF in the case of the pre-rotated approach) as in Figs.~\ref{fig:controlled_state_prep} and \ref{fig:controlled_state_prep_prerotated}. 

Our counts can be verified using the circuit diagrams in Figs.~\ref{fig:optimized_state_prep_circuit} and \ref{fig:state_prep_prerotated}, along with the following additional observations, which reference gate decompositions from App.~\ref{app:decompositions}. 
\begin{itemize}
	\item Each of the $n$ multiply-controlled-$R_y$ gate in Fig.~\ref{fig:optimized_state_prep_circuit} is accomplished by $t$ singly-controlled-$R_y$ rotations in series, by the fixed rotation angles $\pi, \pi 2^{-1}, \pi 2^{-2}, \ldots, \pi2^{-t+1}$. These are implemented with the construction from Fig.~\ref{fig:controlled-ry}, which involves two applications of a single-qubit rotation. These single-qubit rotations are synthesized with a Clifford+$T$ gate sequence with $T$-count denoted by the quantity $R_y$ in the table. The total $T$-depth and $T$-count is thus $2n R_y$.

	\item Each $S_p$ gate with $p = 2,3,\ldots,n, \pm$ in Figs.~\ref{fig:optimized_state_prep_circuit}, \ref{fig:optimized_state_prep_circuit_prerotated}, and \ref{fig:FLAG} is a parallel controlled-swap gate with a single mutual control and many target pairs. For fixed-precision, each $S_p$ is implemented with the construction from Fig.~\ref{fig:multi_cswap} along with the phase-incorrect decomposition in Fig.~\ref{fig:cswap-phase} yielding $T$-depth 4. The total number of controlled-swap gates that occur during $S_2$,\ldots,$S_n$ is $(2^n-n-1)t + (2^n-1)$, where the first term comes from shuffling the $t$-bit angle data and the second term comes from shuffling the sign bit data.  For pre-rotated, the parallel controlled-swaps are implemented with the construction from Fig.~\ref{fig:cswap2} (costing $(N-2)/2$ ancillas), where Toffolis are implemented via the depth-1 count-4 construction of Ref.~\cite{jones2013} (costing another $(N-2)/2$ ancillas). The number of controlled-swaps is $2^n-n-1$. 

	\item In Fig.~\ref{fig:state_prep_prerotated}, the FLAG gate can be performed in parallel with the $S^\dagger$ gate (last gate) of Fig.~\ref{fig:optimized_state_prep_circuit_prerotated}. Note that each requires $N-2$ ancillas (see previous bullet). The depth of the controlled-swap portion of the circuit is $2(n-1)$ for SPF and $n-1$ for FLAG. 

\end{itemize}

%%%%%%%%%%%% START TABLE %%%%%%%%%%%%%%%%%%

\begin{table}[h!]
\begin{center}
\caption{Resource counts for two approaches to state preparation as depicted in Fig.~\ref{fig:optimized_state_prep_circuit} and Fig.~\ref{fig:state_prep_prerotated}. The parameter $R_y$ is the number of $T$ gates needed to synthesize a single qubit rotation about the $Y$ axis by an arbitrary angle, and $t$ is the number of bits precision to store the classical data, where both $t$ and $R_y$ scale as $\bigo{\log(1/\epsilon)}$, where $\epsilon$ is the error on the state prepared by the protocol.}
\label{tab:state-prep-compare}
\renewcommand{\arraystretch}{1.4}
\begin{tabular}{l|l|l}
\textbf{Resource}   & \textbf{Fixed-Precision}                                                                                       & \textbf{Pre-rotated} \\
\hhline{=|=|=}                                                                                                                           
\textbf{\# Qubits}  & $(t+1) \cdot 2^n + n -t$                                                                                     & $4 \cdot 2^n + n - 6$ \\
\hline                                                                                                                               
\textbf{$T$-Depth}    & $2tnR_y + 8n$                                                                                             & $3n+ 4 R_{y} - 3$ \\
\hline                                                                       
\textbf{$T$-Count}    & $ 8(t+1)(2^{n} - 1) +2 t n R_y - 8 tn $ & $ (4 R_y + 16) \cdot 2^n - 4 R_y -16 n - 16$
\end{tabular}
\end{center}
\end{table}

%%%%%%%%%%%% END TABLE %%%%%%%%%%%%%%%%%%

Finally, we remark here that one could also utilize the swap networks of the bucket-brigade \ac{QRAM} to perform the $S_p$ gates in the state preparation routine, which could confer some amount of natural noise resilience. However, due to the interleaved $R_y$ rotations, the parallelism that was previously exploited to ensure log-depth scaling is now broken, so the bucket-brigade state preparation circuit can only achieve a minimum $T$-depth of $\bigo{n^2}$ depth.  Furthermore, the constant factors are higher for qubit and $T$-count. For these reasons, we do not consider a bucket-brigade style state preparation approach, and all of our resource estimates use the select-swap versions presented here.

%%%%%%%%%%%%%%%%%%%%%%%%%%%%%%%%%%%%%%%%%%%%
%                                          %
%     BLOCK-ENCODING RESOURCE ESTIMATE     %
%                                          %
%%%%%%%%%%%%%%%%%%%%%%%%%%%%%%%%%%%%%%%%%%%%

\section{Block-Encoding Resource Estimates}\label{sec:resource_estimates}

\subsection{Overview}

We now have all the necessary ingredients to estimate the full resources required to block-encode a dense matrix of classical data using the circuit shown in Fig.~\ref{fig:U_A}. This is the portion labeled ``Block-Encoding'' in Fig. \ref{fig:block_diagram}. A variety of choices can be made with respect to exactly how the matrix is block-encoded. These options are shown in the upper right side of Fig. \ref{fig:block_diagram}. We outline the controlled version in the next subsection. We provide methods to implement the other options symmetric and q-Norm in App.~\ref{app:alternate-encodings}.

For the standard Frobenius encoding, the block-encoding procedure reduces to two applications of controlled-state preparation requiring both a \ac{QRAM}-like data-loading operation and state preparation routine (see Figs.~\ref{fig:controlled_state_prep} and \ref{fig:controlled_state_prep_prerotated}). We can make one optimization to reduce the resource count: note that the state $\ket{\phi_k}$ defined in Eq.~\eqref{eq:fro-states} is independent of the control register $k$. Therefore, this state can be prepared with standard state preparation. We provide the resource counts for the fixed-precision case in Tab.~\ref{tab:block_encoding_fixed} and pre-rotated block-encoding in Tab.~\ref{tab:block_encoding_pre} for the two versions of \ac{QRAM} that we consider.

%%%%%%%%%%%%%%%%%%%%%%%%%%%%%%%%%%%%%%%%%%%%%%%%%%%%%%%%%%%%%%%%%%%%%%%%%%%%%%%%

\subsection{Fixed precision resources}

For the fixed-precision case, we leave the resource counts in terms of the parameter $\lambda \in \{0,1,\ldots,n\}$ that allows one to trade $T$-depth for circuit width and overall $T$-count. For the select-swap \ac{QRAM}, the minimal $T$-count is achieved by choosing $\lambda = 0$, which yields a $T$-count and $T$-depth proportional to $\bigo{N}$ on $\bigo{N}$ total qubits. The minimal $T$-depth of $\bigo{\log (N)}$ is achieved when $\lambda = n$ at the expense of requiring a $T$-count of $\bigo{N^2}$. 

The fixed-precision bucket-brigade approach cannot achieve the same scaling with respect to $T$-count; in all cases the count is at least $\Omega(N^2)$. The minimal-depth case can achieve the same asymptotic scaling as select-swap, but we note that the constant factors for all resources are higher. Whether this approach can achieve an overall physical resource reduction will depend upon the quantum architecture, error correcting code, and error requirements, which are all beyond the scope of this paper.

%%%%%%%%%%%% START TABLE %%%%%%%%%%%%%%%%%%

\begin{table}[h!]
\begin{center}
\caption{Block-encoding resource requirements for fixed-precision implementation. Taking $\lambda = n$ yields the minimal depth circuit at the cost of needing $\bigo{N^2}$ $T$-count for both \ac{QRAM} implementations. The select-swap \ac{QRAM} can achieve a $T$-count of $\bigo{N}$ by taking $\lambda = 0$.}
\label{tab:block_encoding_fixed}
\renewcommand{\arraystretch}{1.4}
\begin{tabular}{l|l|l}
\textbf{Resource}   & \textbf{Select-Swap}                                                                                                                                                         & \textbf{Bucket-Brigade} \\
\hhline{=|=|=}                                                                                                                                                                                                                 
\textbf{\# Qubits}  & $\begin{aligned}[t] &(t+1) \cdot 2^{n+\lambda} - t \cdot 2^\lambda \\
					&+ 3n -\lambda + 1\end{aligned}$                                                                                   & $(2t+2) \cdot 2^{n + \lambda} -(2t-1) \cdot 2^{\lambda} + 3 n-2$ \\
\hline                                                                                                                                                                                                                         
$T$\textbf{-Depth}    & $8 \cdot 2^{n-\lambda} + 4tnR_y + 16n+ 8 \lambda - 8$                                                      & $(96 \lambda-72) \cdot 2^{n - \lambda} + 4 R_{y} n t + 16 n - 8$ \\
\hline
$T$\textbf{-Count}    & $\begin{aligned}[t] &8(t+1)( 2^{n+\lambda} +2^n)- 8t \cdot 2^{\lambda} \\
&+ 8 \cdot 2^{n-\lambda} + 4 t n R_y \\
&-16tn - 8t -24 \end{aligned}$ & $\begin{aligned}[t] &32(t+1)\cdot 2^{2n} -16(t+1)\cdot 2^{2n-\lambda}  \\
																&+2^{n-\lambda}(-16\lambda + 16t -24) \\
 																&- (16t-48)\cdot 2^n+4tnR_y \\
 																&- 16tn - 16 t -24 \end{aligned}$
\end{tabular}
\end{center}
\end{table}

%%%%%%%%%%%% END TABLE %%%%%%%%%%%%%%%%%%

%%%%%%%%%%%%%%%%%%%%%%%%%%%%%%%%%%%%%%%%%%%%%%%%%%%%%%%%%%%%%%%%%%%%%%%%%%%%%%%%

\subsection{Pre-rotated resources}

For the pre-rotated case, we only consider the minimal-depth circuit, that is, we make no attempt to trade width for depth. Note that any manifestation of the pre-rotated idea would need to satisfy a $T$-count lower bound of $\Omega(N^2)$ due to the need to have controlled-$Ry$ rotations for all $N(N-1)$ angles somewhere in the circuit. However, the benefits of the pre-rotated approach can be seen by both an improvement in asymptotic scaling, and constant factor improvements to both $T$-depth and qubit count. This technique allows us to achieve $T$-depth of $\bigo{\log N + Ry} \sim \bigo{\log (N/\epsilon)}$, which we contrast with the fixed precision approach that scales as $\bigo{R_y t \log N} \sim \bigo{\log N \log^2(1/\epsilon)}$. 

The number of qubits required for this approach is approximately four per classical matrix entry (not counting additional qubits needed for routing operations). One qubit is required for the data and one for the flag, and an additional two ancilla are needed to perform the parallel controlled-swap operations (see Fig. \ref{fig:cswap2} in App.~\ref{app:decompositions}). One could exploit the fact that the parallel controlled-swap gate can use dirty qubits for the ancilla to reduce this to just a single extra ancilla for the $T$-depth one Toffoli gate \cite{selinger2013quantum}. However, the first round of controlled-swap gates in the select-swap \ac{QRAM} circuit operates in parallel across all \ac{QRAM} data registers, so there are no available qubits to do this. One could split the first set of controlled-swap gates into two rounds to utilize the other data qubits as dirty ancilla, but we choose the shortest possible depth approach and counted the cost of the additional ancilla qubit in our resource analysis.

Note that since all the flags are set to 1 at the beginning of the controlled-state preparation protocol, the Toffolis that appear within the gate $V$ of LOADF in Fig.~\ref{fig:LOADF} can be replaced with Clifford CNOTs, saving $T$-depth 2 and $T$-count $2N(N-1)$.

%%%%%%%%%%%% START TABLE %%%%%%%%%%%%%%%%%%

\begin{table}[h!]
\begin{center}
\caption{Block-Encoding resource requirements for pre-rotated method. To compare to Tab.~\ref{tab:block_encoding_fixed} take $\lambda = n$. Loading the classical data in pre-rotated form with flag qubits allows us to achieve asymptotic depth scaling of $\bigo{\log N + Ry} \sim \bigo{\log (N/\epsilon)}$. }
\label{tab:block_encoding_pre}
\renewcommand{\arraystretch}{1.4}
\begin{tabular}{l|l}
\textbf{Resource}   & \textbf{Pre-rotated}           \\                                                                             
\hhline{=|=}                                                                                                                                              
\textbf{\# Qubits}  & $4 \cdot 2^{2n} - 3 \cdot 2^{n} + 2 n-1$        \\
\hline                                                                                                                                                                    
$T$\textbf{-Depth}    & $10n+8R_y -4$                            \\
\hline                                                                                                   
$T$\textbf{-Count}    & $(4R_y + 32)2^{2n} - 24 \cdot 2^n -4 R_y- 32n -8 $   
\end{tabular}
\end{center}
\end{table}

%%%%%%%%%%%% END TABLE %%%%%%%%%%%%%%%%%%

%%%%%%%%%%%%%%%%%%%%%%%%%%%%%%%%%%%%%%%%%%%%%%%%%%%%%%%%%%%%%%%%%%%%%%%%%%%%%%%%

\subsection{Controlled block-encodings}\label{sec:controlled_block_encoding}

\begin{figure}[ht!]
\centering
\scalebox{1.2}{
\mbox{
\Qcircuit @C=0.55em @R=0.8em {
                 &    &                                            &    &                                                                         &     & & &                           & & & & & &                  &\qw &                                        \qw & \qw & \qw                                                                      & \ctrl{2} & \qw & \\
                 &    &                                            &    &                                                                         &     & & &                           & & & & & & \lstick{\ket{0}} &\qw & {\raisebox{0.18cm}{/${}^{D}$}}         \qw & \qw & \qw                                                                      & \qswap   & \qw & \\
\lstick{\ket{0}} &\qw & {\raisebox{0.18cm}{/${}^{D}$}}         \qw &\qw &\multigate{1}{\substack{\text{LOAD}\\\{\bar{\Theta}^{(k)}, s^{(k)}\}}} & \qw & & &                             & & & & & & \lstick{\ket{0}} &\qw & {\raisebox{0.18cm}{/${}^{D}$}}         \qw & \qw & \multigate{1}{\substack{\text{LOAD}\\\{\bar{\Theta}^{(k)}, s^{(k)}\}}} & \qswap   & \qw & \\
\lstick{\ket{0}} &\qw & {\raisebox{0.18cm}{/${}^{Q_\lambda}$}} \qw &\qw &\ghost{\substack{\text{LOAD}\\\{\bar{\Theta}^{(k)}, s^{(k)}\}}}        & \qw & & & \rb{0.5cm}{$\Longrightarrow$} & & & & & & \lstick{\ket{0}} &\qw & {\raisebox{0.18cm}{/${}^{Q_\lambda}$}} \qw & \qw & \ghost{\substack{\text{LOAD}\\\{\bar{\Theta}^{(k)}, s^{(k)}\}}}        & \qw      & \qw & \\
\lstick{\ket{k}} &\qw & {\raisebox{0.18cm}{/${}^{n}$}}         \qw &\qw &\ctrlslash{-1}                                                         & \qw & & &                             & & & & & & \lstick{\ket{k}} &\qw & {\raisebox{0.18cm}{/${}^{n}$}}         \qw & \qw & \ctrlslash{-1}                                                           & \qw      & \qw & \\ 
}
}
}
\caption{\label{fig:controlled_block_encoding} A possible modification to the LOAD operation that allows a controlled-block-encoding ${CU_A = \ket{0} \bra{0} \otimes I + \ket{1}\bra{1} \otimes U_A}$ to be efficiently performed. The LOAD operation can be either $\LOADSS$ or $\LOADBB$ from Sec.~\ref{sec:qram}, and brings $D$ bits of classical data into the third register, where $D = (N-1)t+N$. Controlled on the first register, this data is swapped into the second register, which acts as the input to state preparation, as in Fig.~\ref{fig:controlled_state_prep}. A similar idea could be implemented for the LOADF operation with respect to Fig.~\ref{fig:controlled_state_prep_prerotated}. If the control is $\ket{0}$, the input to state preparation is the trivial state. The cost of this construction is $D$ ancilla qubits and $D$ controlled-swap gates with a common control, which can be performed in $\bigo{1}$ $T$-depth and $\bigo{D}$ $T$-count (in cases that the LOAD operation does not produce garbage, the LOAD ancillas could be reused, in which case no additional ancillas are needed for controlled block encoding). Adding additional controls onto the controlled-swap gate yields a multiply-controlled block-encoding with an arbitrary number of controls, without any additional ancillas.
}
\end{figure}
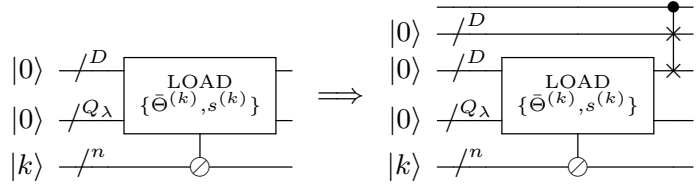

A controlled block-encoding is useful in certain applications, for example in solving linear systems of equations \cite{costa2021optimal}. In particular, we wish to implement the unitary
\begin{equation}\label{eq:CUA}
CU_A = \ket{0}\bra{0}\otimes I + \ket{1}\bra{1}\otimes U_A,
\end{equation}
where $U_A$ is the standard form of the block-encoded matrix $A$ given in Eq. \eqref{eq:block_encoding}. One possible implementation is to add zeros to the \ac{QRAM} and just select these values if the appropriate control bit is 0, but this approach is highly inefficient in qubit count for our assumed case of $N$ being an exact power of two, since the number of qubits in the \ac{QRAM} must be doubled. Instead, we simply select zeros by adding a single binary tree of all zeros, which will be selected using a single register controlled-swap to replace the loaded register if the control bit is one, as shown in Fig.~\ref{fig:controlled_block_encoding}. This method requires one extra qubit for the control, $D$ extra qubits for the $\ket{0}$ state in \ac{QRAM}, and one extra controlled-swap between $D$ qubit registers.

%%%%%%%%%%%% START FIGURE %%%%%%%%%%%%%%%%%%

%%%%%%%%%%%% END FIGURE %%%%%%%%%%%%%%%%%%

%%%%%%%%%%%%%%%%%%%%%%%%%%%%%%%%%%%%%%%%%%%%
%                                          %
%              ERROR ANALYSIS              %
%                                          %
%%%%%%%%%%%%%%%%%%%%%%%%%%%%%%%%%%%%%%%%%%%%

\section{Finite Precision Error Analysis}\label{sec:error}

\subsection{Overview}

The unitary implemented by our circuit is $\tilde{U}_A$, which is different from the exact block-encoding unitary $U_A$ in two respects: (a) the rounding error from the representation of the angles $\theta$ using $t$-qubit registers and (b) the error from the gate synthesis of the $R_y$ rotations. Both (a) and (b) affect the fixed-precision block-encoding; only (b) affects the pre-rotated block-encoding.

%%%%%%%%%%%%%%%%%%%%%%%%%%%%%%%%%%%%%%%%%%%%%%%%%%%%%%%%%%%%%%%%%%%%%%%%%%%%%%%%

\subsection{Rounding error}

For the fixed-precision case, rounding errors cause the ideal angles $\theta \in [0,\pi]$ to be approximated by the angle $\tilde{\theta}$ that is the nearest exact multiple of $2\pi/2^{t}$, i.e.~$|\theta - \tilde{\theta}| \leq \pi 2^{-t}$. We can, therefore, represent $1-(\tilde{\theta}/2\pi)$ exactly in binary with $t$ bits $(\theta_1,\ldots,\theta_{t-1})$, where $\theta_i \in \{0,1\}$ and

\begin{equation}
\label{eqn:fixed_point_repr}
1-\frac{\tilde{\theta}}{2\pi} = 2^{-\theta_1}2^{-2\theta_2}2^{-2\theta_3} \ldots 2^{-t\theta_t}.
%\underbrace{\theta_{t-1}\cdots \theta_{t-p}}_p.\underbrace{\theta_{t-p-1} \cdots \theta_0}_{t-p},
\end{equation}
The rotation $R_y(\theta)$ sends $\ket{0} \mapsto \cos(\theta/2)\ket{0} + \sin(\theta/2) \ket{1}$; we need only cover the space $\theta \in [0,\pi]$, since we may assume the matrix entries are all positive (signs are applied later), and hence $\sin(\theta/2) \in [0,1]$ is sufficient. 
Note that $R_y(\eta)$ has eigenvalues $e^{\pm i \eta/2}$ and hence $||R_y(\eta)-I|| \leq | e^{i\eta/2}-1| \leq \eta/2$, which implies that
\begin{equation}
||R_y(\theta)-R_y(\tilde{\theta})|| = ||R_y(\theta)R_y^{\dagger}(\tilde{\theta})-I|| = ||R_y(\theta-\tilde{\theta})-I || \leq |\theta-\tilde{\theta}|/2 \leq \pi 2^{-t-1}.
\end{equation}
We now argue that the above implies the \emph{controlled}-rotations are close to their rounded versions. The controlled rotations in the circuit for $U_A$ perform a different rotation angle, depending on the setting of several control qubits. Suppose that, for some integer $p$, we have a collection of $2^p$ unitaries $V_i$ and approximations $\tilde{V}_i$ such that $||V_i-\tilde{V}_i||\leq \delta$ for all $i \in [2^p]$. Let $CV$ and $C\tilde{V}$ be the operations that, controlled on a $p$-qubit register being in the state $i$, perform the operations $V_i$ and $\tilde{V}_i$, respectively. It is then easy to verify that $||CV-C\tilde{V}|| \leq \delta$, as $CV$ and $C\tilde{V}$ are each block diagonal matrices with the same block structure. Hence, as long as all angles are correct up to error $\pi 2^{-t-1}$, the controlled-$R_y(\theta)$ operation in the circuit is $(\pi 2^{-t-1})$-close to controlled-$R_y(\tilde{\theta})$.

Let $\tilde{U}'_A$ denote the unitary for which all controlled-$R_y(\theta)$ gates are replaced by an exact implementation of controlled-$R_y(\tilde{\theta})$. As the circuit has $2\log(N)$ controlled-rotations ($\log(N)$ each for $U_R^\dagger$ and $U_L$), by the triangle inequality, we have that
\begin{equation} \label{eq:rounding_error}
||U_A-\tilde{U}'_A|| \leq 2\log(N) ||R_y(\theta) -R_y(\tilde{\theta})|| \leq \pi \log(N) 2^{-t}.
\end{equation}

%%%%%%%%%%%%%%%%%%%%%%%%%%%%%%%%%%%%%%%%%%%%%%%%%%%%%%%%%%%%%%%%%%%%%%%%%%%%%%%%

\subsection{Gate synthesis error}

In the fixed-precision approach, the unitary $\tilde{U}'_A$ exactly implements controlled-$R_y(\tilde{\theta})$ by loading the bits $(\theta_1,\ldots,\theta_t)$ of $\tilde{\theta}$ into $t$ ancilla registers, and then exactly performing controlled-$R_y(\pi2^{-j+1})$ operations (with one control qubit) for $j=1,\ldots,t$ with these ancilla registers acting as the controls. As seen in Fig.~\ref{fig:controlled-ry}, a controlled-$R_y(\pi2^{-j+1})$ is accomplished by decomposition into two CNOTs and two $R_y(\pi2^{-j})$ operations.  The actual circuit $\tilde{U}_A$ differs from $\tilde{U}'_A$ only in that these $R_y(\pi2^{-j})$ operations are performed approximately using a decomposition of the $R_y(\pi2^{-j})$ gate into Clifford+$T$. Denote the unitary enacted by this decomposition by $\tilde{R}_y(\pi2^{-j})$. The decomposition error \cite{ross2016optimal} is then bounded as
\begin{equation}
||R_y(\pi2^{-j}) - \tilde{R}_y(\pi2^{-j}) || \leq \delta_{\text{decomp}},
\end{equation}
as long as we choose a gate sequence with $T$-count (note $T$-depth $=$ $T$-count for a single-qubit operation) at least equal to $R_y$ with 
\begin{equation}
R_y = 3 \log(1/\delta_{\text{decomp}}) + \bigo{\log(\log(1/\delta_{\text{decomp}}))}\,.
\end{equation}

Again using the triangle inequality, we find that replacing the $4t \log(N)$ appearances of a $R_y(\pi2^{-j})$ gate with the approximate $\tilde{R}_y(\pi 2^{-j})$, we incur error bounded as
\begin{equation}\label{eq:synthesis_error}
||\tilde{U}_A - \tilde{U}'_A|| \leq 4t \log(N) \delta_{\text{decomp}}.
\end{equation}

In the pre-rotated approach, there are $2N-2$ controlled-$R_y$ rotations (actually, they are doubly-controlled $R_y$ rotations), but nearly all of them (the ones that do not get injected) are exactly undone. Moreover, by choosing a gate decomposition of $R_y(\theta)$ of the form $H_\theta^*H_\theta$ for some gate sequence $H_\theta$ (where $H_\theta^*$ denotes complex conjugate) the construction of Fig.~\ref{fig:controlled-ry} with $H$ and $H^{\dagger}$ in place of $R_y(\theta/2)$ and $R_y(\-\theta/2)$ guarantees that the action on the target is \emph{exactly} trivial when the control is set to $\ket{0}$. Ultimately, the error on the state prepared by the protocol is exactly the same as it would be in the fixed-precision method using the decomposition $H_\theta^*H_\theta$ for all the controlled-rotations (rather than a product of $t$ gate decompositions). As there are $2\log(N)$ rotations, and thus $4\log(N)$ applications of some sequence $H_\theta$, the overall error is $4\log(N)\delta_\text{decomp}$. Note that there is no rounding error for this approach.

%%%%%%%%%%%%%%%%%%%%%%%%%%%%%%%%%%%%%%%%%%%%%%%%%%%%%%%%%%%%%%%%%%%%%%%%%%%%%%%%

\subsection{Overall block-encoding error}

For the fixed-precision approach, from Eqs.~\eqref{eq:rounding_error} and \eqref{eq:synthesis_error} and the triangle inequality, we have that
\begin{equation}
||U_A - \tilde{U}_A || \leq 4t \log(N) \delta_{\text{decomp}} + \pi \log(N) 2^{-t}\,.
\end{equation}

Further, note that by the definition of block-encoding and sub-multiplicativity of the spectral norm we have
\begin{align}
\lVert A - \alpha(\bra{0} \otimes I)\tilde{U}_A(\ket{0} \otimes I) \rVert &= \alpha\lVert (\bra{0} \otimes I)(U_A-\tilde{U}_A)(\ket{0} \otimes I) \rVert \\
&= \alpha\lVert (\bra{0} \otimes I)(U_A-\tilde{U}_A)(\ket{0} \otimes I) \rVert \\
&\leq \alpha \lVert \bra{0} \otimes I \rVert \cdot \lVert U_A-\tilde{U}_A \rVert \cdot \lVert \ket{0} \otimes I \rVert = \alpha \lVert U_A-\tilde{U}_A \rVert \\
&\leq 4t \alpha \log(N) \delta_{\text{decomp}} + \pi \alpha \log(N) 2^{-t}.
\end{align} 

If we wish for this error to be smaller than $\epsilon$, it suffices to choose $t = \log(\alpha\pi/\epsilon) + \log(\log(N))+1$, and $\delta_{\text{decomp}} = \epsilon/(8t\alpha\log(N))$, which is achieved with
\begin{align}
R_y &= \lceil 3\log(\alpha/\epsilon) + 3\log(\log(N)) + 9 + \bigo{\log(\log(\alpha/\epsilon))} + \bigo{\log(\log(\log(N)))} \rceil  \label{eq:Ry} \\
t &= \lceil \log(\alpha/\epsilon) + \log(\pi) + \log(\log(N)) +1 \rceil. \label{eq:t} 
\end{align}

These equations should be substituted in the fixed-precision resource counts whenever one encounters them, if one wishes to have resources in terms of the final block-encoding error rate $\epsilon$.
%For the Q-Norm block-encoding, there is one additional rotation to ensure proper normalization as shown in Eq.~\eqref{eq:blockencode_pnorm}. In that case one should take $\log(N)\to \log(N)+1$ in Eq.~\eqref{eq:Ry} and \eqref{eq:t}.
 
For the pre-rotated approach, the same analysis gives 
\begin{equation}
\lVert A - \alpha(\bra{0} \otimes I)\tilde{U}_A(\ket{0} \otimes I) \rVert \leq 4 \alpha \log(N) \delta_{\text{decomp}}\,,
\end{equation}
meaning it suffices to choose $\delta_{\text{decomp}} = \epsilon / (4\alpha\log(N))$, which is achieved with a gate sequence of length
\begin{equation}
R_y = \lceil 3\log(\alpha/\epsilon) + 3\log(\log(N)) + 6 + \bigo{\log(\log(\alpha/\epsilon))} + \bigo{\log(\log(\log(N)))} \rceil  \label{eq:Ry_prerotated} \,.
\end{equation}
Our numerical estimates use these formulas for estimating exact resources, but we ignore the $\bigo{\cdot}$ terms which are all doubly logarithmic or smaller. 

%%%%%%%%%%%%%%%%%%%%%%%%%%%%%%%%%%%%%%%%%%%%
%                                          %
%               DISCUSSION                 %
%                                          %
%%%%%%%%%%%%%%%%%%%%%%%%%%%%%%%%%%%%%%%%%%%%

\section{Conclusion}\label{sec:conclusion}

It is well known that loading classical data into a quantum computer is a challenging problem. However, for many problems of practical interest, one assumes access to such classical data without necessarily considering the full cost to encode the data in the quantum computer. Our results provide concrete resource counts and system sizes required to perform this task for two different \ac{QRAM} models. 

Combining \ac{QRAM} with a state preparation routine, we provide detailed circuit descriptions and resource estimates for a commonly used algorithmic primitive: a block-encoding. Our modular implementation also allows one the freedom to consider resource counts for different \ac{QRAM} models that allow for differing optimizations. We provide two such choices: one that minimizes $T$-count or $T$-depth and the other that minimizes the impact of noise. Our results address the practical feasibility of quantum algorithms that require large amounts of classical data, and we note that in fault-tolerant implementations of these algorithms, \ac{QRAM} implementations can be seen as a bottleneck.

The details of our circuits also elucidate the ingredients that would be necessary in a circuit architecture optimized for block-encoding classical data. The \ac{QRAM} circuits we describe use a large number of controlled-swap gates that we assume can be applied in parallel. Realizing such parallelism requires applying $T$ gates across many qubits at once and to achieve $\polylog(N)$ depth (which is necessary in any scenario where an exponential speedup is sought), our constructions require $\bigo{N^2}$ parallel $T$ gates. Even if a processor with so many qubits were available, it could be challenging to implement these parallel operations at large $N$ due to the overhead required for magic-state-distillation and decoding latencies, which could significantly decrease the rate at which layers can be applied. Furthermore, we assume that fanout-CNOT gates with arbitrarily long range can be applied in one time step. In surface-code lattice-surgery architectures, this parallelism is possible, but it requires additional communication qubits for performing the required lattice-surgery operations, thereby increasing the qubit resources beyond what we have considered here. Therefore, although we do not prove any rigorous lower bounds on block-encoding, our resource estimates might be viewed as maximally optimistic for block-encoding of dense classical data, and they would increase as architectural constraints are added.

%%%%%%%%%%%%%%%%%%%%%%%%%%%%%%%%%%%%%%%%%%%%
%                                          %
%               DISCUSSION                 %
%                                          %
%%%%%%%%%%%%%%%%%%%%%%%%%%%%%%%%%%%%%%%%%%%%

\acknowledgements\label{sec:acknowledgements}
We thank David Bader, Thom Bohdanowicz, Paul Burchard, Connor Hann, Helmut Katzgraber, Rajiv Krishnakumar, Cedric Lin, Shantu Roy, Martin Schuetz, and James Tarantino for helpful discussions. We are especially grateful to Earl Campbell for early collaboration during an initial phase of this project.

%%%%%%%%%%%%%%%%%%%%%%%%%%%%%%%%%%%%%%%%%%%%%%%%%%%%%%%%%%%%%%%%%%%%%%%%%%%%%%%%

\bibliographystyle{apsrev4-1_custom}
\bibliography{references}

\clearpage

%%%%%%%%%%%%%%%%%%%%%%%%%%%%%%%%%%%%%%%%%%%%
%                                          %
%               APPENDICES                 %
%                                          %
%%%%%%%%%%%%%%%%%%%%%%%%%%%%%%%%%%%%%%%%%%%%

\appendix

\section{Alternative Block-Encoding Strategies}\label{app:alternate-encodings}

\subsection{Block-encoding off-diagonal hermitian matrices}\label{app:off-diagonal_matrices}

In cases where the matrix $A$ is not square or when one requires a Hermitian block-encoding, rather than block-encode $A$ directly, one can instead block-encode the matrix 
\begin{equation}
\label{eq:expandedA}
\bar{A} = \begin{pmatrix} 0 & A \\ A^{T} & 0\end{pmatrix} \quad \text{where} \quad \bar{A} \in \mathbb{R}^{(M+N)\times (M+N)}.
\end{equation} 
Block-encoding $\bar{A}$ ensures that the matrix is Hermitian and square at the cost of one extra qubit.

We assume that $M \ge N$ without loss of generality (we could block encode the transpose to ensure this, if necessary). For the Frobenius encoding we also take $M=2^m$ and $N=2^n$, which can be enforced by padding the matrix. This implies that $\ell = m+1$. We now define the $2^{2\ell}$ dimensional operators $U_L$ and $U_R$ as the following:
\begin{align}
\label{eq:URUL}
U_R\ket{i}_\ell\ket{0}_\ell & = \ket{\psi_i}_{2\ell} \\ \nonumber
U_L\ket{i}_\ell\ket{0}_\ell & = \ket{\phi_i}_{2\ell}
\end{align}
where 
\begin{subequations}\label{eq:statedefs_psi}
\begin{align}
\ket{\psi_j} & = \sum_{k\in [N]} \frac{A_{j,k}}{||A_{j, \cdot}||}\ket{j, M+k}, \label{eq:statedefs_psij} \\ 
\ket{\psi_{M+k}} & = \sum_{j\in [M]} \frac{||A_{j, \cdot}||}{||A||_F}\ket{M+k, j}, \label{eq:statedefs_psiMk}
\end{align}
\end{subequations}
\begin{subequations}\label{eq:statedefs_phi}
\begin{align}
\ket{\phi_j} & = \sum_{k\in [N]} \frac{A_{j,k}}{||A_{j, \cdot}||}\ket{M+k, j}, \label{eq:statedefs_phij} \\ 
\ket{\phi_{M+k}} & = \sum_{j\in [M]} \frac{||A_{j, \cdot}||}{||A||_F}\ket{j, M+k}, \label{eq:statedefs_phiMk}
\end{align}
\end{subequations}
with $j \in [M]$ and $k \in [N]$. We use the notation $[M] = [1, 2, \cdots , M]$,  $[N] = [1,2, \cdots, N]$, and $||A_{j, \cdot}||$ denotes the norm of the $j^{th}$ row of $A$ such that $\sum_{j\in[M]}||A_{j, \cdot}||^2 = ||A||_F^2.$ From these definitions it is straightforward to show that 
\begin{equation}
\label{eq:stateidentities}
\begin{gathered}
\braket{\psi_j | \phi_{j^\prime}} = \braket{\psi_{M+k} | \phi_{M+k^\prime}} = 0, \\
\braket{\psi_j | \phi_{M+k}} = \frac{A_{j,k}}{||A||_F}, \quad \braket{\psi_{M+k} | \phi_j} = \frac{A_{k,j}^T}{||A||_F},
\end{gathered}
\end{equation}
and
\begin{equation}
\label{eq:blockderivation}
(\bra{0}_\ell\otimes I_{M+N})U_R^\dagger U_L(I_{M+N}\otimes \ket{0}_\ell) = \sum_{k \in M+[N]}\sum_{j\in [M]} \ket{k}\bra{j}\frac{A^T_{k,j}}{||A||_F} + \sum_{j \in [M]}\sum_{k \in M+[N]} \ket{j}\bra{k} \frac{A_{j, k}}{||A||_F},
\end{equation}
where the notation $M+[N] = [M+1, \cdots M+N]$, and $I_{M+N}$ is the projector onto the space spanned by basis states $\ket{j}$ with $j \in [M+N]$. This shows that the unitary $U_R^\dagger U_L$ is the block encoding of Eq. \eqref{eq:expandedA} as desired.

Both $U_R$ and $U_L$ are controlled-state preparation operators and can be implemented as described in the main text. The upshot of this construction, compared to a direct application of the construction in the main text to the off-diagonal Hermitian matrix, is that the normalization factor for the block-encoding is $||A||_F$ rather than $2||A||_F$.

%%%%%%%%%%%%%%%%%%%%%%%%%%%%%%%%%%%%%%%%%%%%%%%%%%%%%%%%%%%%%%%%%%%%%%%%%%%%%%%%

\subsection{Q-norm block-encoding}\label{app:q-norm_encoding}

For some algorithms, it might be advantageous to use a slightly modified block-encoding that we call the q-norm block-encoding. This is an $(\mu_p(A), \lceil\log (N+1)\rceil, \epsilon)$ block-encoding of $A$ where
\begin{equation}
\label{eq:blockencode_pnorm}
U^{(q)} = (U_R^{(q)})^\dagger U_L^{(q)} = \begin{pmatrix} A/\mu_p(A) & \cdot \\ \cdot & \cdot \end{pmatrix}
\end{equation}
such that 
\begin{equation}
\label{eq:blockencodeerror_pnorm}
||A - \mu_p(A)(\bra{0}_\ell \otimes I_{N+1})U(\ket{0}_\ell \otimes I_{N+1})|| \le \epsilon,
\end{equation}
with $\mu_p(A) = \sqrt{S_{2p}(A)S_{2(1-p)}(A^T)}$, $p\in [0, 1]$, and $S_{q}(A) = \max_j ||A_{j, \cdot}||_q^q$ is the $q^{\text{th}}$ power of the maximum q-norm of any row. As in the main text, we take $N=2^n$, which can be enforced by padding the matrix, if necessary. This restriction ensures that certain required rotations can be implemented without the need for basis state permutations or alternative encodings \cite{Vartiainen2004}. Just as shown in Eq.~\eqref{eq:URUL}, we define two operators $U_R^{(q)}$ and $U_L^{(q)}$, but now with the superscript $(q)$ to denote that this is the q-norm version. These are $2^{2\ell+1}$ dimensional operators that prepare the states $U_R^{(q)}\ket{i}_\ell\ket{0}_{\ell+1} = \ket{\psi_i^{(q)}}_{2\ell+1}$ and $U_L^{(q)}\ket{i}_\ell\ket{0}_{\ell+1} = \ket{\phi_i^{(q)}}_{2\ell+1}$ where 
\begin{subequations}\label{eq:statedefs_pnorm}
\begin{align}
\ket{\psi_j^{(q)}} & = \sum_{k\in [N]} \frac{\sgn (A_{j,k})|A_{j,k}|^p}{\sqrt{||A_{j, \cdot}||_{2p}}}\big[\cos \chi_j \ket{j, k} + \sin \chi_j\ket{j,N+k}\big], \label{eq:statedefs_pnorm_psij} \\
\ket{\phi_{k}^{(q)}} & = \sum_{j\in [N]} \frac{|A_{j, \cdot}|^{1-p}}{\sqrt{||A_{\cdot, k}||_{2(p-1)}}}\big[ \cos \chi_{k}\ket{j, k} + \sin \chi_{k}\ket{N+j, k}\big], \label{eq:statedefs_pnorm_phiMk}
\end{align}
\end{subequations}
with $j\in [N]$ and $k \in [N]$ for the control registers as before, $\sgn$ the signum function, and 
\begin{equation}
\label{eq:chi_def}
\begin{split}
\cos \chi_j &= \sqrt{\frac{||A_{j, \cdot}||_{2p}}{S_{2p}(A)}} \\
\cos \chi_{k} &= \sqrt{\frac{||A_{\cdot, k}||_{2(p-1)}}{S_{2(1-p)}(A^T)}}.
\end{split}
\end{equation}
Similar to the Frobenius encoding, it is straightforward to show that 
\begin{equation}
\label{eq:blockderivation_pnorm}
(\bra{0}_{\ell+1}\otimes I_{N})(U_R^{(q)})^\dagger U_L^{(q)}(I_{N}\otimes \ket{0}_{\ell+1}) = \sum_{j \in [N]}\sum_{k \in [N]} \ket{j}\bra{k} \frac{A_{j, k}}{\mu_p(A)}.
\end{equation}
As with the Frobenius case shown in the main text, the block-encoding procedure is reduced to showing how to create circuits to prepare the states given in Eq. \eqref{eq:statedefs_pnorm}. With the encoding presented here, this task is nearly identical to the Frobenius case with just minor changes. Specifically, one must perform a controlled-rotation on an ancilla qubit by the angle $\chi_j$ controlled by the index $j$, which requires LOAD and its adjoint, as well as a controlled-rotation. This controlled-rotation can be performed either in fixed precision, or using pre-rotated states with flags. In addition, both controlled-state preparations require a complete LOAD and $\text{LOAD}^\dagger$ operation (recall that for the Frobenius case, the second state preparation was independent of the control register). 

%%%%%%%%%%%%%%%%%%%%%%%%%%%%%%%%%%%%%%%%%%%%%%%%%%%%%%%%%%%%%%%%%%%%%%%%%%%%%%%%

\subsection{Symmetrized q-norm block-encoding}\label{app:q-norm_encoding-sym}

We can also perform a symmetrized block-encoding for the q-norm. A similar block-encoding was given in Refs. \cite{chakraborty2019, gilyen2019}, but with different state definitions. We prefer the block-encodings below, as they allow for simpler circuit implementations. We obtain a $(\mu_p(A), \lceil\log (M+N+1)\rceil, \epsilon)$ block-encoding of $\bar{A}$ such that 
\begin{equation}
\label{eq:blockencodeerror_pnorm-sym}
||\bar{A} - \mu_p(\bar{A})(\bra{0}_\ell \otimes I_{M+N+1})U(\ket{0}_\ell \otimes I_{M+N+1})|| \le \epsilon,
\end{equation}
with $\mu_p(A) = \sqrt{S_{2p}(A)S_{2(1-p)}(A^T)}$, $p\in [0, 1]$, and $S_{q}(A) = \max_j ||A_{j, \cdot}||_q^q$ is the $q^{\text{th}}$ power of the maximum q-norm of any row. The states to prepare in this case are given by
\begin{subequations}\label{eq:statedefs_pnorm_psi-sym}
\begin{align}
\ket{\psi_j^{(q)}} & = \sum_{k\in [N]} \frac{\sgn (A_{j,k})|A_{j,k}|^p}{\sqrt{||A_{j, \cdot}||_{2p}}}\big[\cos \chi_j \ket{j, M+k} + \sin \chi_j\ket{j,3M+k}\big], \label{eq:statedefs_pnorm_psij-sym}\\
\ket{\psi_{M+k}^{(q)}} & = \sum_{j\in [M]} \frac{|A_{j, k}|^{1-p}}{\sqrt{||A_{\cdot, k}||_{2(1-p)}}}\big[\cos \chi_{M+k} \ket{M+k, j} + \sin \chi_{M+k}\ket{M+k,2M+j}\big], \label{eq:statedefs_pnorm_psiMk-sym}
%\ket{\psi_j^{(q)}} & = \sum_{k\in [N]} \frac{\sgn (A_{j,k})|A_{j,k}|^p}{\sqrt{S_{2p}(A)}}\ket{j, M+k} + \sqrt{1-\frac{\sum_{k\in[N]}|A_{j,k}|^{2p}}{S_{2p}(A)}}\ket{j,N+M+1}, \label{eq:statedefs_pnorm_psij}\\
%\ket{\psi_{M+k}^{(q)}} & = \sum_{j\in [M]} \frac{|A_{j, k}|^{1-p}}{\sqrt{S_{2(1-p)}(A^T)}}\ket{M+k, j} + \sqrt{1-\frac{\sum_{j\in[M]}|A_{j,k}|^{2(1-p)}}{S_{2(1-p)}(A^T)}}\ket{M+k,N+M+1}, \label{eq:statedefs_pnorm_psiMk}
\end{align}
\end{subequations}
\begin{subequations}\label{eq:statedefs_pnorm_phi-sym}
\begin{align}
\ket{\phi_j^{(q)}} & = \sum_{k\in [N]} \frac{\sgn (A_{j,k})|A_{j,k}|^p}{\sqrt{||A_{j, \cdot}||_{2p}}}\big[\cos \chi_j\ket{M+k, j} + \sin \chi_j\ket{3M+k,j}\big], \label{eq:statedefs_pnorm_phij-sym} \\
\ket{\phi_{M+k}^{(q)}} & = \sum_{j\in [M]} \frac{|A_{j, \cdot}|^{1-p}}{\sqrt{||A_{\cdot, k}||_{2(p-1)}}}\big[ \cos \chi_{M+k}\ket{j, M+k} + \sin \chi_{M+k}\ket{2M+j, M+k}\big], \label{eq:statedefs_pnorm_phiMk-sym}
\end{align}
\end{subequations}
with $j\in [M]$ and $k \in [N]$ for the control registers. From these states, it is straightforward to show that
\begin{equation}
\label{eq:blockderivation_pnorm-sym}
(\bra{0}_{\ell+1}\otimes I_{M+N})(U_R^{(q)})^\dagger U_L^{(q)}(I_{M+N}\otimes \ket{0}_{\ell+1})
= \sum_{k \in M+[N]}\sum_{j\in [M]} \ket{k}\bra{j}\frac{A^T_{k,j}}{\mu_p(A)} + \sum_{j \in [M]}\sum_{k \in M+[N]} \ket{j}\bra{k} \frac{A_{j, k}}{\mu_p(A)}.
\end{equation}
As noted above, these state definitions vary slightly from those given in Refs. \cite{chakraborty2019, gilyen2019}, as the subspaces rotated by the angle $\chi_i$ are different. Our description allows for a much simpler circuit implementation, as one just needs to implement a controlled-rotation followed by the state preparation procedure outlined in the main text.

%%%%%%%%%%%%%%%%%%%%%%%%%%%%%%%%%%%%%%%%%%%%%%%%%%%%%%%%%%%%%%%%%%%%%%%%%%%%%%%%

\section{QRAM Details}\label{app:qram_details}

\subsection{Select-swap}\label{app:select_swap}

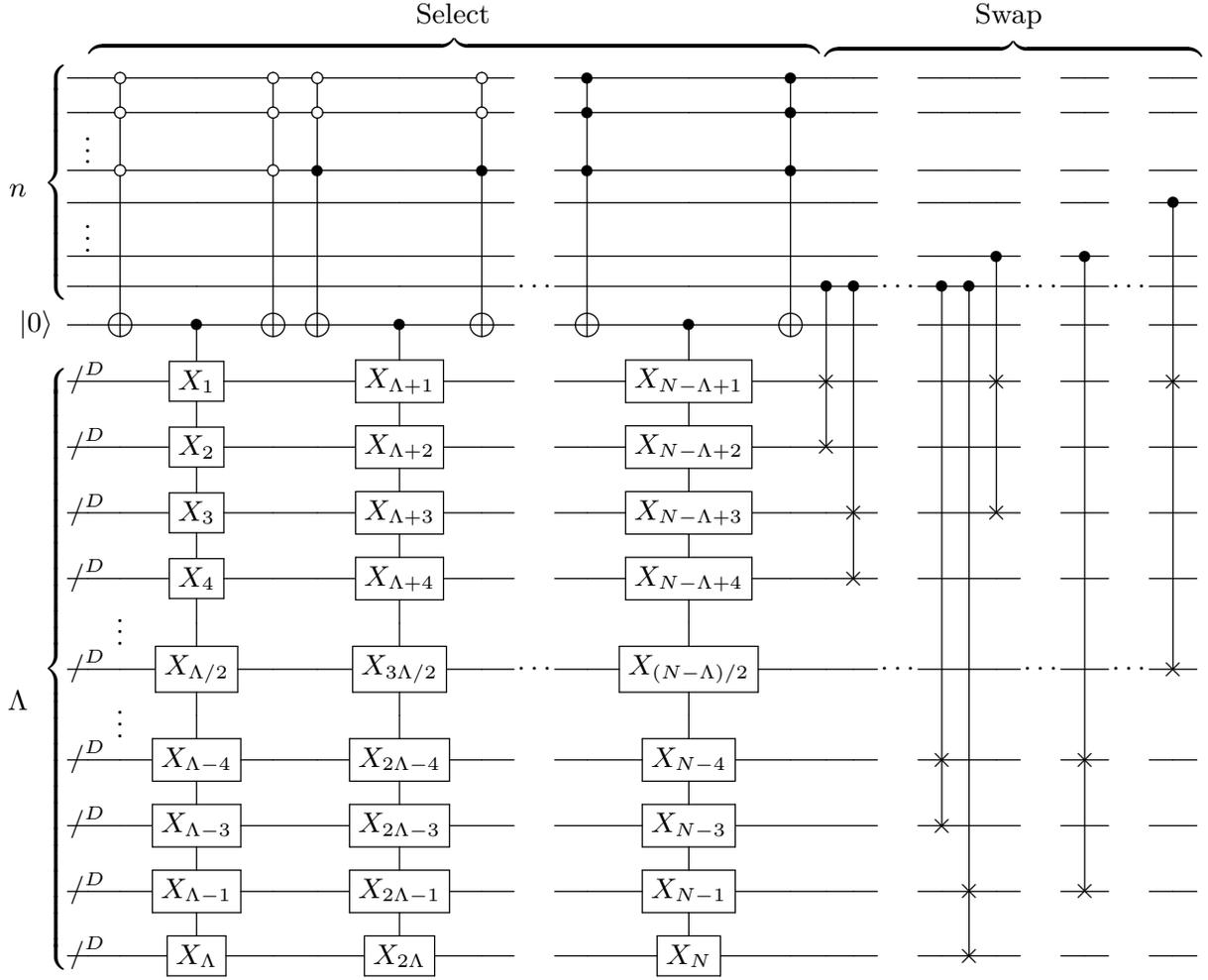
\begin{figure}[ht!]
\centering
\scalebox{1.2}{
\mbox{
\Qcircuit @C=0.7em @R=0.8em {
                 &                                  &           &                           &           &           & \raisebox{0.8cm}{\hspace{1.2cm}Select} &           &     &        & &           &                               &           &          &          &     &        & &           &           &          &          &        & & \raisebox{0.8cm}{\hspace{-1.7cm}Swap} &     &        & &           &     & \\
				 & \qw                              & \ctrlo{1} & \qw                       & \ctrlo{1} & \ctrlo{1} & \qw                                    & \ctrlo{1} & \qw &        & & \ctrl{1}  & \qw                           & \ctrl{1}  & \qw      & \qw      & \qw &        & & \qw       & \qw       & \qw      & \qw      &        & & \qw                                   & \qw &        & & \qw       & \qw & \\
				 & \qw                              & \ctrlo{2} & \qw                       & \ctrlo{2} & \ctrlo{2} & \qw                                    & \ctrlo{2} & \qw &        & & \ctrl{2}  & \qw                           & \ctrl{2}  & \qw      & \qw      & \qw &        & & \qw       & \qw       & \qw      & \qw      &        & & \qw                                   & \qw &        & & \qw       & \qw & \\
				 & \vdots                           &           &                           &           &           &                                        &           &     &        & &           &                               &           &          &          &     &        & &           &           &          &          &        & &                                       &     &        & &           &     & \\
				 & \qw                              & \ctrlo{5} & \qw                       & \ctrlo{5} & \ctrl{5}  & \qw                                    & \ctrl{5}  & \qw &        & & \ctrl{5}  & \qw                           & \ctrl{5}  & \qw      & \qw      & \qw &        & & \qw       & \qw       & \qw      & \qw      &        & & \qw                                   & \qw &        & & \qw       & \qw & \\
                 & \qw                              & \qw       & \qw                       & \qw       & \qw       & \qw                                    & \qw       & \qw &        & & \qw       & \qw                           & \qw       & \qw      & \qw      & \qw &        & & \qw       & \qw       & \qw      & \qw      &        & & \qw                                   & \qw &        & & \ctrl{10} & \qw & \\
                 & \vdots                           &           &                           &           &           &                                        &           &     &        & &           &                               &           &          &          &     &        & &           &           &          &          &        & &                                       &     &        & &           &     & \\
                 & \qw                              & \qw       & \qw                       & \qw       & \qw       & \qw                                    & \qw       & \qw &        & & \qw       & \qw                           & \qw       & \qw      & \qw      & \qw &        & & \qw       & \qw       & \ctrl{5} & \qw      &        & & \ctrl{12}                             & \qw &        & & \qw       & \qw & \\
                 & \qw                              & \qw       & \qw                       & \qw       & \qw       & \qw                                    & \qw       & \qw & \cdots & & \qw       & \qw                           & \qw       & \ctrl{3} & \ctrl{5} & \qw & \cdots & & \ctrl{10} & \ctrl{12} & \qw      & \qw      & \cdots & & \qw                                   & \qw & \cdots & & \qw       & \qw & \\
\lstick{\ket{0}} & \qw                              & \targ     & \ctrl{1}                  & \targ     & \targ     & \ctrl{1}                               & \targ     & \qw &        & & \targ     & \ctrl{1}                      & \targ     & \qw      & \qw      & \qw &        & & \qw       & \qw       & \qw      & \qw      &        & & \qw                                   & \qw &        & & \qw       & \qw & \\
				 & {\raisebox{0.15cm}{$/^D$}} \qw   & \qw       & \gate{X_1}                & \qw       & \qw       & \gate{X_{\Lambda+1}}                   & \qw       & \qw &        & & \qw       & \gate{X_{N-\Lambda+1}}        & \qw       & \qswap   & \qw      & \qw &        & & \qw       & \qw       & \qswap   & \qw      &        & & \qw                                   & \qw &        & & \qswap    & \qw & \\ 
                 & {\raisebox{0.15cm}{$/^D$}} \qw   & \qw       & \gate{X_2} \qwx           & \qw       & \qw       & \gate{X_{\Lambda+2}} \qwx              & \qw       & \qw &        & & \qw       & \gate{X_{N-\Lambda+2}} \qwx   & \qw       & \qswap   & \qw      & \qw &        & & \qw       & \qw       & \qw      & \qw      &        & & \qw                                   & \qw &        & & \qw       & \qw & \\ 
                 & {\raisebox{0.15cm}{$/^D$}} \qw   & \qw       & \gate{X_3} \qwx           & \qw       & \qw       & \gate{X_{\Lambda+3}} \qwx              & \qw       & \qw &        & & \qw       & \gate{X_{N-\Lambda+3}} \qwx   & \qw       & \qw      & \qswap   & \qw &        & & \qw       & \qw       & \qswap   & \qw      &        & & \qw                                   & \qw &        & & \qw       & \qw & \\ 
                 & {\raisebox{0.15cm}{$/^D$}} \qw   & \qw       & \gate{X_4} \qwx           & \qw       & \qw       & \gate{X_{\Lambda+4}} \qwx              & \qw       & \qw &        & & \qw       & \gate{X_{N-\Lambda+4}} \qwx   & \qw       & \qw      & \qswap   & \qw &        & & \qw       & \qw       & \qw      & \qw      &        & & \qw                                   & \qw &        & & \qw       & \qw & \\ 
                 &                                  & \vdots    & \qwx                      &           &           & \qwx                                   &           &     &        & &           & \qwx                          &           &          &          &     &        & &           &           &          &          &        & &                                       &     &        & &           &     & \\
                 & {\raisebox{0.15cm}{$/^D$}} \qw   & \qw       & \gate{X_{\Lambda/2}} \qwx & \qw       & \qw       & \gate{X_{3\Lambda/2}} \qwx             & \qw       & \qw & \cdots & & \qw       & \gate{X_{(N-\Lambda)/2}} \qwx & \qw       & \qw      & \qw      & \qw & \cdots & & \qw       & \qw       & \qw      & \qw      & \cdots & & \qw                                   & \qw & \cdots & & \qswap    & \qw & \\ 
                 &                                  & \vdots    & \qwx                      &           &           & \qwx                                   &           &     &        & &           & \qwx                          &           &          &          &     &        & &           &           &          &          &        & &                                       &     &        & &           &     & \\
                 & {\raisebox{0.15cm}{$/^D$}} \qw   & \qw       & \gate{X_{\Lambda-4}} \qwx & \qw       & \qw       & \gate{X_{2\Lambda-4}} \qwx             & \qw       & \qw &        & & \qw       & \gate{X_{N-4}} \qwx           & \qw       & \qw      & \qw      & \qw &        & & \qswap    & \qw       & \qw      & \qw      &        & & \qswap                                & \qw &        & & \qw       & \qw & \\ 
                 & {\raisebox{0.15cm}{$/^D$}} \qw   & \qw       & \gate{X_{\Lambda-3}} \qwx & \qw       & \qw       & \gate{X_{2\Lambda-3}} \qwx             & \qw       & \qw &        & & \qw       & \gate{X_{N-3}} \qwx           & \qw       & \qw      & \qw      & \qw &        & & \qswap    & \qw       & \qw      & \qw      &        & & \qw                                   & \qw &        & & \qw       & \qw & \\ 
                 & {\raisebox{0.15cm}{$/^D$}} \qw   & \qw       & \gate{X_{\Lambda-1}} \qwx & \qw       & \qw       & \gate{X_{2\Lambda-1}} \qwx             & \qw       & \qw &        & & \qw       & \gate{X_{N-1}} \qwx           & \qw       & \qw      & \qw      & \qw &        & & \qw       & \qswap    & \qw      & \qw      &        & & \qswap                                & \qw &        & & \qw       & \qw & \\ 
                 & {\raisebox{0.15cm}{$/^D$}} \qw   & \qw       & \gate{X_\Lambda}     \qwx & \qw       & \qw       & \gate{X_{2\Lambda}}   \qwx             & \qw       & \qw &        & & \qw       & \gate{X_{N}}   \qwx           & \qw       & \qw      & \qw      & \qw &        & & \qw       & \qswap    & \qw      & \qw      &        & & \qw                                   & \qw &        & & \qw       & \qw & \\                                                                                                                                                 
{\inputgroupv{2}{9}{0.8em}{1.0em}{\raisebox{-2.0cm}{$n$}}}
{\inputgroupv{11}{21}{0.8em}{1.0em}{\raisebox{-6.7cm}{$\Lambda$}}}
{\gategroup{2}{3}{21}{14}{1.7em}{^\}}}
{\gategroup{2}{16}{21}{30}{1.9em}{^\}}}
}}
}
\caption{\label{fig:select-swap-simple}Select-Swap implementation with variable circuit depth and width. The $n=s+\lambda$ qubit control register is divided into $s$ qubits for the select control and $\lambda$ qubits for the swap control. The select circuit requires $2^s$ fanout CNOT gates as well as $2^{s+1}$ multi-controlled Toffoli gates that can be optimally decomposed using the unary iteration procedure, which requires $s$ additional ancillas (not shown) \cite{Babbush2018unary}. The swap circuit requires $\Lambda-1$ controlled-swap gates between $D$-qubit registers.}
\end{figure}

The select-swap circuit implementation \cite{low2018trading} is shown in Fig.~\ref{fig:select-swap-simple}. The first portion, labelled ``Select'', utilizes a unary iteration \cite{Babbush2018unary} to select from $\Lambda$ out of $N$ possible classical data-loading operations of $D$-qubit registers. For fixed-precision, this requires a fanout CNOT gate to load the classical data, which is Clifford and can be implemented efficiently in surface-code architectures \cite{Litinski2018latticesurgery}. For pre-rotated gates, this presents an issue, as we must control on both the select control register as well as the flag register, requiring one to always use $\bigo{ND}$ Toffoli gates. For this reason, we only use the pre-rotated gates in the minimal-depth cases.

The ``Swap'' portion uses the remaining qubits in the control register to swap the desired state into the top register. This circuit realizes the data-loading with garbage query 
$$
\text{LOAD}_\text{ss}\left(\sum_{j=0}^{2^n-1}\alpha_j \ket{j}_n\ket{0}_D\ket{0}_{(\Lambda-1)D}\right) = \sum_{j=0}^{2^n-1}\alpha_j \ket{j}_n\ket{b_j}_D\ket{g_j}_{(\Lambda-1)D},
$$
as discussed in Eq. \eqref{eq:qram_query_garbage} in the main text. The select-swap architecture enables one to achieve an optimal $T$-count of $\bigo{N}$ by taking $\Lambda = \bigo{\sqrt{N}}$ \cite{low2018trading}.

%%%%%%%%%%%% START FIGURE %%%%%%%%%%%%%%%%%%

%%%%%%%%%%%% END FIGURE %%%%%%%%%%%%%%%%%%

%%%%%%%%%%%%%%%%%%%%%%%%%%%%%%%%%%%%%%%%%%%%%%%%%%%%%%%%%%%%%%%%%%%%%%%%%%%%%%%%

\subsection{Bucket-brigade}\label{app:bucket_brigade_qram}

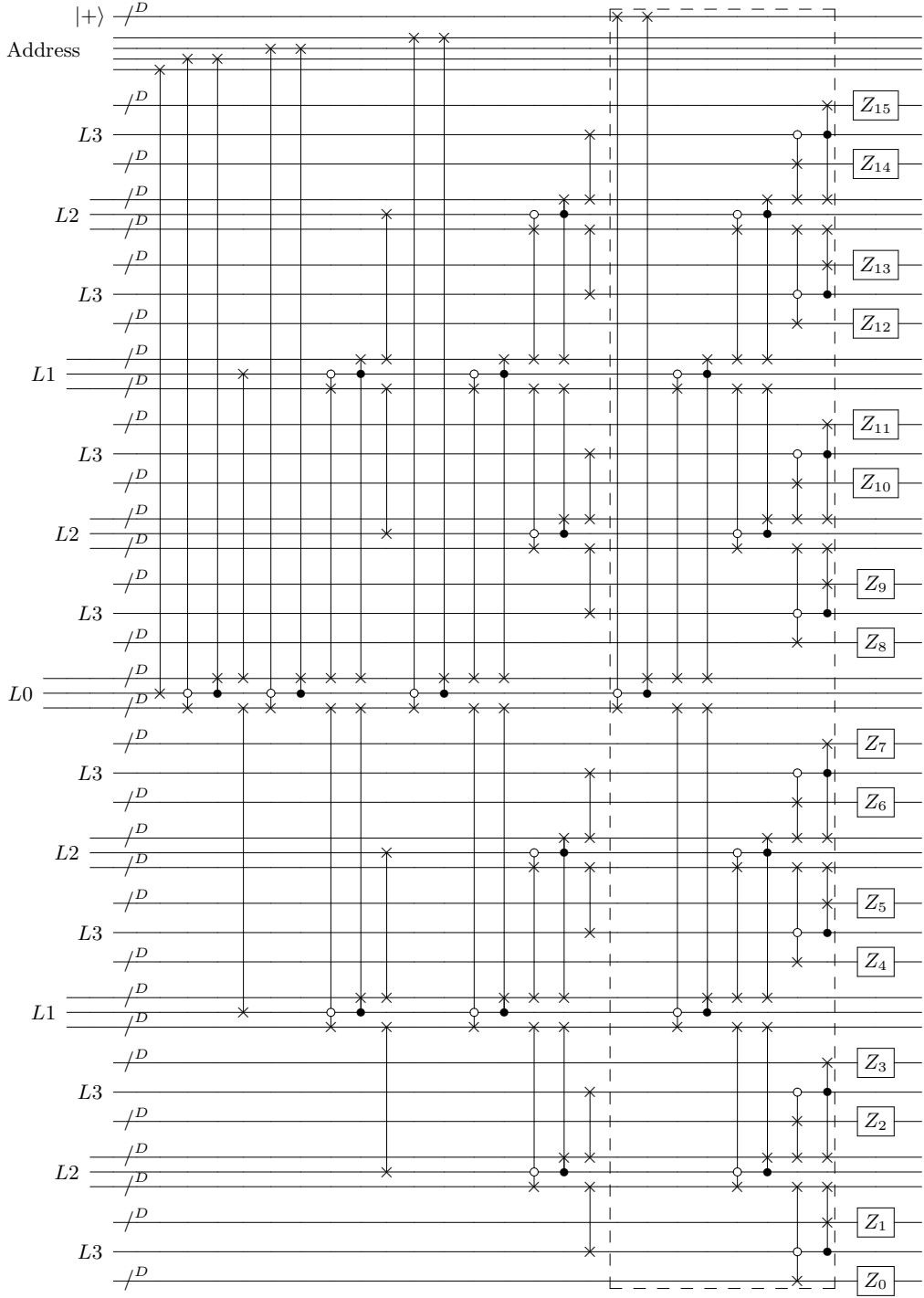
\begin{figure}
\centering
\scalebox{0.95}{
\mbox{
\Qcircuit @C=1.1em @R=0.5em {
                          &     & \ket{+} &     & {\raisebox{0.15cm}{$/^D$}} \qw & \qw             & \qw             & \qw             & \qw             & \qw             & \qw             & \qw             & \qw             & \qw             & \qw             & \qw              & \qw             & \qw             & \qw             & \qw             & \qw             & \qswap \qwx[36] & \qswap \qwx[35] & \qw             & \qw             & \qw             & \qw             & \qw             & \qw             & \qw           & \qw & \\
                          &     &         &     &                                &                 &                 &                 &                 &                 &                 &                 &                 &                 &                 &                  &                 &                 &                 &                 &                 &                 &                 &                 &                 &                 &                 &                 &                 &               &     & \\
                          &     &         &     & \qw                            & \qw             & \qw             & \qw             & \qw             & \qw             & \qw             & \qw             & \qw             & \qw             & \qswap \qwx[34] & \qswap \qwx[33]  & \qw             & \qw             & \qw             & \qw             & \qw             & \qw             & \qw             & \qw             & \qw             & \qw             & \qw             & \qw             & \qw             & \qw           & \qw & \\
                          &     &         &     & \qw                            & \qw             & \qw             & \qw             & \qw             & \qswap \qwx[33] & \qswap \qwx[32] & \qw             & \qw             & \qw             & \qw             & \qw              & \qw             & \qw             & \qw             & \qw             & \qw             & \qw             & \qw             & \qw             & \qw             & \qw             & \qw             & \qw             & \qw             & \qw           & \qw & \\
\raisebox{0.3cm}{Address} &     &         &     & \qw                            & \qw             & \qswap \qwx[32] & \qswap \qwx[32] & \qw             & \qw             & \qw             & \qw             & \qw             & \qw             & \qw             & \qw              & \qw             & \qw             & \qw             & \qw             & \qw             & \qw             & \qw             & \qw             & \qw             & \qw             & \qw             & \qw             & \qw             & \qw           & \qw & \\
                          &     &         &     & \qw                            & \qswap \qwx[31] & \qw             & \qw             & \qw             & \qw             & \qw             & \qw             & \qw             & \qw             & \qw             & \qw              & \qw             & \qw             & \qw             & \qw             & \qw             & \qw             & \qw             & \qw             & \qw             & \qw             & \qw             & \qw             & \qw             & \qw           & \qw & \\
                          &     &         &     &                                &                 &                 &                 &                 &                 &                 &                 &                 &                 &                 &                  &                 &                 &                 &                 &                 &                 &                 &                 &                 &                 &                 &                 &                 &               &     & \\
                          &     &         &     & {\raisebox{0.15cm}{$/^D$}} \qw & \qw             & \qw             & \qw             & \qw             & \qw             & \qw             & \qw             & \qw             & \qw             & \qw             & \qw              & \qw             & \qw             & \qw             & \qw             & \qw             & \qw             & \qw             & \qw             & \qw             & \qw             & \qw             & \qw             & \qswap \qwx[1]  & \gate{Z_{15}} & \qw & \\
                          &     & L3      &     & \qw                            & \qw             & \qw             & \qw             & \qw             & \qw             & \qw             & \qw             & \qw             & \qw             & \qw             & \qw              & \qw             & \qw             & \qw             & \qw             & \qswap \qwx[3]  & \qw             & \qw             & \qw             & \qw             & \qw             & \qw             & \ctrlo{1}       & \ctrl{3}        & \qw           & \qw & \\
                          &     &         &     & {\raisebox{0.15cm}{$/^D$}} \qw & \qw             & \qw             & \qw             & \qw             & \qw             & \qw             & \qw             & \qw             & \qw             & \qw             & \qw              & \qw             & \qw             & \qw             & \qw             & \qw             & \qw             & \qw             & \qw             & \qw             & \qw             & \qw             & \qswap \qwx[2]  & \qw             & \gate{Z_{14}} & \qw & \\
                          &     &         &     &                                &                 &                 &                 &                 &                 &                 &                 &                 &                 &                 &                  &                 &                 &                 &                 &                 &                 &                 &                 &                 &                 &                 &                 &                 &               &     & \\
                          &     &         & \qw & {\raisebox{0.15cm}{$/^D$}} \qw & \qw             & \qw             & \qw             & \qw             & \qw             & \qw             & \qw             & \qw             & \qw             & \qw             & \qw              & \qw             & \qw             & \qw             & \qswap \qwx[8]  & \qswap          & \qw             & \qw             & \qw             & \qw             & \qw             & \qswap \qwx[8]  & \qswap          & \qswap          & \qw           & \qw & \\
                          & L2  &         & \qw & \qw                            & \qw             & \qw             & \qw             & \qw             & \qw             & \qw             & \qw             & \qw             & \qswap \qwx[7]  & \qw             & \qw              & \qw             & \qw             & \ctrlo{1}       & \ctrl{-1}       & \qw             & \qw             & \qw             & \qw             & \qw             & \ctrlo{1}       & \ctrl{-1}       & \qw             & \qw             & \qw           & \qw & \\
                          &     &         & \qw & {\raisebox{0.15cm}{$/^D$}} \qw & \qw             & \qw             & \qw             & \qw             & \qw             & \qw             & \qw             & \qw             & \qw             & \qw             & \qw              & \qw             & \qw             & \qswap \qwx[6]  & \qw             & \qswap \qwx[3]  & \qw             & \qw             & \qw             & \qw             & \qswap \qwx[6]  & \qw             & \qswap \qwx[3]  & \qswap \qwx[2]  & \qw           & \qw & \\
                          &     &         &     &                                &                 &                 &                 &                 &                 &                 &                 &                 &                 &                 &                  &                 &                 &                 &                 &                 &                 &                 &                 &                 &                 &                 &                 &                 &               &     & \\
                          &     &         &     & {\raisebox{0.15cm}{$/^D$}} \qw & \qw             & \qw             & \qw             & \qw             & \qw             & \qw             & \qw             & \qw             & \qw             & \qw             & \qw              & \qw             & \qw             & \qw             & \qw             & \qw             & \qw             & \qw             & \qw             & \qw             & \qw             & \qw             & \qw             & \qswap          & \gate{Z_{13}} & \qw & \\
                          &     & L3      &     & \qw                            & \qw             & \qw             & \qw             & \qw             & \qw             & \qw             & \qw             & \qw             & \qw             & \qw             & \qw              & \qw             & \qw             & \qw             & \qw             & \qswap          & \qw             & \qw             & \qw             & \qw             & \qw             & \qw             & \ctrlo{1}       & \ctrl{-1}       & \qw           & \qw & \\
                          &     &         &     & {\raisebox{0.15cm}{$/^D$}} \qw & \qw             & \qw             & \qw             & \qw             & \qw             & \qw             & \qw             & \qw             & \qw             & \qw             & \qw              & \qw             & \qw             & \qw             & \qw             & \qw             & \qw             & \qw             & \qw             & \qw             & \qw             & \qw             & \qswap          & \qw             & \gate{Z_{12}} & \qw & \\
                          &     &         &     &                                &                 &                 &                 &                 &                 &                 &                 &                 &                 &                 &                  &                 &                 &                 &                 &                 &                 &                 &                 &                 &                 &                 &                 &                 &               &     & \\
                          &     & \qw     & \qw & {\raisebox{0.15cm}{$/^D$}} \qw & \qw             & \qw             & \qw             & \qw             & \qw             & \qw             & \qw             & \qswap \qwx[1]  & \qswap          & \qw             & \qw              & \qw             & \qswap \qwx[1]  & \qswap          & \qswap          & \qw             & \qw             & \qw             & \qw             & \qswap \qwx[1]  & \qswap          & \qswap          & \qw             & \qw             & \qw           & \qw & \\
L1                        &     & \qw     & \qw & \qw                            & \qw             & \qw             & \qw             & \qswap \qwx[15] & \qw             & \qw             & \ctrlo{1}       & \ctrl{15}       & \qw             & \qw             & \qw              & \ctrlo{1}       & \ctrl{15}       & \qw             & \qw             & \qw             & \qw             & \qw             & \ctrlo{1}       & \ctrl{15}       & \qw             & \qw             & \qw             & \qw             & \qw           & \qw & \\
                          &     & \qw     & \qw & {\raisebox{0.15cm}{$/^D$}} \qw & \qw             & \qw             & \qw             & \qw             & \qw             & \qw             & \qswap \qwx[14] & \qw             & \qswap \qwx[7]  & \qw             & \qw              & \qswap \qwx[14] & \qw             & \qswap \qwx[7]  & \qswap \qwx[7]  & \qw             & \qw             & \qw             & \qswap \qwx[14] & \qw             & \qswap \qwx[7]  & \qswap \qwx[7]  & \qw             & \qw             & \qw           & \qw & \\
                          &     &         &     &                                &                 &                 &                 &                 &                 &                 &                 &                 &                 &                 &                  &                 &                 &                 &                 &                 &                 &                 &                 &                 &                 &                 &                 &                 &               &     & \\
                          &     &         &     & {\raisebox{0.15cm}{$/^D$}} \qw & \qw             & \qw             & \qw             & \qw             & \qw             & \qw             & \qw             & \qw             & \qw             & \qw             & \qw              & \qw             & \qw             & \qw             & \qw             & \qw             & \qw             & \qw             & \qw             & \qw             & \qw             & \qw             & \qw             & \qswap \qwx[1]  & \gate{Z_{11}} & \qw & \\
                          &     & L3      &     & \qw                            & \qw             & \qw             & \qw             & \qw             & \qw             & \qw             & \qw             & \qw             & \qw             & \qw             & \qw              & \qw             & \qw             & \qw             & \qw             & \qswap \qwx[3]  & \qw             & \qw             & \qw             & \qw             & \qw             & \qw             & \ctrlo{1}       & \ctrl{3}        & \qw           & \qw & \\
                          &     &         &     & {\raisebox{0.15cm}{$/^D$}} \qw & \qw             & \qw             & \qw             & \qw             & \qw             & \qw             & \qw             & \qw             & \qw             & \qw             & \qw              & \qw             & \qw             & \qw             & \qw             & \qw             & \qw             & \qw             & \qw             & \qw             & \qw             & \qw             & \qswap \qwx[2]  & \qw             & \gate{Z_{10}} & \qw & \\
                          &     &         &     &                                &                 &                 &                 &                 &                 &                 &                 &                 &                 &                 &                  &                 &                 &                 &                 &                 &                 &                 &                 &                 &                 &                 &                 &                 &               &     & \\
                          &     &         & \qw & {\raisebox{0.15cm}{$/^D$}} \qw & \qw             & \qw             & \qw             & \qw             & \qw             & \qw             & \qw             & \qw             & \qw             & \qw             & \qw              & \qw             & \qw             & \qw             & \qswap          & \qswap          & \qw             & \qw             & \qw             & \qw             & \qw             & \qswap          & \qswap          & \qswap          & \qw           & \qw & \\
                          & L2  &         & \qw & \qw                            & \qw             & \qw             & \qw             & \qw             & \qw             & \qw             & \qw             & \qw             & \qswap          & \qw             & \qw              & \qw             & \qw             & \ctrlo{1}       & \ctrl{-1}       & \qw             & \qw             & \qw             & \qw             & \qw             & \ctrlo{1}       & \ctrl{-1}       & \qw             & \qw             & \qw           & \qw & \\
                          &     &         & \qw & {\raisebox{0.15cm}{$/^D$}} \qw & \qw             & \qw             & \qw             & \qw             & \qw             & \qw             & \qw             & \qw             & \qw             & \qw             & \qw              & \qw             & \qw             & \qswap          & \qw             & \qswap \qwx[3]  & \qw             & \qw             & \qw             & \qw             & \qswap          & \qw             & \qswap \qwx[3]  & \qswap \qwx[2]  & \qw           & \qw & \\
                          &     &         &     &                                &                 &                 &                 &                 &                 &                 &                 &                 &                 &                 &                  &                 &                 &                 &                 &                 &                 &                 &                 &                 &                 &                 &                 &                 &               &     & \\
                          &     &         &     & {\raisebox{0.15cm}{$/^D$}} \qw & \qw             & \qw             & \qw             & \qw             & \qw             & \qw             & \qw             & \qw             & \qw             & \qw             & \qw              & \qw             & \qw             & \qw             & \qw             & \qw             & \qw             & \qw             & \qw             & \qw             & \qw             & \qw             & \qw             & \qswap          & \gate{Z_{9}}  & \qw & \\
                          &     & L3      &     & \qw                            & \qw             & \qw             & \qw             & \qw             & \qw             & \qw             & \qw             & \qw             & \qw             & \qw             & \qw              & \qw             & \qw             & \qw             & \qw             & \qswap          & \qw             & \qw             & \qw             & \qw             & \qw             & \qw             & \ctrlo{1}       & \ctrl{-1}       & \qw           & \qw & \\
                          &     &         &     & {\raisebox{0.15cm}{$/^D$}} \qw & \qw             & \qw             & \qw             & \qw             & \qw             & \qw             & \qw             & \qw             & \qw             & \qw             & \qw              & \qw             & \qw             & \qw             & \qw             & \qw             & \qw             & \qw             & \qw             & \qw             & \qw             & \qw             & \qswap          & \qw             & \gate{Z_{8}}  & \qw & \\
                          &     &         &     &                                &                 &                 &                 &                 &                 &                 &                 &                 &                 &                 &                  &                 &                 &                 &                 &                 &                 &                 &                 &                 &                 &                 &                 &                 &               &     & \\
                          & \qw & \qw     & \qw & {\raisebox{0.15cm}{$/^D$}} \qw & \qw             & \qw             & \qswap          & \qswap          & \qw             & \qswap          & \qswap          & \qswap          & \qw             & \qw             & \qswap           & \qswap          & \qswap          & \qw             & \qw             & \qw             & \qw             & \qswap          & \qswap          & \qswap          & \qw             & \qw             & \qw             & \qw             & \qw           & \qw & \\
\lstick{L0}               & \qw & \qw     & \qw & \qw                            & \qswap          & \ctrlo{1}       & \ctrl{-1}       & \qw             & \ctrlo{1}       & \ctrl{-1}       & \qw             & \qw             & \qw             & \ctrlo{1}       & \ctrl{-1}        & \qw             & \qw             & \qw             & \qw             & \qw             & \ctrlo{1}       & \ctrl{-1}       & \qw             & \qw             & \qw             & \qw             & \qw             & \qw             & \qw           & \qw & \\
                          & \qw & \qw     & \qw & {\raisebox{0.15cm}{$/^D$}} \qw & \qw             & \qswap          & \qw             & \qswap \qwx[15] & \qswap          & \qw             & \qswap \qwx[15] & \qswap \qwx[15] & \qw             & \qswap          & \qw              & \qswap \qwx[15] & \qswap \qwx[15] & \qw             & \qw             & \qw             & \qswap          & \qw             & \qswap \qwx[15] & \qswap \qwx[15] & \qw             & \qw             & \qw             & \qw             & \qw           & \qw & \\
                          &     &         &     &                                &                 &                 &                 &                 &                 &                 &                 &                 &                 &                 &                  &                 &                 &                 &                 &                 &                 &                 &                 &                 &                 &                 &                 &                 &               &     & \\
                          &     &         &     & {\raisebox{0.15cm}{$/^D$}} \qw & \qw             & \qw             & \qw             & \qw             & \qw             & \qw             & \qw             & \qw             & \qw             & \qw             & \qw              & \qw             & \qw             & \qw             & \qw             & \qw             & \qw             & \qw             & \qw             & \qw             & \qw             & \qw             & \qw             & \qswap \qwx[1]  & \gate{Z_{7}}  & \qw & \\
                          &     & L3      &     & \qw                            & \qw             & \qw             & \qw             & \qw             & \qw             & \qw             & \qw             & \qw             & \qw             & \qw             & \qw              & \qw             & \qw             & \qw             & \qw             & \qswap \qwx[3]  & \qw             & \qw             & \qw             & \qw             & \qw             & \qw             & \ctrlo{1}       & \ctrl{3}        & \qw           & \qw & \\
                          &     &         &     & {\raisebox{0.15cm}{$/^D$}} \qw & \qw             & \qw             & \qw             & \qw             & \qw             & \qw             & \qw             & \qw             & \qw             & \qw             & \qw              & \qw             & \qw             & \qw             & \qw             & \qw             & \qw             & \qw             & \qw             & \qw             & \qw             & \qw             & \qswap \qwx[2]  & \qw             & \gate{Z_{6}}  & \qw & \\
                          &     &         &     &                                &                 &                 &                 &                 &                 &                 &                 &                 &                 &                 &                  &                 &                 &                 &                 &                 &                 &                 &                 &                 &                 &                 &                 &                 &               &     & \\
                          &     &         & \qw & {\raisebox{0.15cm}{$/^D$}} \qw & \qw             & \qw             & \qw             & \qw             & \qw             & \qw             & \qw             & \qw             & \qw             & \qw             & \qw              & \qw             & \qw             & \qw             & \qswap \qwx[8]  & \qswap          & \qw             & \qw             & \qw             & \qw             & \qw             & \qswap \qwx[8]  & \qswap          & \qswap          & \qw           & \qw & \\
                          & L2  &         & \qw & \qw                            & \qw             & \qw             & \qw             & \qw             & \qw             & \qw             & \qw             & \qw             & \qswap \qwx[7]  & \qw             & \qw              & \qw             & \qw             & \ctrlo{1}       & \ctrl{-1}       & \qw             & \qw             & \qw             & \qw             & \qw             & \ctrlo{1}       & \ctrl{-1}       & \qw             & \qw             & \qw           & \qw & \\
                          &     &         & \qw & {\raisebox{0.15cm}{$/^D$}} \qw & \qw             & \qw             & \qw             & \qw             & \qw             & \qw             & \qw             & \qw             & \qw             & \qw             & \qw              & \qw             & \qw             & \qswap \qwx[6]  & \qw             & \qswap \qwx[3]  & \qw             & \qw             & \qw             & \qw             & \qswap \qwx[6]  & \qw             & \qswap \qwx[3]  & \qswap \qwx[2]  & \qw           & \qw & \\
                          &     &         &     &                                &                 &                 &                 &                 &                 &                 &                 &                 &                 &                 &                  &                 &                 &                 &                 &                 &                 &                 &                 &                 &                 &                 &                 &                 &               &     & \\
                          &     &         &     & {\raisebox{0.15cm}{$/^D$}} \qw & \qw             & \qw             & \qw             & \qw             & \qw             & \qw             & \qw             & \qw             & \qw             & \qw             & \qw              & \qw             & \qw             & \qw             & \qw             & \qw             & \qw             & \qw             & \qw             & \qw             & \qw             & \qw             & \qw             & \qswap          & \gate{Z_{5}}  & \qw & \\
                          &     & L3      &     & \qw                            & \qw             & \qw             & \qw             & \qw             & \qw             & \qw             & \qw             & \qw             & \qw             & \qw             & \qw              & \qw             & \qw             & \qw             & \qw             & \qswap          & \qw             & \qw             & \qw             & \qw             & \qw             & \qw             & \ctrlo{1}       & \ctrl{-1}       & \qw           & \qw & \\
                          &     &         &     & {\raisebox{0.15cm}{$/^D$}} \qw & \qw             & \qw             & \qw             & \qw             & \qw             & \qw             & \qw             & \qw             & \qw             & \qw             & \qw              & \qw             & \qw             & \qw             & \qw             & \qw             & \qw             & \qw             & \qw             & \qw             & \qw             & \qw             & \qswap          & \qw             & \gate{Z_{4}}  & \qw & \\
                          &     &         &     &                                &                 &                 &                 &                 &                 &                 &                 &                 &                 &                 &                  &                 &                 &                 &                 &                 &                 &                 &                 &                 &                 &                 &                 &                 &               &     & \\
                          &     & \qw     & \qw & {\raisebox{0.15cm}{$/^D$}} \qw & \qw             & \qw             & \qw             & \qw             & \qw             & \qw             & \qw             & \qswap          & \qswap          & \qw             & \qw              & \qw             & \qswap          & \qswap          & \qswap          & \qw             & \qw             & \qw             & \qw             & \qswap          & \qswap          & \qswap          & \qw             & \qw             & \qw           & \qw & \\
L1                        &     & \qw     & \qw & \qw                            & \qw             & \qw             & \qw             & \qswap          & \qw             & \qw             & \ctrlo{1}       & \ctrl{-1}       & \qw             & \qw             & \qw              & \ctrlo{1}       & \ctrl{-1}       & \qw             & \qw             & \qw             & \qw             & \qw             & \ctrlo{1}       & \ctrl{-1}       & \qw             & \qw             & \qw             & \qw             & \qw           & \qw & \\
                          &     & \qw     & \qw & {\raisebox{0.15cm}{$/^D$}} \qw & \qw             & \qw             & \qw             & \qw             & \qw             & \qw             & \qswap          & \qw             & \qswap \qwx[7]  & \qw             & \qw              & \qswap          & \qw             & \qswap \qwx[7]  & \qswap \qwx[7]  & \qw             & \qw             & \qw             & \qswap          & \qw             & \qswap \qwx[7]  & \qswap \qwx[7]  & \qw             & \qw             & \qw           & \qw & \\
                          &     &         &     &                                &                 &                 &                 &                 &                 &                 &                 &                 &                 &                 &                  &                 &                 &                 &                 &                 &                 &                 &                 &                 &                 &                 &                 &                 &               &     & \\
                          &     &         &     & {\raisebox{0.15cm}{$/^D$}} \qw & \qw             & \qw             & \qw             & \qw             & \qw             & \qw             & \qw             & \qw             & \qw             & \qw             & \qw              & \qw             & \qw             & \qw             & \qw             & \qw             & \qw             & \qw             & \qw             & \qw             & \qw             & \qw             & \qw             & \qswap \qwx[1]  & \gate{Z_{3}}  & \qw & \\
                          &     & L3      &     & \qw                            & \qw             & \qw             & \qw             & \qw             & \qw             & \qw             & \qw             & \qw             & \qw             & \qw             & \qw              & \qw             & \qw             & \qw             & \qw             & \qswap \qwx[3]  & \qw             & \qw             & \qw             & \qw             & \qw             & \qw             & \ctrlo{1}       & \ctrl{3}        & \qw           & \qw & \\
                          &     &         &     & {\raisebox{0.15cm}{$/^D$}} \qw & \qw             & \qw             & \qw             & \qw             & \qw             & \qw             & \qw             & \qw             & \qw             & \qw             & \qw              & \qw             & \qw             & \qw             & \qw             & \qw             & \qw             & \qw             & \qw             & \qw             & \qw             & \qw             & \qswap \qwx[2]  & \qw             & \gate{Z_{2}}  & \qw & \\
                          &     &         &     &                                &                 &                 &                 &                 &                 &                 &                 &                 &                 &                 &                  &                 &                 &                 &                 &                 &                 &                 &                 &                 &                 &                 &                 &                 &               &     & \\
                          &     &         & \qw & {\raisebox{0.15cm}{$/^D$}} \qw & \qw             & \qw             & \qw             & \qw             & \qw             & \qw             & \qw             & \qw             & \qw             & \qw             & \qw              & \qw             & \qw             & \qw             & \qswap          & \qswap          & \qw             & \qw             & \qw             & \qw             & \qw             & \qswap          & \qswap          & \qswap          & \qw           & \qw & \\
                          & L2  &         & \qw & \qw                            & \qw             & \qw             & \qw             & \qw             & \qw             & \qw             & \qw             & \qw             & \qswap          & \qw             & \qw              & \qw             & \qw             & \ctrlo{1}       & \ctrl{-1}       & \qw             & \qw             & \qw             & \qw             & \qw             & \ctrlo{1}       & \ctrl{-1}       & \qw             & \qw             & \qw           & \qw & \\
                          &     &         & \qw & {\raisebox{0.15cm}{$/^D$}} \qw & \qw             & \qw             & \qw             & \qw             & \qw             & \qw             & \qw             & \qw             & \qw             & \qw             & \qw              & \qw             & \qw             & \qswap          & \qw             & \qswap \qwx[3]  & \qw             & \qw             & \qw             & \qw             & \qswap          & \qw             & \qswap \qwx[3]  & \qswap \qwx[2]  & \qw           & \qw & \\
                          &     &         &     &                                &                 &                 &                 &                 &                 &                 &                 &                 &                 &                 &                  &                 &                 &                 &                 &                 &                 &                 &                 &                 &                 &                 &                 &                 &               &     & \\
                          &     &         &     & {\raisebox{0.15cm}{$/^D$}} \qw & \qw             & \qw             & \qw             & \qw             & \qw             & \qw             & \qw             & \qw             & \qw             & \qw             & \qw              & \qw             & \qw             & \qw             & \qw             & \qw             & \qw             & \qw             & \qw             & \qw             & \qw             & \qw             & \qw             & \qswap          & \gate{Z_{1}}  & \qw & \\
                          &     & L3      &     & \qw                            & \qw             & \qw             & \qw             & \qw             & \qw             & \qw             & \qw             & \qw             & \qw             & \qw             & \qw              & \qw             & \qw             & \qw             & \qw             & \qswap          & \qw             & \qw             & \qw             & \qw             & \qw             & \qw             & \ctrlo{1}       & \ctrl{-1}       & \qw           & \qw & \\
                          &     &         &     & {\raisebox{0.15cm}{$/^D$}} \qw & \qw             & \qw             & \qw             & \qw             & \qw             & \qw             & \qw             & \qw             & \qw             & \qw             & \qw              & \qw             & \qw             & \qw             & \qw             & \qw             & \qw             & \qw             & \qw             & \qw             & \qw             & \qw             & \qswap          & \qw             & \gate{Z_{0}}  & \qw & \\
{\gategroup{1}{22}{66}{29}{.7em}{--}}                                            
}}
}
\caption{\label{fig:bucket-brigade}Bucket Brigade QRAM with an address space of four bits, shown here since it can help visualize the general case. Note that when single qubits are swapped into a $D$-qubit register (e.g., the first gate in the circuit), it is assumed that they occupy the first position. As discussed in Fig.~10 of \cite{hann2021}, additional parallelization is possible (but hard to depict in the drawing), which reduces the depth of a QRAM with $n$ address qubits from $\bigo{n^2}$ to $\bigo{n}$. An example of this effect is seen in the figure by noting that the region in the dashed box can be shifted left by three layers. The total $T$-count of the $n$-qubit version of this circuit is $16(D+1)(2^n-1)$ over $T$-depth of $48(n-1)$ assuming that controlled-swap gates are implemented with the phase-incorrect construction of Fig.~\ref{fig:cswap-phase}. The total number of qubits is $n + D + (2D+1)(2^n-1)-1$ (where we save a qubit by omitting the first swap gate and identifying the first address qubit directly with the $L0$ router setting).}
\end{figure}

The original formulation \cite{giovannetti2008qram} of the bucket-brigade idea assumed access to qutrits. Most modern quantum hardware implements qubits, so we analyze the qubit version of the bucket-brigade QRAM, as laid out and analyzed in \cite{hann2021}. While it is difficult to draw an $m$ address qubit version of the BB-QRAM, the four-qubit version is sufficient to understand how the circuit can be generalized. This circuit is shown in Fig.~\ref{fig:bucket-brigade}. The address qubits are routed from the address space into individual routers. The bottom qubit in the address space is routed to the level-0 or $L0$ router. The next qubit is routed to the $L1$ router in either the bottom or top branch, depending on whether the bottom address qubit is a $\ket{0}$ or $\ket{1}$, respectively. The rest of the address qubits follow in the same manner, being sent to the appropriate router depending upon the value of the preceding address qubits. Once all address qubits have been routed, the circuit swaps the bus register (top of figure) into the appropriate memory location where $Z$ (identity) gates are applied to qubits in the register if the corresponding classical bit is a 1 (0). The routing portion of the circuit is then reversed to swap the bus and address qubits back to their initial locations (not shown). This formulation assumes that the router qubits are initialized to the $\ket{0}$ state and the bus register in initialized to the $\ket{+}_D$ state. Similarly to the SS-QRAM, a version that can utilize routing qubits in an arbitrary initial state is given in \cite{hann2021} at the cost of additional queries. We do not consider that case here.

A naive counting of the circuit depth of the BB-QRAM leads to a depth complexity of $\bigo{m^2}$ controlled swaps. However, this complexity does not take into account additional parallelism that can be exploited. The dashed box in Fig. \ref{fig:bucket-brigade} can be shifted to the left such that the first two controlled swaps can be done in parallel with the eight controlled swaps being used to route qubits from $L1$ to $L2$. This parallelism persists as the address space increases and, as a result, each additional address qubit will only add a total depth of $6$ controlled swaps to the circuit, leading to a total depth of of $\bigo{m}$ controlled-swap gates for the circuit.

%%%%%%%%%%%% START FIGURE %%%%%%%%%%%%%%%%%%

%%%%%%%%%%%% END FIGURE %%%%%%%%%%%%%%%%%%

%%%%%%%%%%%%%%%%%%%%%%%%%%%%%%%%%%%%%%%%%%%%%%%%%%%%%%%%%%%%%%%%%%%%%%%%%%%%%%%%

%%%%%%%%%%%%%%%%%%%%%%%%%%%%%%%%%%%%%%%%%%%%%%%%%%%%%%%%%%%%%%%%%%%%%%%%%%%%%%%%

\section{Gate Decompositions}\label{app:decompositions}

For completeness we show the gate decompositions that we use for the resource counts, which can also be found in Refs. \cite{barenco1995, low2018trading, Amy2013}. For the fixed-precision state preparation, the phases of the swap gates are not important, whicht can provide some resource savings. We denote circuits that apply the correct bit transformations but add incorrect phases by $\simeq$, whereas exact decompositions have the $=$ sign. We consistently use the phase-incorrect versions of the swaps presented here for the fixed-precision resource estimates, which allows us to achieve the minimal $T$-count. We consistently use the phase-correct version of the swaps for the pre-rotated case, since the phase is important in that case. We use the $T\text{-depth }= 1$ version of the Toffoli \cite{selinger2013quantum, jones2013} for our resource counts. This version requires an additional ancilla, but it achieves minimal $T$-depth, which was the motivation behind the state preparation with flags procedure that we introduced in the main text.

%%%%%%%%%%%% START FIGURE %%%%%%%%%%%%%%%%%%

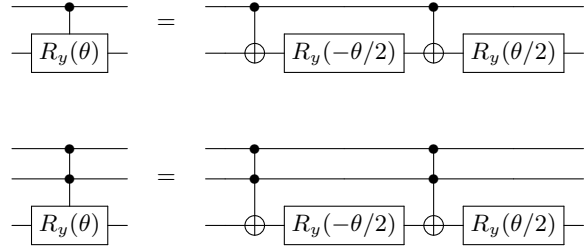
\begin{figure}[ht!]
\centering
\mbox{
\Qcircuit @C=0.8em @R=1.0em {
& \ctrl{1}           & \qw & &                       & & & \qw & \ctrl{1} & \qw                   & \ctrl{1} & \qw                  & \qw & \\
& \gate{R_y(\theta)} & \qw & & {\raisebox{0.8cm}{=}} & & & \qw & \targ    & \gate{R_y(-\theta/2)} & \targ    & \gate{R_y(\theta/2)} & \qw & \\
&                    &     & &                       & & &     &          &                       &          &                      &     & \\
&                    &     & &                       & & &     &          &                       &          &                      &     & \\
& \ctrl{1}           & \qw & &                       & & & \qw & \ctrl{1} & \qw                   & \ctrl{1} & \qw                  & \qw & \\
& \ctrl{1}           & \qw & & {\raisebox{0cm}{=}}   & & & \qw & \ctrl{1} & \qw                   & \ctrl{1} & \qw                  & \qw & \\
& \gate{R_y(\theta)} & \qw & &                       & & & \qw & \targ    & \gate{R_y(-\theta/2)} & \targ    & \gate{R_y(\theta/2)} & \qw & \\
}}
\caption{\label{fig:controlled-ry} Decomposition of controlled-$R_y$ rotation and controlled-controlled-$R_y$ rotation used for resource counts. Since $R_y(\theta)$ is equivalent to $R_y(\theta + 2\pi)$ up to a global phase, controlled-$R_y(\theta)$ is equivalent to controlled-$R_y(\theta+2\pi)$ up to a controlled-phase, i.e.~a $Z$ gate. Thus, for angles $\theta \in [\pi,2\pi]$, a $Z$ gate should additionally be applied to the first qubit in the top diagram, and a $CZ$ gate should be applied between the top two qubits of the bottom diagram.}
\end{figure}

%%%%%%%%%%%% END FIGURE %%%%%%%%%%%%%%%%%%

%%%%%%%%%%%% START FIGURE %%%%%%%%%%%%%%%%%%

\begin{figure}[ht!]
\centering
\mbox{
\Qcircuit @C=0.6em @R=0.8em {
                      & \qw    & \ctrl{6} & \qw       & \qw       & \qw       & \ctrl{6}  & \qw & \\
\lstick{\ket{\psi_1}} & \qw    & \qw      & \ctrl{1}  & \targ     & \ctrl{1}  & \qw       & \qw & \rstick{\ket{\phi_1}} \\
\lstick{\ket{\phi_1}} & \qw    & \qw      & \targ     & \ctrl{-1} & \targ     & \qw       & \qw & \rstick{\ket{\psi_1}} \\
\lstick{\ket{0}}      & \qw    & \targ    & \qw       & \ctrl{-1} & \qw       & \targ     & \qw & \rstick{\ket{0}}      \\ 
\lstick{\ket{\psi_2}} & \qw    & \qw      & \ctrl{1}  & \targ     & \ctrl{1}  & \qw       & \qw & \rstick{\ket{\phi_2}} \\ 
\lstick{\ket{\phi_2}} & \qw    & \qw      & \targ     & \ctrl{-1} & \targ     & \qw       & \qw & \rstick{\ket{\psi_2}} \\
\lstick{\ket{0}}      & \qw    & \targ    & \qw       & \ctrl{-1} & \qw       & \targ     & \qw & \rstick{\ket{0}}      \\
%{\inputgroupv{1}{4}{0.8em}{1.0em}{\raisebox{-2.5cm}{$n$}}}
%{\inputgroupv{5}{15}{0.8em}{1.0em}{\raisebox{-7.3cm}{$N$}}}
}}
\caption{\label{fig:cswap2}Phase-correct parallel controlled-swap between two qubit registers in arbitrary states with a single control use for pre-rotated resource estimates. Two clean ancillas are required (see Ref.~\cite{low2018trading} for how dirty ancillas can be used at the expense of two additional Toffoli gates). With an additional clean ancilla (not shown), the Toffoli gates can be constructed with a $T$-depth of 1 and a $T$-count of 4 \cite{selinger2013quantum, jones2013}. The output state is shown assuming the input control is $\ket{1}$. If the control is $\ket{0}$ the states are not swapped. To perform parallel Toffolis (instead of controlled-swaps), simply omit the CNOT gates in the second and fourth layers. }
\end{figure}
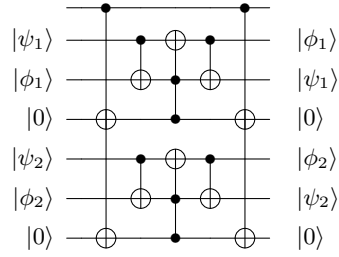

%%%%%%%%%%%% END FIGURE %%%%%%%%%%%%%%%%%%

%%%%%%%%%%%% START FIGURE %%%%%%%%%%%%%%%%%%

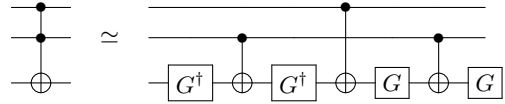
\begin{figure}[ht!]
\centering
\mbox{
\Qcircuit @C=0.8em @R=1.0em {
& \ctrl{1} & \qw & &        & & & \qw              & \qw      & \qw              & \ctrl{2} & \qw      & \qw      & \qw      & \\
& \ctrl{1} & \qw & & \simeq & & & \qw              & \ctrl{1} & \qw              & \qw      & \qw      & \ctrl{1} & \qw      & \\
& \targ    & \qw & &        & & & \gate{G^\dagger} & \targ    & \gate{G^\dagger} & \targ    & \gate{G} & \targ    & \gate{G} & \\
}}
\caption{\label{fig:Toffoli} Toffoli gate decomposed into a $T$-depth 4 circuit used for fixed-precision resource estimates. The gate $G=S^\dagger H T H S$. The decomposition is exact up to a phase that takes $\ket{100}\to -\ket{100}$.}
\end{figure}

%%%%%%%%%%%% END FIGURE %%%%%%%%%%%%%%%%%%

%%%%%%%%%%%% START FIGURE %%%%%%%%%%%%%%%%%%

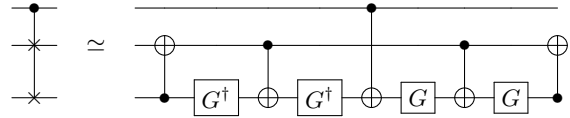
\begin{figure}[ht!]
\centering
\mbox{
\Qcircuit @C=0.8em @R=1.0em {
& \ctrl{1}    & \qw & &        & & & \qw       & \qw              & \qw      & \qw              & \ctrl{2} & \qw      & \qw      & \qw      & \qw       & \\
& \qswap      & \qw & & \simeq & & & \targ     & \qw              & \ctrl{1} & \qw              & \qw      & \qw      & \ctrl{1} & \qw      & \targ     & \\
& \qswap \qwx & \qw & &        & & & \ctrl{-1} & \gate{G^\dagger} & \targ    & \gate{G^\dagger} & \targ    & \gate{G} & \targ    & \gate{G} & \ctrl{-1} & \\
}}
\caption{\label{fig:cswap-phase} Controlled-swap gate decomposed into a $T$-depth and $T$-count 4 circuit used for fixed-precision resource estimates. The gate $G=S^\dagger H T H S$. The decomposition is exact up to a phase that takes $\ket{100}\to -\ket{100}$.}
\end{figure}

%%%%%%%%%%%% END FIGURE %%%%%%%%%%%%%%%%%%

%%%%%%%%%%%% START FIGURE %%%%%%%%%%%%%%%%%%

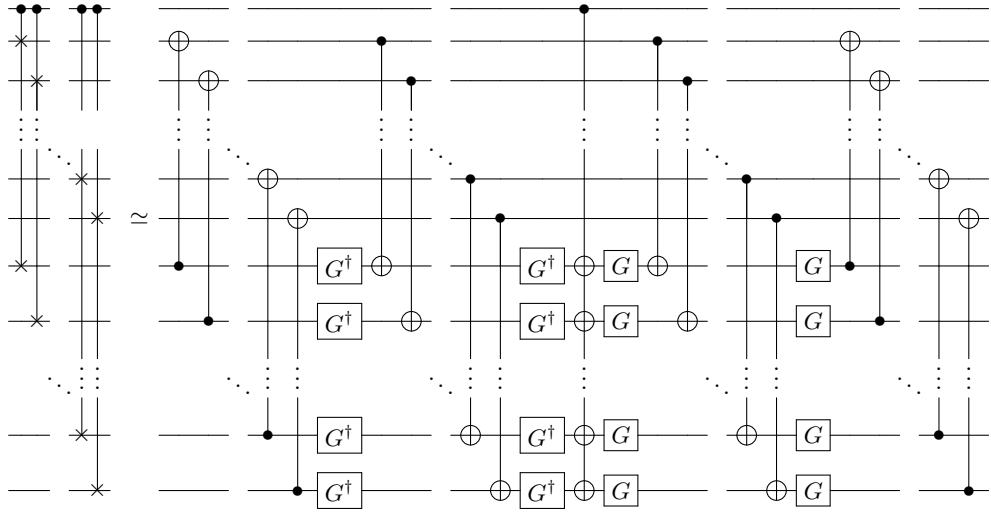
\begin{figure}[ht!]
\centering
\mbox{
\Qcircuit @C=0.4em @R=0.8em {
& \ctrl{3}   & \ctrl{3}    & \qw &        & & \ctrl{3}    & \ctrl{3}    & \qw & & &        & & & \qw       & \qw       & \qw &        & & \qw       & \qw       & \qw              & \qw        & \qw        & \qw &        & & \qw        & \qw        & \qw              & \ctrl{3}   & \qw      & \qw        & \qw        & \qw &        & & \qw        & \qw        & \qw      & \qw       & \qw       & \qw &        & & \qw       & \qw       & \qw & \\
& \qswap     & \qw         & \qw &        & & \qw         & \qw         & \qw & & &        & & & \targ     & \qw       & \qw &        & & \qw       & \qw       & \qw              & \ctrl{1}   & \qw        & \qw &        & & \qw        & \qw        & \qw              & \qw        & \qw      & \ctrl{1}   & \qw        & \qw &        & & \qw        & \qw        & \qw      & \targ     & \qw       & \qw &        & & \qw       & \qw       & \qw & \\
& \qw        & \qswap      & \qw &        & & \qw         & \qw         & \qw & & &        & & & \qw \qwx  & \targ     & \qw &        & & \qw       & \qw       & \qw              & \qw \qwx   & \ctrl{1}   & \qw &        & & \qw        & \qw        & \qw              & \qw        & \qw      & \qw \qwx   & \ctrl{1}   & \qw &        & & \qw        & \qw        & \qw      & \qw \qwx  & \targ     & \qw &        & & \qw       & \qw       & \qw & \\
&            & \qwx        &     &        & &             &             &     & & &        & & & \qwx      & \qwx      &     &        & &           &           &                  & \qwx       & \qwx       &     &        & &            &            &                  &            &          & \qwx       & \qwx       &     &        & &            &            &          & \qwx      & \qwx      &     &        & &           &           &     & \\
& \qvdots     & \qvdots      &     &        & &             &             &     & & &        & & & \qvdots    & \qvdots    &     &        & &           &           &                  & \qvdots     & \qvdots     &     &        & &            &            &                  & \qvdots     &          & \qvdots     & \qvdots     &     &        & &            &            &          & \qvdots    & \qvdots    &     &        & &           &           &     & \\
&            &             &     & \ddots & &             &             &     & & &        & & &           &           &     & \ddots & &           &           &                  &            &            &     & \ddots & &            &            &                  &            &          &            &            &     & \ddots & &            &            &          &           &           &     & \ddots & &           &           &     & \\
& \qw \qwx   & \qw \qwx    & \qw &        & & \qswap \qwx & \qw    \qwx & \qw & & &        & & & \qw \qwx  & \qw \qwx  & \qw &        & & \targ     & \qw       & \qw              & \qw \qwx   & \qw \qwx   & \qw &        & & \ctrl{1}   & \qw        & \qw              & \qw \qwx   & \qw      & \qw \qwx   & \qw \qwx   & \qw &        & & \ctrl{1}   & \qw        & \qw      & \qw \qwx  & \qw \qwx  & \qw &        & & \targ     & \qw       & \qw & \\
& \qw \qwx   & \qw \qwx    & \qw &        & & \qw \qwx    & \qswap \qwx & \qw & & & \simeq & & & \qw \qwx  & \qw \qwx  & \qw &        & & \qw \qwx  & \targ     & \qw              & \qw \qwx   & \qw \qwx   & \qw &        & & \qw \qwx   & \ctrl{1}   & \qw              & \qw \qwx   & \qw      & \qw \qwx   & \qw \qwx   & \qw &        & & \qw \qwx   & \ctrl{1}   & \qw      & \qw \qwx  & \qw \qwx  & \qw &        & & \qw \qwx  & \targ     & \qw & \\
& \qswap\qwx & \qw \qwx    & \qw &        & & \qw \qwx    & \qw \qwx    & \qw & & &        & & & \ctrl{-1} & \qw \qwx  & \qw &        & & \qw \qwx  & \qw \qwx  & \gate{G^\dagger} & \targ \qwx & \qw \qwx   & \qw &        & & \qw \qwx   & \qw \qwx   & \gate{G^\dagger} & \targ \qwx & \gate{G} & \targ \qwx & \qw \qwx   & \qw &        & & \qw \qwx   & \qw \qwx   & \gate{G} & \ctrl{-1} & \qw \qwx  & \qw &        & & \qw \qwx  & \qw \qwx  & \qw & \\
& \qw        & \qswap \qwx & \qw &        & & \qw \qwx    & \qw \qwx    & \qw & & &        & & & \qw       & \ctrl{-1} & \qw &        & & \qw \qwx  & \qw \qwx  & \gate{G^\dagger} & \qw        & \targ \qwx & \qw &        & & \qw \qwx   & \qw \qwx   & \gate{G^\dagger} & \targ \qwx & \gate{G} & \qw        & \targ \qwx & \qw &        & & \qw \qwx   & \qw \qwx   & \gate{G} & \qw       & \ctrl{-1} & \qw &        & & \qw \qwx  & \qw \qwx  & \qw & \\
&            &             &     &        & & \qwx        & \qwx        &     & & &        & & &           &           &     &        & & \qwx      & \qwx      &                  &            &            &     &        & & \qwx       & \qwx       &                  & \qwx       &          &            &            &     &        & & \qwx       & \qwx       &          &           &           &     &        & & \qwx      & \qwx      &     & \\
&            &             &     & \ddots & & \qvdots      & \qvdots      &     & & &        & & &           &           &     & \ddots & & \qvdots    & \qvdots    &                  &            &            &     & \ddots & & \qvdots     & \qvdots     &                  & \qvdots     &          &            &            &     & \ddots & & \qvdots    & \qvdots     &          &           &           &     & \ddots & & \qvdots    & \qvdots    &     & \\
&            &             &     &        & &             &             &     & & &        & & &           &           &     &        & &           &           &                  &            &            &     &        & &            &            &                  &            &          &            &            &     &        & &            &            &          &           &           &     &        & &           &           &     & \\
& \qw        & \qw         & \qw &        & & \qswap \qwx & \qw \qwx    & \qw & & &        & & & \qw       & \qw       & \qw &        & & \ctrl{-1} & \qw       & \gate{G^\dagger} & \qw        & \qw        & \qw &        & & \targ \qwx & \qw \qwx   & \gate{G^\dagger} & \targ \qwx & \gate{G} & \qw        & \qw        & \qw &        & & \targ \qwx & \qw \qwx   & \gate{G} & \qw       & \qw       & \qw &        & & \ctrl{-1} & \qw       & \qw & \\
& \qw        & \qw         & \qw &        & & \qw         & \qswap \qwx & \qw & & &        & & & \qw       & \qw       & \qw &        & & \qw       & \ctrl{-2} & \gate{G^\dagger} & \qw        & \qw        & \qw &        & & \qw        & \targ \qwx & \gate{G^\dagger} & \targ \qwx & \gate{G} & \qw        & \qw        & \qw &        & & \qw        & \targ \qwx & \gate{G} & \qw       & \qw       & \qw &        & & \qw       & \ctrl{-2} & \qw & \\
}}
\caption{\label{fig:multi_cswap} Controlled-swap between multi-qubit registers decomposed into a set of $T$ gates, a fanout CNOT, and $G=S^\dagger H T H S$ gates \cite{barenco1995, low2018trading}. The $G$ (or equivalently $T$) gates and CNOT gates can all be implemented in parallel. The decomposition on the right adds a phase to certain basis states, so it can only be used in situations where that phase is irrelevant \cite{barenco1995}. We only use this version for fixed-precision resource estimates. The entire circuit has a $T$-depth of 4. The $T$-count is $4t$ where $t$ is the size of the registers being swapped.}
\end{figure}

%%%%%%%%%%%% END FIGURE %%%%%%%%%%%%%%%%%%

\end{document}